\documentclass[floats,floatfix,showpacs,amssymb,prd,twocolumn,superscriptaddress,nofootinbib,nolongbibliography,reprint]{revtex4-1}

\usepackage{amssymb,amsmath,verbatim,mathtools,needspace,enumitem,etoolbox,graphicx,physics,microtype,afterpage,xspace,tabularx,lmodern,multirow,bm}
\usepackage{gensymb}
\usepackage{microtype} 
\usepackage[dvipsnames, usenames]{xcolor}
\definecolor{linkcolor}{rgb}{0.0,0.3,0.5}
\usepackage[unicode, colorlinks=true, linkcolor=linkcolor, citecolor=linkcolor, filecolor=linkcolor, urlcolor=linkcolor, linktocpage, breaklinks]{hyperref}
\usepackage[all]{hypcap}
\usepackage[T1]{fontenc}
\usepackage[utf8]{inputenc}
\usepackage[usenames,dvipsnames]{xcolor}
\usepackage{csquotes}
\usepackage{booktabs}
\hypersetup{colorlinks=true,citecolor=romared,linkcolor=romared,urlcolor=romared}

\definecolor{romared}{RGB}{142,0,28}

\newcommand{\be}{\begin{equation}}
\newcommand{\ee}{\end{equation}}

\def\be{\begin{equation}}
\def\ee{\end{equation}}
\newcommand{\beq}{\begin{eqnarray}}
\newcommand{\eeq}{\end{eqnarray}}
\usepackage{makecell}
\usepackage{soul}

\newcolumntype{Y}{>{\centering\arraybackslash}X}

\usepackage{ulem}

\usepackage{bm}

\let\Re\undefined
\let\Im\undefined
\DeclareMathSymbol{\Re}{\mathord}{symbols}{"3C}
\DeclareMathSymbol{\Im}{\mathord}{symbols}{"3D}

\makeatletter
\def\l@subsubsection#1#2{}
\makeatother

\begin{document}

\title{Relativistic effects in extreme-mass-ratio inspirals within scalar clouds:\\ Eccentric and inclined orbits}

\author{Qi-Xuan~Xu}
\email{qixuan.xu@tecnico.ulisboa.pt}
\affiliation{CENTRA, Departamento de Física, Instituto Superior Técnico – IST,\\
Universidade de Lisboa – UL, Avenida Rovisco Pais 1, 1049-001 Lisboa, Portugal}

\author{Richard~Brito}
\affiliation{CENTRA, Departamento de Física, Instituto Superior Técnico – IST,\\
Universidade de Lisboa – UL, Avenida Rovisco Pais 1, 1049-001 Lisboa, Portugal}

\author{Riccardo~Della~Monica}
\affiliation{CENTRA, Departamento de Física, Instituto Superior Técnico – IST,\\
Universidade de Lisboa – UL, Avenida Rovisco Pais 1, 1049-001 Lisboa, Portugal}

\author{Rodrigo~Vicente}
\affiliation{Gravitation Astroparticle Physics Amsterdam (GRAPPA),\\
University of Amsterdam, 1098 XH Amsterdam, The Netherlands}

\author{Chen~Yuan}
\affiliation{CENTRA, Departamento de Física, Instituto Superior Técnico – IST,\\
Universidade de Lisboa – UL, Avenida Rovisco Pais 1, 1049-001 Lisboa, Portugal}

\date{\today}

\begin{abstract}
We study extreme-mass-ratio inspirals (EMRIs) evolving in a scalar cloud environment that may form through superradiant instabilities, using a fully relativistic perturbative framework that allows for eccentric and inclined orbits. EMRIs, consisting of a stellar-mass compact object inspiraling into a supermassive black hole, are key sources for space-based gravitational-wave detectors such as LISA. Previous relativistic studies of EMRIs in scalar clouds have been restricted to circular, equatorial motion. Here, instead, we focus on a Schwarzschild black hole background to incorporate eccentricity and orbital inclination. By computing the scalar energy and angular momentum scattered off to spatial infinity and absorbed at the event horizon, we show that orbital eccentricity can induce a dense spectrum of resonances near the last stable orbit, associated with strong relativistic apsidal precession. We further find that orbital inclination can significantly modify the orbital energy and angular momentum losses. In particular, we identify a critical inclination angle below which, at sufficiently small orbital radii, there is a net transfer of energy from the scalar cloud to the orbit. Moreover, for sufficiently large eccentricities, resonances associated with relativistic apsidal precession persist across the full range of inclinations, although their structure changes significantly between prograde and retrograde orbits. These results provide a foundation for future studies of EMRIs in scalar cloud environments on fully generic orbits around spinning black holes.
\end{abstract}

\maketitle 

\tableofcontents

\newpage
\section{Introduction}\label{sec:Introduction}
Gravitational-wave (GW) observations have opened a direct channel for studying the dynamics of compact objects in the strong-field regime, establishing GW astronomy as a unique tool for astrophysics and fundamental physics~\cite{LIGOScientific:2026sit,LIGOScientific:2025brd,LIGOScientific:2025wao}. In the coming decades, this programme will be extended by next-generation ground-based detectors, such as the Einstein Telescope~\cite{ET:2025xjr} and Cosmic Explorer~\cite{Evans:2021gyd}, and by space-based interferometers such as LISA~\cite{LISA:2024hlh}, TianQin~\cite{TianQin:2015yph}, and Taiji~\cite{Luo:2021qji}. Space-based detectors will be sensitive to compact binaries in the milliHertz band, including intermediate- and extreme-mass-ratio inspirals (I/EMRIs)~\cite{LISA:2024hlh}. In an EMRI, a stellar-mass compact object slowly inspirals into a massive black hole (BH), remaining in the detector band for tens to hundreds of thousands of orbital cycles and probing regions close to the central object~\cite{LISAConsortiumWaveformWorkingGroup:2023arg,LISA:2024hlh}. Given their long observational duration and complex waveform morphology, EMRIs are ideal probes of the spacetime geometry of massive BHs~\cite{Ryan:1995wh,Moore:2017lxy,Destounis:2021mqv}, of beyond-GR emission channels~\cite{Maselli:2021men,Barsanti:2022vvl,Speri:2024qak}, and of possible matter distributions in their vicinity~\cite{Barausse:2014tra,Cardoso:2019rou,Cardoso:2022whc,Vicente:2025gsg,Duque:2025yfm,Dyson:2026ddd,Macedo:2026jfv,Lui:2026uai}.

One theoretically well-motivated class of environments is provided by ultralight bosonic fields. If sufficiently light bosons exist, they can form macroscopic condensates around spinning BHs through superradiant instabilities~\cite{Arvanitaki:2009fg,Brito:2014wla,East:2017ovw,Brito:2015oca}. In binary systems, these boson clouds can affect the orbital evolution and leave imprints on the emitted GW signal~\cite{Baumann:2018vus,Hannuksela:2018izj,Berti:2019wnn,Baumann:2019ztm,Baumann:2021fkf,Baumann:2022pkl,Tomaselli:2023ysb,Duque:2023seg,Brito:2023pyl,Boskovic:2024fga,Tomaselli:2024dbw,Tomaselli:2024bdd,Tomaselli:2025jfo,Dyson:2025dlj,Li:2025ffh,Boskovic:2025ixx,Roy:2025qaa,Keijzer:2026vul,Xu:2026qrg}. In this work, we focus on condensates formed by a massive scalar field and refer to them as \textit{scalar clouds}. Importantly, the effects studied here do not require modifications of General Relativity or an intrinsic scalar charge for the secondary object~\cite{Maselli:2021men,Barsanti:2022vvl,Speri:2024qak}. Rather, they arise from the gravitational interaction between a standard compact object and the scalar cloud environment.

The dynamics of compact binaries inside bosonic clouds has been studied extensively using Newtonian approximations~\cite{Baumann:2018vus,Hannuksela:2018izj,Berti:2019wnn,Baumann:2019ztm,Baumann:2021fkf,Baumann:2022pkl,Tomaselli:2023ysb,Boskovic:2024fga,Tomaselli:2024dbw,Tomaselli:2024bdd,DellaMonica:2025zby,Tomaselli:2025jfo,Boskovic:2025ixx}. These approaches capture key physical mechanisms, including resonant and non-resonant transitions between cloud states, emission into unbound radiative scalar modes, and the associated orbital backreaction. However, they are not designed to describe the late stages of the inspiral, where relativistic orbital dynamics and strong-field effects become essential. Relativistic perturbative frameworks for modelling EMRIs in scalar clouds have recently been developed, but have so far focused on circular orbits, either around Schwarzschild BHs~\cite{Duque:2023seg,Brito:2023pyl,Keijzer:2026vul} or Kerr BHs~\cite{Dyson:2025dlj,Li:2025ffh}. Realistic EMRIs are expected to exhibit nonzero eccentricity and orbital inclination~\cite{Mancieri:2025cmx,Sun:2025lbr}, both of which significantly influence their waveform morphology and secular evolution~\cite{Hughes:2021exa,Isoyama:2021jjd}. Building on these developments, the present work provides the first relativistic analysis of eccentric and inclined EMRIs evolving inside scalar clouds, complementing Newtonian studies of generic orbits in such environments~\cite{Berti:2019wnn,Tomaselli:2023ysb,Boskovic:2024fga,Tomaselli:2024dbw,Tomaselli:2024bdd,Tomaselli:2025jfo,Boskovic:2025ixx}.

To this end, we develop a fully relativistic framework for computing scalar radiation from eccentric and inclined EMRIs evolving inside scalar clouds in a Schwarzschild background. Following Refs.~\cite{Brito:2023pyl,Dyson:2025dlj,Li:2025ffh,Keijzer:2026vul}, we treat the cloud as a perturbative environment and use a two-parameter expansion. The first parameter is the mass ratio $q:=m_p/M\ll1$, where $M$ is the mass of the primary BH and $m_p$ is the mass of the secondary compact object, modelled as a point particle. The second is the scalar-field amplitude $\epsilon\ll1$, which controls the cloud's backreaction on the background geometry. At leading order in this expansion, the particle follows geodesics of the background spacetime, while the gravitational perturbations it induces source perturbations of the scalar cloud. The resulting scalar radiation carries energy and angular momentum to infinity and through the BH horizon, contributing to the secular evolution of the orbit.

A particularly important lesson from Newtonian studies is that the companion's orbital motion can induce transitions between different cloud levels, with resonant enhancement whenever an orbital harmonic matches the energy splitting between them~\cite{Baumann:2018vus,Baumann:2019ztm,Tomaselli:2024bdd,Boskovic:2024fga,Boskovic:2025ixx}. In purely Newtonian treatments, most of these resonances are associated with low orbital frequencies and therefore would typically lie outside the most relevant LISA band for typical EMRI parameters. Our relativistic calculation shows that eccentricity qualitatively enriches this picture. Because relativistic bound orbits are characterized by distinct radial and azimuthal frequencies, the scalar source contains harmonics involving both frequencies. This opens additional resonant channels between cloud states, especially in the strong-field regime, where relativistic precession makes the separation between radial and azimuthal motion significant. In a companion Letter~\cite{Xu:2026cky}, we highlight the phenomenological consequences of these eccentricity-induced resonances. The present paper provides the detailed derivation underlying those results and extends the analysis to inclined orbits.

The remainder of this paper is organized as follows.
In Sec.~\ref{Sec:Framework}, we review the theoretical framework and perturbative scheme employed to study EMRIs in the presence of a scalar cloud, as well as the basic properties of quasi-bound states of a massive scalar field in a BH spacetime, specializing to a Schwarzschild background.
In Sec.~\ref{Sec:Point Particle in Eccentric, Equatorial Motion}, we analyse the leading-order response of the scalar cloud to a point particle on a prograde eccentric, equatorial orbit. We show that the orbital motion acts as a source of scalar perturbations and compute the corresponding orbital energy and angular momentum loss rates due to scalar waves propagating through the horizon and to infinity, presenting explicit results for the case of a non-axisymmetric prograde dipolar scalar cloud. In Sec.~\ref{Sec:Point Particle in Generic Orbits} we show how to generalize the results to generically inclined orbits when considering a Schwarzschild BH background, and show how the orbital loss rates are affected by inclination. Finally, in Sec.~\ref{Sec:Conclusions and Outlook}, we summarize our main findings and discuss future extensions of our work. Additional details of our computations are also shown in the Appendices. In particular, to verify our results, in App.~\ref{App:Revisiting Newtonian fluxes}, we develop a Newtonian framework that closely follows the relativistic calculations, showing how our perturbative scheme is fully equivalent to previous Newtonian studies that employed quantum-mechanical perturbation theory tools to perform the same type of calculations~\cite{Tomaselli:2023ysb}.

Throughout this work, we use geometrized units in which $G=c=1$ and adopt a mostly-plus metric signature. Unless otherwise stated, we use Schwarzschild coordinates $x^{\mu}=(t,r,\theta^A)$, where $\theta^{A}=(\theta, \varphi)$ denote the usual spherical angular coordinates.

\section{Framework}\label{Sec:Framework}
\subsection{Action and equations of motion}

As a starting point, we consider the action for a complex scalar field $\mathbf{\Phi}$ with mass $\hbar\mu$ minimally coupled to gravity in the presence of a point-like perturber with mass $m_p$:
\begin{equation}
S=\int{d}^{4}x\sqrt{-\mathbf{g}}\left(\frac{\mathbf{R}}{16\pi}-\mathfrak{L}[\mathbf{\Phi}]\right)-m_{p}\int_{\gamma}{d}\tau,
\end{equation}
where $\mathfrak{L}[\mathbf{\Phi}]=\partial_{\mu} \mathbf{\Phi}\partial^{\mu} \mathbf{\Phi}^*+\mu^2|\mathbf{\Phi}|^2$ is the scalar field's Lagrangian density, $\gamma$ is the worldline of the point-particle, and $\tau$ its proper time. 

Varying this action with respect to the metric and the scalar field, we obtain the Einstein-Klein-Gordon (EKG) field equations:
\begin{equation}
\begin{aligned}\label{eq: EKG field equation}
    \mathbf{G}_{\mu\nu} = 8\pi(T^{\mathbf{\Phi}}_{\mu\nu}+T^p_{\mu\nu}),\qquad
    \Box \mathbf{\Phi} = \mu^2 \mathbf{\Phi},
\end{aligned}
\end{equation}
where $\mathbf{G}_{\mu\nu}:= \mathbf{R}_{\mu\nu}-\mathbf{g}_{\mu\nu}\mathbf{R}/2$ and $\Box:=\mathbf{g}_{\mu\nu}\nabla^\mu\nabla^{\nu}$.  
The stress-energy tensor of the scalar field, $T_{\mu\nu}^\mathbf{\Phi}$, reads
\begin{equation}\label{eq: stress-energy tensor definition}
T_{\mu\nu}^\mathbf{\Phi}=2\partial_{(\mu}\mathbf{\Phi}\partial_{\nu)}\mathbf{\Phi}^*-\mathbf{g}_{\mu\nu}\left(\partial_\alpha\mathbf{\Phi}\partial^\alpha\mathbf{\Phi}^*+\mu^2|\mathbf{\Phi}|^2\right),
\end{equation}
whereas the stress-energy tensor of the point-particle (describing a small secondary object), $T_{\mu\nu}^p$, is given by
\begin{equation}\label{eq: stress-energy tensor of the particle}
    T^{p}_{\mu\nu}[\mathbf{g}]=m_p\int_{\gamma} u_\mu u_\nu\frac{\delta^{(4)}\left(x^\mu-\mathbf{x}_p^\mu(\tau)\right)}{\sqrt{-\mathbf{g}}}d\tau,
 \end{equation}
with $u^{\mu}=d\mathbf{x}_p^{\mu}/d\tau$ the four-velocity along the particle's worldline.

The complex scalar field possesses a global $U(1)$ symmetry which implies the existence of a conserved current given by
\begin{equation}\label{eq: current}
j^\mu=-i\left(\mathbf{\Phi}^*\partial^\mu\mathbf{\Phi}-\mathbf{\Phi}\partial^\mu\mathbf{\Phi}^*\right).
\end{equation}
In the absence of dissipation in the system (such as scalar waves escaping to infinity or being absorbed at the horizon), the conservation of the current implies the existence of a conserved Noether charge:
\begin{equation}
Q_\Sigma = \int_\Sigma d \Sigma_\mu \,j^\mu
\end{equation}
where $\Sigma$ is a spacelike hypersurface. 
Physically, the conserved charge $Q$ can be interpreted as describing the total number of scalar \textit{particles} or quanta composing the configuration. As we discuss below, it also provides a useful diagnostic for tracking the exchange of energy-momentum between the scalar field and the point-like perturber.

\subsection{Perturbation scheme}\label{sec:Perturbation scheme}
In addition to considering small mass ratios $q\ll1$, we also treat the backreaction of the scalar field on the BH spacetime perturbatively by introducing as a second expansion parameter $\epsilon \ll 1$, which characterizes the scalar field amplitude. 
The gravitational and scalar fields are expanded as in Ref.~\cite{Dyson:2025dlj}:
\begin{align}
\mathbf{g}_{\mu\nu} =&g_{\mu\nu}+\epsilon^{2}h_{\mu\nu}^{(2,0)}+q\,h_{\mu\nu}^{(0,1)}+\epsilon^{2}q\,h_{\mu\nu}^{(2,1)} \nonumber\\
 & +q^2h_{\mu\nu}^{(0,2)}+\epsilon^2 q^2h_{\mu\nu}^{(2,2)}+\cdots, \label{eq:metric_expansion}\\
\mathbf{\Phi}  =&\epsilon\,\phi^{(1,0)}+\epsilon\, q\,\phi^{(1,1)}+\epsilon\, q^2\phi^{(1,2)}+\cdots\,.\label{eq:scalar_expansion}
\end{align}
Inserting these into the EKG field equations~\eqref{eq: EKG field equation}, we can solve the problem order-by-order. In Eqs.~\eqref{eq:metric_expansion} and~\eqref{eq:scalar_expansion} we factor out the explicit powers of $\epsilon$ and $q$, so that the superscripts $(i,j)$ label the order $\epsilon^i q^j$ at which each $\mathcal{O}(1)$ coefficient enters and the power counting remains explicit. In what follows, we also use a superscript $^{(0)}$ to denote background quantities.

At order $\mathcal{O}(\epsilon^0 q^0)$, we obtain the vacuum Einstein field equations $\mathbf{G}_{\mu\nu}[g_{\mu\nu}]= 0$. For simplicity, in this work we will restrict to the spherically symmetric Schwarzschild solution:
\begin{equation}
ds^2=-f(r)dt^2+f(r)^{-1}dr^2+r^2d\theta^2+r^2\sin^2\theta d\varphi^2,
\end{equation}
with $f(r) = 1-2M/r$. 

Since the stress–energy tensor of the scalar field is quadratic in the field amplitude, the spacetime metric is only deformed at order $\mathcal{O}(\epsilon^{2})$~\cite{Brito:2023pyl}. Consequently, at order $\mathcal{O}(\epsilon^1 q^0)$, the only relevant equation is the Klein–Gordon (KG) equation,
\begin{align}
\Box^{(0)}\phi^{(1,0)} & =\mu^{2}\phi^{(1,0)},\label{eq: (1,0) order KG equation}
\end{align}
where $\Box^{(0)}:=g_{\mu\nu}\nabla^{(0)\mu}\nabla^{(0)\nu}$ denotes the d’Alembertian operator constructed from the background Schwarzschild metric.

Continuing this procedure, we can consider the impact of the point particle. Considering first the corrections at order $\mathcal{O}(\epsilon^0 q^1)$, the only relevant equation is the Einstein equation:
\begin{equation}
    \delta \mathbf{G}_{\mu\nu}[h^{(0,1)}_{\mu\nu}] =8\pi T_{\mu\nu}^{p}[g],\label{eq: (0,1) order Einstein equation}
\end{equation}
which can be solved to obtain the metric perturbation $h^{(0,1)}$ using standard vacuum BH perturbation theory. At this order, the point particle follows a geodesic of the background Schwarzschild metric.

Finally, the influence of the point particle on the scalar field enters at order $\mathcal{O}(\epsilon^1 q^1)$. The corresponding KG equation reads
\begin{equation}\label{eq: (1,1) order KG equation}
      \left(\Box^{(0)}-\mu^2\right)\phi^{(1,1)}= S^\phi\left[h^{(0,1)}, \phi^{(1,0)}\right],
\end{equation}
where $S^\phi$ is given by
\begin{equation}
    S^\phi\left[h^{(0,1)}, \phi^{(1,0)}\right] = \nabla_{\mu}^{(0)}\bar h_{(0,1)}^{\mu\nu}\partial_{\nu}\phi^{(1,0)} + h_{(0,1)}^{\mu\nu}\nabla_{\mu}^{(0)}\partial_{\nu}\phi^{(1,0)},
\end{equation}
with the trace-reversed metric perturbation given by
\begin{equation}
    \bar h_{\mu\nu}^{(0,1)}=h_{\mu\nu}^{(0,1)}-\frac{1}{2}g_{\mu\nu}h^{(0,1)},
\end{equation}
and $h^{(0,1)} :=g^{\mu\nu}h_{\mu\nu}^{(0,1)}$ is the trace of the metric perturbation.

This programme can be continued to compute $h^{(0,2)}_{\mu\nu}$ and $h^{(2,1)}_{\mu\nu}$, as well as higher-order terms. However, in this work we will focus on obtaining the orbital loss rates due to the scalar field scattered off to infinity and through the horizon, as captured by the perturbation $\phi^{(1,1)}$. These have been computed in the case of EMRIs in equatorial, circular orbits in previous works~\cite{Brito:2023pyl,Duque:2023seg,Dyson:2025dlj,Li:2025ffh}; here, we extend these results to generic orbits in Schwarzschild.


More generically, the same formalism can be applied to a Kerr background, as was done in Refs.~\cite{Dyson:2025dlj,Li:2025ffh} for equatorial, circular EMRIs. The calculation requires solving Eq.~\eqref{eq: (0,1) order Einstein equation} in Kerr for generic orbits, which is a highly non-trivial task. While (as argued below) a Schwarzschild BH may be a good proxy for low-spin Kerr BHs---a generic outcome of superradiant instabilities for $M\mu\ll 1$---, our work also serves as preparation for when ongoing efforts to compute $h^{(0,1)}_{\mu\nu}$ on a Kerr background succeed.

\subsection{Scalar cloud}\label{sec: Scalar cloud}
The scalar field solution at order $\mathcal{O}(\epsilon^1 q^0)$ yields the quasi-bound states characterized by an eigenfrequency~$\omega$. These quasi-bound states are analogous to the bound states in a hydrogen atom and can be labelled by three quantum numbers $\{n_i, \ell_i, m_i\}$. Around a spinning BH, modes with $m_i>0$ can be superradiantly unstable~\cite{Detweiler:1980uk, Dolan:2007mj,Baumann:2019eav} and lead to the formation of a scalar cloud, which is often also referred to as a gravitational atom~\cite{Brito:2015oca}. For clouds that grow solely from a superradiant instability, the backreaction of the cloud on the metric is generically small~\cite{Brito:2014wla,East:2017ovw} justifying our assumption of treating the scalar field perturbatively. 

While the BH spin is essential to form scalar clouds through superradiance, the most unstable mode, $\{0,1,1\}$ is always characterized by $M\mu \lesssim 1/2$. The radial profiles of the bound states peak at a radius that scales with $M/(M\mu)^2$, which implies that for small $M\mu$ the cloud is localized far from the horizon where effects related to the BH spin are small. Additionally, for small $M\mu$ the superradiant extraction drives the BH toward a small saturation spin. These considerations imply that using a Schwarzschild background metric may provide a good approximation of the result for scalar clouds formed around Kerr BHs through a superradiant instability, as recently confirmed in Ref.~\cite{Dyson:2025dlj}. Contrary to the Kerr case, where true bound states (with zero decay rate) exist, quasi-bound states for Schwarzschild BHs always decay in time due to absorption at the horizon; still, they can be extremely long-lived when $M\mu \ll 1$~\cite{Detweiler:1980uk, Dolan:2007mj, Baumann:2019eav, Bao:2022hew}.

Since the Schwarzschild metric is spherically symmetric, the scalar field can be decomposed as
\begin{equation}\label{eq: scalar field ansatz}
\phi^{(1,0)}(t,r,\theta^A)=R^{b}_{n_i\ell_i}(r)Y_{\ell_im_i}(\theta^A)e^{-i\omega t},
\end{equation}
where $Y_{\ell_im_i}(\theta^A)$ are scalar spherical harmonics.
Inserting Eq.~\eqref{eq: scalar field ansatz} into the KG equation, one can show that the radial function $R^{b}_{n_i\ell_i}(r)$ satisfies the differential equation
\begin{equation}\label{eq: radial KG bound state}
\frac{d^2(rR^{b}_{n_i\ell_i})}{dr_*^2}+\left(\omega^2-V_{\ell_i}\right)rR^{b}_{n_i\ell_i}=0,
\end{equation}
where $r_{*}$ is the Regge-Wheeler \textit{tortoise} coordinate, defined by $dr_{*}/dr = 1/f(r)$, and the effective potential reads
\begin{equation}
V_{\ell_i}=f(r)\left[\mu^2+\frac{\ell_i(\ell_i+1)}{r^2}+\frac{2M}{r^3}\right].
\end{equation}
Notice that, due to the spherical symmetry of the background, both $V_{\ell_i}$ and the radial function are independent of $m_i$. This also implies that quasi-bound states with the same overtone number~$n_i$ and angular number $ \ell_i$, but different azimuthal number~$m_i$, have a degenerate spectrum. 

Imposing appropriate boundary conditions, solutions to Eq.~\eqref{eq: (1,0) order KG equation} can be obtained by using a continued-fraction method~\cite{Cardoso:2005vk, Dolan:2007mj}, which we employ throughout this work. For quasi-bound state solutions, one imposes boundary conditions in which the field decays exponentially at spatial infinity, whereas close to the event horizon only ingoing waves are present. This pair of boundary conditions is satisfied for an infinite, discrete spectrum of complex eigenfrequencies that can be labelled according to the two quantum numbers $\omega:=\omega_{n_i\ell_i}$.\footnote{In a Kerr background $\omega$ also depends on the azimuthal number~$m_i$.}

It is worth noting that the metric perturbations generated by the orbiting object and by the self-gravity of the cloud can introduce small corrections to the quasi-bound state eigenfrequencies~\cite{Cannizzaro:2023jle}. These shifts are expected to be subdominant and have negligible impact on the leading-order orbital loss rates by a point particle within the scalar cloud. Accordingly, we neglect such effects in the present analysis.

From $\phi^{(1,0)}$ we can define the scalar cloud mass as
\begin{equation}
M_b=-\int d\Omega\int_{2M}^{\infty}dr\,r^2
(g^{\rm Schw})^{t\mu}
T_{\mu t}\left[\epsilon\phi^{(1,0)},g^{\rm Schw}\right].
\end{equation}
Here,
$d\Omega\equiv\sin\theta\,d\theta\,d\varphi$, and
$T_{\mu\nu}[\epsilon\phi^{(1,0)},g^{\rm Schw}]$
is the scalar stress-energy tensor defined in
Eq.~\eqref{eq: stress-energy tensor definition},
evaluated using the background scalar field
$\epsilon\phi^{(1,0)}$ and the Schwarzschild metric.
The index is raised with the inverse background metric
$(g^{\rm Schw})^{\mu\nu}$.
The arbitrary amplitude of $\phi^{(1,0)}$ can be fixed
such that $\epsilon\equiv\sqrt{M_b/M}\ll1$.

Finally, we note that interpreting our setup as a proxy for a superradiantly generated Kerr $\{0,1,1\}$ cloud near saturation, the subsequent $\{0,2,2\}$ instability and, for a real scalar field, GW-annihilation depletion occur on timescales much longer than the year-scale strong-field EMRI evolution. In the small-$M\mu$ limit, the respective relevant timescales for these two processes are~\cite{Detweiler:1980uk,Brito:2017zvb}
\begin{equation}
\tau^{022}_{\rm inst}\approx
1.5\times10^{7}
\left(\frac{M}{10^6M_{\odot}}\right)
\left(\frac{0.2}{M\mu}\right)^{14}\,{\rm yr},
\end{equation}
and
\begin{equation}
\tau_{\rm gw}\approx
8\times10^{5}
\left(\frac{0.05}{M_{b}/M}\right)
\left(\frac{M}{10^6M_{\odot}}\right)
\left(\frac{0.2}{M\mu}\right)^{14}\,{\rm yr}\,.
\end{equation}
Numerically evaluated timescales for $M\mu=0.2$, $M=10^6M_\odot$, and $M_{b}/M=0.05$ give very similar results $\tau_{\rm inst}^{022}\simeq1.87\times10^7\,{\rm yr}$ and $\tau_{\rm gw}\simeq3.80\times10^6\,{\rm yr}$. The very strong dependence on $(M \mu)^{-14}$, makes these timescales highly sensitive to $M\mu$; going from $M\mu=0.2$ to $M\mu=0.1$ increases these timescales by four orders of magnitude, while still allowing the $\{0,1,1\}$ state to grow sufficiently fast compared to other typical astrophysical scales; for example, for $M=10^6M_{\odot}$, $M\mu=0.1$ and initial black-hole spin $a/M=0.9$, the $\{0,1,1\}$ state grows on a timescale $\mathcal{O}(10^4)$ yr and even faster for larger $M\mu$. Therefore, although a full understanding of how likely it is to find an EMRI in such a cloud requires further work, we expect that for $M\mu\lesssim 0.2$ a cloud could survive long enough in order to produce detectable effects in EMRIs.

\section{Point Particle in Eccentric, Equatorial Motion}\label{Sec:Point Particle in Eccentric, Equatorial Motion}
Using the formalism presented in the previous section, we now study the leading-order perturbations generated by a point particle in equatorial motion around a Schwarzschild BH surrounded by a scalar cloud. Specifically, we consider perturbations to both the metric and the scalar field up to order $\mathcal{O}(\epsilon^1 q^1)$. As discussed above, these are governed by Eqs.~\eqref{eq: (0,1) order Einstein equation} and \eqref{eq: (1,1) order KG equation}. Accordingly, our first task is to determine the metric perturbations and reconstruct the perturbed geometry by solving Eq.~\eqref{eq: (0,1) order Einstein equation}; these results are then substituted into the source term of the scalar field equation, Eq.~\eqref{eq: (1,1) order KG equation}, to obtain the corresponding scalar field perturbations.

\subsection{Bound orbits}
In what follows we will focus on generic eccentric orbits. Before determining the metric perturbations, it is therefore useful to introduce the relevant orbital parameters used in our analysis.

We consider bound timelike geodesics in a Schwarzschild geometry; these can be parametrized through $\mathbf{x}_p^{\mu}(\tau)=\{t_p,r_{p},\theta_p,\varphi_p\}(\tau)$ with proper time $\tau$, where the subscript $p$ denotes quantities evaluated along the worldline of the point-particle. 
For the moment we exploit the spherical symmetry of Schwarzschild to place the orbit in the equatorial plane, i.e. $\theta_p = \pi/2$.\footnote{Generic inclinations will be incorporated in Sec.~\ref{Sec:Point Particle in Generic Orbits} through a relative rotation between the orbital plane and the scalar cloud.}
The geodesics can be completely described in terms of two conserved quantities—the specific energy $\mathcal{E}$ and angular momentum $\mathcal{L}$. These quantities can be expressed in terms of the (dimensionless) semi-latus rectum $p$ and the eccentricity $e$~\cite{Cutler:1994pb,Barack:2010tm}:
\begin{equation}
\mathcal{E}^2=\frac{(p-2-2e)(p-2+2e)}{p(p-3-e^2)},\quad\mathcal{L}^2=\frac{p^2M^2}{p-3-e^2},
\end{equation}
where $p$ and $e$ can be defined in terms of the periapsis $r_{\rm min}$ and apoapsis $r_{\rm max}$ (i.e., the turning points where $dr_p/d\tau=0$) through
\begin{equation}
r_{\rm max}=\frac{pM}{1-e},\qquad r_{\rm min}=\frac{pM}{1+e}.
\end{equation}
The four-velocity ${u}^{\mu}=d\mathbf{x}_p^{\mu}/d\tau$ along the geodesic can then be expressed as
\begin{align}
u^t=\frac{\mathcal{E}}{f_p},\qquad~u^\varphi=\frac{\mathcal{L}}{r_p^2},\qquad~u^r=\sqrt{\mathcal{E}^2-U_p^2},   
\end{align}
where
\begin{equation}
    U(r) := \sqrt{f(r)\left(1+\frac{\mathcal{L}^2}{r^2}\right)}.
\end{equation}

On a Schwarzschild background, eccentric geodesic orbits are typically not closed, and the motion is therefore not strictly periodic when viewed by an asymptotic static observer. To parametrize this behaviour, \citet{1959RSPSA.249..180D} introduced a phase angle, the so-called {\it relativistic anomaly}~$\chi$, defined in such a way that, as $\chi$ increases monotonically from $0$ to $2\pi$, the radial separation $r_p$ evolves from $r_{\rm min}$ to $r_{\rm max}$ and back to $r_{\rm min}$. The anomaly is related to the radial position through the Keplerian-like expression:
\begin{equation}
    r_{p} = \frac{pM}{1+e\cos\chi}.
\end{equation}
In terms of $\chi$, the remaining geodesic position coordinates $t_p$ and $\varphi_p$ are given by
\begin{align}
    \frac{dt_p}{d\chi}&=\frac{p^2M}{(p-2-2e\cos\chi)(1+e\cos\chi)^2}\sqrt{\frac{(p-2)^2-4e^2}{p-6-2e\cos\chi}},\label{eq:t(chi)}\\
\frac{d\varphi_p}{d\chi}&=\sqrt{\frac{p}{p-6-2e\cos\chi}}.
\end{align}

The radial motion executes a periodic libration between periapsis and apoapsis with a frequency and a period given, respectively, by:
\begin{equation}
\Omega_r:=\frac{2\pi}{T_r},\qquad T_r:=\int_0^{2\pi}\left(\frac{dt_p}{d\chi}\right)d\chi.
\end{equation}
Although eccentric Schwarzschild orbits are not closed, one can still define a mean azimuthal rate  at which the azimuthal angle advances by averaging the instantaneous angular frequency $d\varphi_p/dt$ over a radial period:
\begin{equation}
\Omega_\varphi:=\frac{1}{T_r}\int_0^{T_r}\left(\frac{d\varphi_p}{dt}\right)dt = \frac{1}{T_r}\int_0^{2\pi}\left(\frac{d\varphi_p}{d\chi}\right)d\chi.
\end{equation}
The azimuthal phase can then be written as $\varphi_p(t) = \Omega_\varphi t + \Delta\varphi(t)$, where $\Delta\varphi(t)$ captures the periodic modulation driven by the radial oscillation~\cite{Cutler:1994pb}. Since this modulation arises entirely from the oscillatory radial motion, $\Delta\varphi(t)$ is periodic with fundamental frequency $\Omega_r$.

The motion therefore contains two intrinsic frequencies: the mean azimuthal frequency $\Omega_\varphi$ and the radial libration frequency $\Omega_r$.  
Accordingly, this restricts the metric perturbation spectrum to~\cite{Cutler:1994pb,Hopper:2010uv}
\begin{equation}
    \sigma_{mn} = m\Omega_\varphi + n\Omega_r, \qquad m,n\in\mathbb{Z}.
\end{equation}
The index $m$ counts the number of times a given term winds around the azimuth, i.e., it labels the azimuthal harmonic of the orbital motion. As we discuss below, the integer $m$ naturally matches the azimuthal index appearing in a spherical-harmonic decomposition of metric perturbations. In contrast, $n$ labels the harmonics of the radial oscillation.

For completeness, we briefly summarize the conditions under which stable bound timelike geodesics exist in a Schwarzschild spacetime.  
Bound eccentric orbits are specified by parameters $0 \le e < 1$ and $p \ge p_{\rm LSO}(e)$, where \(p_{\rm LSO}(e)\) denotes the {\it separatrix}, or last stable orbit (LSO), that separates stable from unstable eccentric motion.  
Explicitly, the separatrix is given by~\cite{Cutler:1994pb}
\begin{equation}
\label{eq:separatrix}
p_{\rm LSO}(e) = 6 + 2e.
\end{equation}
Orbits with $p<p_{\rm LSO}(e)$ lack an inner radial turning point and therefore correspond to plunging trajectories. In the circular limit $(e\to 0)$, the separatrix reduces to the familiar innermost stable circular orbit (ISCO). In this work, we focus on orbits with parameters above the separatrix, corresponding to stable bound motion.

\subsection{Metric perturbations in Regge-Wheeler gauge}
Perturbations of a Schwarzschild BH are most conveniently described within the Regge-Wheeler-Zerilli (RWZ) framework. This formalism, pioneered by~\citet{Regge:1957td} and subsequently extended by~\citet{Zerilli:1970se,Zerilli:1970wzz} and~\citet{Vishveshwara:1970cc}, decomposes metric perturbations into spherical-harmonic modes and separates them into odd (or axial) and even (or polar) parity sectors. At each multipolar order, the radiative gravitational degrees of freedom can be encoded in two {\it gauge-invariant master functions}, one for each parity, that satisfy linear inhomogeneous wave equations on the Schwarzschild background. Later developments by~\citet{Moncrief:1974am}, and further refinements in gauge-invariant and covariant formulations~\cite{Cunningham:1978zfa, Martel:2005ir}, have greatly streamlined the treatment of these perturbations. 

While the master functions are gauge-invariant, reconstructing the full metric perturbations requires choosing a specific gauge. In this work, we will use the Regge-Wheeler gauge~\cite{Regge:1957td} and closely follow the analysis in Ref.~\cite{Hopper:2010uv}. In this gauge the metric perturbations~$h_{\mu\nu}^{(0,1)}$ can be written as:
\begin{equation}
h_{\mu\nu}^{(0,1)}= \sum_{l,m}\left[h_{\mu\nu}^{\text{axial},lm} +h_{\mu\nu}^{\text{polar},lm}\right],
\end{equation}
where
\begin{equation}h_{\mu\nu}^{\mathrm{axial},lm}=
\begin{pmatrix}
0 & 0 & -h_{0}Y_{,\varphi}/\sin\theta & h_{0}\sin\theta\, Y_{,\theta} \\
* & 0 & -h_{1}Y_{,\varphi}/\sin\theta & h_{1}\sin\theta\, Y_{,\theta} \\
* & * & 0 & 0 \\
* & * & * & 0
\end{pmatrix}\,,
\end{equation}
represent axial perturbations, and
\begin{equation}h_{\mu\nu}^{\mathrm{polar},lm}=
\begin{pmatrix}
H_{0}Y & H_{1}Y & 0 & 0 \\
* & H_{2}Y & 0 & 0 \\
* & * & r^{2}K\,Y & 0 \\
* & * & * & r^{2}\sin^{2}\theta\, K\,Y
\end{pmatrix}\,,
\end{equation}
represent polar perturbations. Here, $Y:=Y_{lm}(\theta^{A})$ denotes a scalar spherical harmonic, $Y_{,\theta/\varphi}:=\partial_{\theta/\varphi} Y_{lm}(\theta^{A})$, and ${H_0,H_1,H_2,K,h_0,h_1}$ are functions of $t$ and $r$ that depend on the angular numbers $l$ and $m$. The entries marked with $*$ represent symmetric components, such that $h_{\mu\nu} = h_{\nu\mu}$. As we discuss below, for the radiative degrees of freedom, corresponding to modes with $l\geq 2$, these functions can be constructed from the even- and odd-parity master functions that we define below. The monopole ($l=0$), and dipole ($l=1$) are non-radiative and need a separate treatment as we will also discuss. 

\subsubsection{The Regge-Wheeler-Zerilli formalism and method of extended homogeneous solutions}\label{sec: RWZ formalism}

Within the RWZ formalism, gravitational perturbations for $l \geq 2$ can be fully described by the even-parity Zerilli-Moncrief master function $\Psi^{\rm e}_{lm}$~\cite{Moncrief:1974am} and the odd-parity Cunningham-Price-Moncrief master function $\Psi^{\rm o}_{lm}$~\cite{Cunningham:1979px}, both satisfying inhomogeneous wave equations of the form
\begin{equation}\label{eq:TD RWZ wave eq}
\left[-\frac{\partial^2}{\partial t^2}+\frac{\partial^2}{\partial r_*^2}-V_l^{\rm e/o}(r)\right]\Psi^{\rm e/o}_{l m}(t,r)=\Pi^{\rm e/o}_{l m}(t,r),
\end{equation}
where $r_*$ is again the Regge-Wheeler \textit{tortoise} coordinate [as in Eq.~\eqref{eq: radial KG bound state}], and the even/odd potentials read
\begin{equation}
\begin{aligned}
    &V^{\rm e}_l = \frac{f \left(18 M^3+18 \lambda  M^2 r+6 \lambda ^2 M
   r^2+2 \lambda ^2 (\lambda +1) r^3\right)}{r^3 (3 M+\lambda  r)^2} \,,\\
    &V^{\rm o}_l= f\left(\frac{l(l+1)}{r^2}-\frac{6M}{r^3}\right)\,,
\end{aligned}
\end{equation}
with $\lambda := (l-1)(l+2)/2$. The source terms of these equations, $\Pi^{\rm e/o}_{l m}(t,r)$, are parity dependent. The superscripts ${\rm e}$ and ${\rm o}$ denote the even- and odd-parity sectors, respectively. For brevity, we will now suppress explicit parity labels, as the following discussion applies uniformly to both parity sectors. The source terms have the following general form~\cite{Hopper:2010uv}
\begin{equation}\label{eq:source term RW}
\Pi_{l m}(t,r)={G}_{l m}(t)\delta[r-r_p(t)]+{F}_{l m}(t)\delta^{\prime}[r-r_p(t)],
\end{equation}
where ${G}_{l m}$ and ${F}_{l m}$ are smooth, unique functions of time only (explicit expressions are given in App.~\ref{app: Source terms}). 

Eq.~\eqref{eq:TD RWZ wave eq} can in principle be solved directly in the time domain (TD). However, TD schemes face well-known challenges, including handling the distributional source on a numerical grid, ensuring stability, and imposing accurate outgoing-wave boundary conditions. In contrast, frequency domain (FD) approaches transform the problem into ordinary differential equations (ODEs) for the FD modes, offering improved accuracy, especially for bound orbits~\cite{Hopper:2010uv}. Motivated by these advantages, we adopt the FD version of the RWZ formalism. 

Applying a Fourier decomposition in time, the TD wave equation~\eqref{eq:TD RWZ wave eq} reduces to a set of ODEs for the radial mode functions:
\begin{equation}\label{eq: RW_radial}
    \left[\frac{d^2}{dr_*^2}-V_l(r)+\sigma_{mn}^2\right]R_{l mn}(r)=J_{l mn}(r),
\end{equation}
where $R_{l mn}(r)$ and $J_{lmn}(r)$ are Fourier harmonic amplitudes
\begin{align}
    R_{l mn}(r)\equiv\frac{1}{T_r}\int_0^{T_r}dt\,\Psi_{l m}(t,r)e^{i\sigma_{mn}t},\\
    \quad J_{l mn}(r)\equiv\frac{1}{T_r}\int_0^{T_r}dt\,\Pi_{l m}(t,r)e^{i\sigma_{mn}t}.\label{source_fourieramp}
\end{align}

To obtain the solutions of Eq.~\eqref{eq: RW_radial}, the first step is to solve the associated homogeneous equation and construct a set of two linearly independent solutions. We choose one solution, denoted $\hat R^-_{lmn}(r)$, to satisfy the \textit{in} boundary condition,
\begin{equation}\label{eq:Rhat-}
\hat{R}_{l{mn}}^-(r_*\to-\infty)=e^{-i\sigma_{mn}r_*}.
\end{equation}
The second solution, denoted $\hat R^+_ {l mn}(r)$, is chosen to satisfy the \textit{up} boundary condition at spatial infinity,
\begin{equation}\label{eq:Rhat+}
\hat{R}_{l{mn}}^+(r_*\to+\infty)=e^{i\sigma_{mn}r_*}.
\end{equation}
With these homogeneous solutions, the Green's function method yields the inhomogeneous solution satisfying \textit{in} and \textit{up} boundary conditions,
\begin{equation}
    R_{l mn}(r) = c_{l mn}^{+}(r)\hat{R}_{l mn}^{+}(r) + c_{l mn}^{-}(r)\hat{R}_{l mn}^{-}(r),
\end{equation}
where the radial coefficients are given by
\begin{subequations}\label{eq:coefficient functions of wave eq}
\begin{align}
c_{lmn}^{+}(r)
&= \frac{1}{\mathcal{W}_{lmn}}
   \int_{r_{\min}}^{r} 
   dr'\,\frac{\hat{R}^{-}_{lmn}(r')\, J_{lmn}(r')}{f(r')},\\[4pt]
c_{lmn}^{-}(r)
&= \frac{1}{\mathcal{W}_{lmn}}
   \int_{r}^{r_{\max}} 
   dr'\,\frac{\hat{R}^{+}_{lmn}(r')\, J_{lmn}(r')}{f(r')}.
\end{align}
\end{subequations}
The quantity
\begin{equation}
\mathcal{W}_{lmn}
:=
\hat{R}_{lmn}^{-}\frac{d\hat{R}_{lmn}^{+}}{dr_*}
- \hat{R}_{lmn}^{+}\frac{d\hat{R}_{lmn}^{-}}{dr_*},
\end{equation}
is the Wronskian associated with the two linearly independent homogeneous solutions; this can be shown to be independent of $r_*$.

Outside the source libration region, the inhomogeneous radial solution reduces to a linear combination of normalized homogeneous solutions, with constant coefficients determined at the boundaries of the source domain. Explicitly,
\begin{align}
    &R_{l mn}^+(r)=C_{l mn}^+\hat{R}_{l mn}^+(r),\quad r\geq r_{\max},\\
    &R_{l{mn}}^-(r)=C_{l{mn}}^-\hat{R}_{l{mn}}^-(r),\quad r\leq r_{\min},
\end{align}
where $C_{l mn}^\pm$ are defined through the values of $c_{l mn}^\pm(r)$ at the boundaries of the source region. Using Eq.~\eqref{eq:coefficient functions of wave eq}, the normalization coefficients can be expressed as
\begin{equation}
C_{l mn}^+=c_{l mn}^+(r_{\rm max})\,,\quad  C_{l mn}^-=c_{l mn}^-(r_{\rm min})\,.
\end{equation}
Since $J_{l mn}$ is given by Eq.~\eqref{source_fourieramp}, the computation of the $C_{l mn}^\pm$ coefficients requires double integration. In order to avoid the singularity at the turning points, it is useful to switch the order of integration and integrate over the relativistic anomaly $\chi$ instead of $t$. Explicitly writing the source in terms of $G_{lm}$ and $F_{lm}$, one gets
\begin{widetext}
\begin{align}
\begin{split}\label{eq:Clmn}
C_{l{mn}}^{\pm}=\frac{1}{\mathcal{W}_{l{mn}}T_{r}}\int_{0}^{2\pi}\left[\frac{1}{f_{p}}\hat{R}_{l{mn}}^{\mp}(r_{p}){G}_{l{m}}(\chi)+\left(\frac{2M}{r_{p}^{2}f_{p}^{2}}\hat{R}_{l{mn}}^{\mp}(r_{p})-\frac{1}{f_{p}}\frac{d\hat{R}_{l{mn}}^{\mp}(r_{p})}{dr}\right){F}_{l{m}}(\chi)\right]e^{i\sigma_{{m}{n}}t(\chi)}\frac{dt}{d\chi}d\chi,
\end{split}
\end{align}
\end{widetext}
where $t(\chi)$ is given by the integral form of Eq.~\eqref{eq:t(chi)}.

After obtaining the Fourier amplitudes one could in principle directly reconstruct $\Psi_{l m}(t,r)$ from its series representation
\begin{equation}
\Psi_{l m}(t,r)=\sum_n R_{l mn}(r)e^{-i\sigma_{{mn}}t}\,.
\end{equation}
However, as discussed in Ref.~\cite{Barack:2008ms,Hopper:2010uv}, this series converges very poorly in the source libration region due to the discontinuity of $\Psi_{l m}(t,r)$ at the location of the particle. This problem is even worse for the radial derivative $\partial_r\Psi_{l m}(t,r)$ which suffers from a Dirac delta function singularity at $r_p(t)$. To solve this problem, Ref.~\cite{Hopper:2010uv} used the {\it method of extended homogeneous solutions} (MEHS), first developed in Ref.~\cite{Barack:2008ms}.
This method uses the Fourier-harmonic modes of the homogeneous equation in the FD to construct homogeneous solutions to the TD wave equation. From the homogeneous solutions one can construct a full (weak) solution to the inhomogeneous master equation. The Fourier convergence of homogeneous solutions is exponentially rapid, ensuring the convergence of the inhomogeneous solution.

From the radial homogeneous solutions $R_{l mn}^\pm(r)$, we can obtain the TD homogeneous solutions by computing the Fourier series
\begin{equation}\label{eq:grav_fourier}
\Psi_{l m}^\pm(t,r)=\sum_n R_{l mn}^\pm(r)e^{-i\sigma_{{mn}}t}.
\end{equation}
The crucial observation of the MEHS is that the homogeneous solutions $\Psi_{l m}^\pm(t,r)$, when extended beyond their natural domains, can be combined to reproduce the full inhomogeneous solution valid everywhere. Specifically, the solution to the inhomogeneous wave equation~\eqref{eq:TD RWZ wave eq} is given by~\cite{Hopper:2010uv}
\begin{equation}
\begin{aligned}\label{eq:TD extended homogeneous solutions}
    \Psi_{l m}(t,r)=\Psi_{l m}^+(t,r)\Theta\left[r-r_p(t)\right]+\Psi_{l m}^-(t,r)\Theta\left[r_p(t)-r\right]\,,
\end{aligned}
\end{equation}
where $\Theta\left[r-r_p(t)\right]$ is the Heaviside function.

\subsubsection{Gravitational-wave flux}
Since only the modes with $l \geq 2$ contribute to gravitational radiation, the master functions $\Psi_{l m}^{\rm e/o}(t,r)$ can be used to evaluate the GW flux carried both out to future null infinity and down into the BH horizon. Once the inhomogeneous radial solution has been decomposed into normalized homogeneous solutions with asymptotic amplitudes $C_{lmn}^\pm$, the mode-by-mode contributions to the radiated energy and angular momentum follow directly from the standard Isaacson stress-energy tensor for linearized GWs~\cite{Thorne:1980ru,Martel:2005ir, Hopper:2010uv}. The GW energy flux $\dot{E}^g$ and angular momentum flux $\dot{L}^g$ at infinity ($+\infty$) and at the horizon ($-\infty$) can be written as~\cite{Hopper:2010uv}
\begin{align}
    \left\langle\dot{E}_{l m}^{g,\pm\infty}\right\rangle=\frac{1}{64\pi}\frac{(l+2)!}{(l-2)!}\sum_{n}\sigma_{mn}^{2}\left|C_{l mn}^{\pm}\right|^{2},\\
    \quad\left\langle\dot{L}_{l m}^{g,\pm\infty}\right\rangle=\frac{m}{64\pi}\frac{(l+2)!}{(l-2)!}\sum_{n}\sigma_{mn}\left|C_{l mn}^{\pm}\right|^{2}\,,
\end{align}
where $\left\langle\cdot\right\rangle$ denotes an average over several radial librations. These expressions show that each Fourier mode contributes independently to the radiative fluxes, with the strength of the contribution governed entirely by the asymptotic amplitude $C_{lmn}^\pm$ determined in the source region. In practical computations, the sums over $n$ converge rapidly for bound orbits, since $\sigma_{mn}$ grows linearly with the harmonic index.

\subsubsection{Metric perturbation reconstruction}\label{Sec: metric reconstruction}
For $l\geq 2$, the metric perturbation components can be reconstructed directly from $\Psi_{l m}^{\rm e/o}(t,r)$~\cite{Hopper:2010uv}. Generically, any metric perturbation amplitude constructed from the master functions inherits the same distributional structure associated with the point-particle source. A generic metric perturbation amplitude may therefore be written in the form~\cite{Hopper:2010uv}
\begin{equation}\label{eq: generic_metric_amp}
\mathcal{M} (t,r)=\mathcal{M}^+(t,r)\Theta(z)+\mathcal{M}^-(t,r)\Theta(-z)+\mathcal{M}^S(t)\delta(z),
\end{equation}
where $z:= r-r_p(t)$, $\mathcal{M}^+~(\mathcal{M}^-)$ represents a smooth function in the region $r>r_p~(r<r_p)$, and $\mathcal{M}^{S}$ is a smooth function of $t$ alone. Explicit expressions for the metric perturbation amplitudes in Regge-Wheeler gauge were computed in Ref.~\cite{Hopper:2010uv} and are explicitly given in App.~\ref{app: metric perturbation amplitudes}.

Finally, we note that, to obtain a complete description of the metric perturbation, we also need to include the non-radiative modes with $l=0$ and $l=1$. Although these modes do not contribute to the GW flux at infinity or into the BH horizon, they are essential for reconstructing the full perturbed metric. For a point particle orbiting a Schwarzschild BH, the monopole and dipole perturbations were first derived in a particular gauge by Zerilli~\cite{Zerilli:1970wzz}.
In this work, we make use of analytic expressions for these low-$l$ modes given in Ref.~\cite{Hopper:2015icj}, and we summarize the relevant formulas in App.~\ref{app: Monopole and dipole perturbations}. These closed-form solutions are required for the subsequent computation of scalar perturbations; in particular, Ref.~\cite{Brito:2023pyl} emphasizes that the dipole mode plays a key role in determining the scalar radiation absorbed by the BH horizon.

\subsection{Scalar perturbations}
At order $\mathcal{O}(\epsilon^1q^1)$, we need to solve the inhomogeneous KG equation~\eqref{eq: (1,1) order KG equation}, which describes perturbations to the scalar cloud configuration. Similarly to Refs.~\cite{Annulli:2020lyc, Brito:2023pyl}, we can decompose the scalar perturbations as:
\begin{equation}
\begin{aligned}\label{eq: (1,1) order scalar field ansatz}
\phi^{(1,1)}= & \frac{1}{r}\sum_{\ell_f,m_f}\left[Z_{\ell_f m_f}^{T}(t, r)Y_{\ell_fm_f}(\theta^{A})\right]e^{-i\omega t}.
\end{aligned}
\end{equation}
Using this \textit{ansatz} in Eq.~\eqref{eq: (1,1) order KG equation} and factorizing the source term as $S^\phi= (r-2M)^{-1}\sum_{l,m}S_{lm}^{T}(t,r,\theta^A)\, e^{-i\omega t}$, we find that the resulting equation for $Z^T_{\ell_fm_f}$ is
\begin{align}\label{eq:inhomogeneous wavefunction}
\sum_{\ell_{f},m_{f}}Y_{\ell_{f}m_{f}}(\theta^A)&\left[\frac{\partial^{2}}{\partial r_{*}^{2}}-\left(\frac{\partial}{\partial t}-i\omega\right)^2-V_f(r)\right]Z_{\ell_{f}m_{f}}^{T}(t,r)\nonumber\\
&=\sum_{l,m}S_{lm}^{T}(t,r,\theta^A),
\end{align}
where
\begin{equation}
    V_f = \left(1-\frac{2M}{r}\right)\left(\mu^{2}+\frac{\ell_{f}(\ell_{f}+1)}{r^{2}}+\frac{2M}{r^{3}}\right),
\end{equation}
and
\begin{equation}
\begin{aligned}\label{eq:source in TD}
  S_{lm}^{T}(t,r,\theta^A)&=P_{lm}^{T}(t,r)Y_{lm}Y_{\ell_im_i} \\
 & +\hat{P}_{lm}^{T}(t, r)\left(Y_{,\theta}^{lm}Y_{,\theta}^{\ell_im_i}+\frac{Y_{,\varphi}^{lm}Y_{,\varphi}^{\ell_im_i}}{\sin^2\theta}\right) \\
 & +A_{lm}^{T}(t,r)\frac{Y_{,\theta}^{lm}Y_{,\varphi}^{\ell_{i}m_{i}}-Y_{,\varphi}^{lm}Y_{,\theta}^{\ell_{i}m_{i}}}{\sin\theta}.
\end{aligned}
\end{equation}
In the Regge-Wheeler gauge, the TD functions appearing in the source term take the form
\begin{widetext}
\begin{align}
\begin{split}\label{eq: radial functions of source}
 P^T_{lm}(t,r) =& -\frac{1}{2 f r} R \Bigg[ r^2 2 \omega^2 H_0 + i r^2 \omega \dot{H}_0 + f \Big\{ -4 i f r \omega H_1 + f \big[ -2 f (\ell_i + \ell_i^2 + \mu^2 r^2) + r^2 2 \omega^2 \big] H_2 \\
& + i f r^2 \omega \dot{H}_2 + 2 f \ell_i(\ell_i+1)K
  + 2 i r^2 \omega \dot{K} - 2 i f r^2 \omega H_1'
  \Big\}\Bigg] \\
& + \frac{R'}{2}\Big\{(f-1)H_0
  + f\big[4 i r \omega H_1 - 2 r \dot{H}_1 + r H_0'
  + f\big((1-f)H_2 + f r H_2' - 2 r K'\big)\big]\Big\},\\
\hat{P}^T_{lm}(t,r)=&\frac{R\left(H_0-f^2H_2\right)}{2r}, \\
A^T_{lm}(t,r)=&\frac{2i\omega rh_0R-r\dot{h}_0R+f\left\{frRh_1^{\prime}+\left[\left(1-f\right)R+2frR^{\prime}\right]h_1 \right\}}{r^2},
\end{split}
\end{align}
\end{widetext}
where $R:= R^b_{n_i\ell_i}(r)$ is the radial function of the scalar cloud and overdots represent time derivatives. To simplify the notation, we also omit the explicit dependence of the metric functions on $l$ and $m$. Notice that the functions $P^T_{lm}$ and $\hat{P}^T_{lm}$ only depend on polar functions, whereas $A^T_{lm}$ only depends on axial functions.

Since the source term in Eq.~\eqref{eq:source in TD} depends linearly on the metric functions, which, as we discussed previously, can be represented as Fourier series in $\sigma_{mn}$, it is natural to also expand $Z_{\ell_f m_f}^{T}$ in a Fourier series representation:
\begin{equation}\label{eq: scalar_pert_amps}
    Z_{\ell_f m_f}^{T}(t, r) = \sum_{n}Z^{\ell_fm_fn}(r)e^{-i\sigma_{mn} t}\,.
\end{equation}
The same can be done for $S_{lm}^{T}$, but, in practice, this amounts to expanding each time-domain function $P^T_{lm}$, $\hat{P}^T_{lm}$ and $A^T_{lm}$ in a Fourier series with Fourier harmonic amplitudes given by
\begin{align}
    P_{lmn}(r) =\frac{1}{T_r}\int_0^{T_r}dt\,P_{lm}^{T}(t,r)e^{i\sigma_{mn}t},\label{eq: Plmn in FD}\\
    \hat{P}_{lmn}(r) =\frac{1}{T_r}\int_0^{T_r}dt\,\hat{P}_{lm}^{T}(t,r)e^{i\sigma_{mn}t},\\
    A_{lmn}(r) =\frac{1}{T_r}\int_0^{T_r}dt\,A_{lm}^{T}(t,r)e^{i\sigma_{mn}t}.
\end{align}

As done in Ref.~\cite{Brito:2023pyl}, we separate the angular part in Eq.~\eqref{eq:inhomogeneous wavefunction} by projecting it onto a spherical harmonics basis. Namely, we multiply Eq.~\eqref{eq:inhomogeneous wavefunction} by $Y^*_{\ell'_f m'_f}$ and integrate over the solid angle. Using the orthogonality properties of the spherical harmonics, we find one radial equation for each pair of angular numbers $\{\ell'_f,m'_f\}$ with a source term that contains the following integrals:
\begin{subequations}\label{eq:spherical harmonics integrals}
\begin{align}
 & \mathcal{P}_{m_f^{\prime},m,m_i}^{\ell_f^{\prime},l,\ell_i}:=\int d\Omega\, Y_{\ell_f^{\prime}m_f^{\prime}}^*Y_{lm}Y_{\ell_im_i},\label{subeq:spherical harmonics integrals P} \\
 & \hat{\mathcal{P}}_{m_f^{\prime},m,m_i}^{\ell_f^{\prime},l,\ell_i}:=\int d\Omega\, Y_{\ell_f^{\prime}m_f^{\prime}}^*\mathbf{Y}_a^{lm}\mathbf{Y}_b^{\ell_im_i}\gamma^{ab},\label{subeq:spherical harmonics integrals Phat} \\
 & \mathcal{A}_{m_f^{\prime},m,m_i}^{\ell_f^{\prime},l,\ell_i}:=\int d\Omega\, Y_{\ell_f^{\prime}m_f^{\prime}}^*\mathbf{X}_a^{lm}\mathbf{Y}_b^{\ell_im_i}\gamma^{ab},\label{subeq:spherical harmonics integrals A}
\end{align}
\end{subequations}
where $\gamma^{ab}:=\text{diag}(1,1/\sin^2\theta)$ and $\mathbf{Y}_a^{lm} := (Y^{lm}_{,\theta}, Y^{lm}_{,\varphi})$, $\mathbf{X}_a^{lm} := \left(-\frac{Y_{,\varphi}^{lm}}{\sin\theta}\text{, sin }\theta Y_{,\theta}^{lm}\right)$ are vector spherical harmonics. As shown in App.~D of Ref.~\cite{Brito:2023pyl}, the integrals in Eq.~\eqref{eq:spherical harmonics integrals} can be computed explicitly in terms of the Wigner 3-j symbols, such that the integrals vanish unless the following selection rules are satisfied:
\begin{itemize}
    \item $m_i+m-m_f=0$;
    \item $|\ell_f-\ell_i|\leq l\leq \ell_f+\ell_i$;
    \item $\ell_f+\ell_i + l$ is even for the integrals~\eqref{subeq:spherical harmonics integrals P} and~\eqref{subeq:spherical harmonics integrals Phat};
    \item $\ell_f+\ell_i + l$ is odd for the integral~\eqref{subeq:spherical harmonics integrals A}.
\end{itemize}
Using these properties allows us to obtain an ODE for the Fourier-domain amplitude $Z^{\ell_fm_fn}$ given by:
\begin{equation}\label{eq:ordinary differential equation}
\left[\frac{d^2}{dr_*^2}+(\omega+\sigma_{mn})^2-V_f\right]Z^{\ell_fm_fn}={S}_{m_f,m_i;n}^{\ell_f,\ell_i}\,,
\end{equation}
where $m=m_f-m_i$ due to the selection rules. In particular, if $\ell_i=m_i=1$, the source term is given by
\begin{equation}
\begin{aligned}
 {S}_{m_f,1;n}^{\ell_f,1}(r) &=
\left[ \right.
\mathcal{P}_{m_f,m,1}^{\ell_f,l,1}P_{lmn}(r)\left(\delta_{l,\ell_f-1}+\delta_{l,\ell_f+1}\right) \\
 & +\hat{\mathcal{P}}_{m_f,m,1}^{\ell_f,l,1}\hat{P}_{lmn}(r)\left(\delta_{l,\ell_f-1}+\delta_{l,\ell_f+1}\right) \\
 & \left.+\mathcal{A}_{m_f,m,1}^{\ell_f,l,1}A_{lmn}(r)\delta_{l,\ell_f}\right]\delta_{m,m_f-1}.
\end{aligned}
\end{equation}

As in the case of metric perturbations, solutions to Eq.~\eqref{eq:ordinary differential equation} can be found using a Green's function approach. For the scalar perturbations, the asymptotic behaviour of the solutions at infinity depends on $\rm sgn(\omega_+^2-\mu^2)$, where $\omega_+ \equiv \omega + \sigma_{mn}$. If $\omega_+^2-\mu^2 > 0$, one can construct one solution that is purely outgoing at spatial infinity and another that is purely ingoing at the horizon:
\begin{align}
& Z_{\mathrm{in}}(r \to 2M) = e^{-i\omega_+r_*},\label{eq:Zin}\\
& Z_{\mathrm{up}}(r \to \infty) = e^{ik_+r_*}r^{-\nu_+},\label{eq:Zup}
\end{align}
where $k_+\equiv\mathrm{~sgn}(\omega_+)\sqrt{\omega_+^2-\mu^2}$ and $\nu_+\equiv-iM\mu^2/k_+$. The factor sgn$(\omega_+)$ ensures that $Z_{\mathrm{up}}$ describes an outgoing wave at infinity. By contrast, if $\omega_+^2-\mu^2 < 0$, waves cannot propagate to infinity, and we instead impose regularity of $Z_{\mathrm{up}}$ there. In this case, we take $k_{+} \equiv \sqrt{\omega_+^2-\mu^2}$ with the principal square root in Eq.~\eqref{eq:Zup}, which yields $Z_{\mathrm{up}} \sim e^{-|k_+|r_{*}}r^{-\nu_+}$ at infinity. The solution of the inhomogeneous equation satisfying the appropriate boundary conditions is then given by (omitting indices for clarity)
\begin{equation}
\begin{aligned}\label{eq:solution of inhomogeneous equation}
Z(r)&= \frac{Z_{\mathrm{up}}(r)}{\mathcal{W}}\int_{2M}^r\frac{Z_{\mathrm{in}}(r^{\prime}){S}(r^{\prime})}{f(r^{\prime})}\,dr^{\prime} \\
 &\qquad +\frac{Z_{\mathrm{in}}(r)}{\mathcal{W}}\int_r^\infty\frac{Z_{\mathrm{up}}(r^{\prime}){S}(r^{\prime})}{f(r^{\prime})}\,dr^{\prime},
\end{aligned}
\end{equation}
where the Wronskian is now given by
\begin{equation}\label{eq:Wronkskian_Z}
\mathcal{W} := Z_{\mathrm{in}}\frac{dZ_{\mathrm{up}}}{dr_{*}}-Z_{\mathrm{up}}\frac{dZ_{\mathrm{in}}}{dr_{*}}.
\end{equation}

\subsubsection{Energy and angular momentum loss rates}\label{sec: scalar fluxes}
In addition to GWs, the perturbations due to the secondary object will induce the emission of scalar waves, both to infinity and through the BH horizon~\cite{Brito:2023pyl}. For the purpose of computing the fluxes related to the scalar perturbations $\phi^{(1,1)}$, we will simply use Eq.~\eqref{eq: stress-energy tensor definition} evaluated in a Schwarzschild BH background~\cite{Brito:2023pyl,Dyson:2025dlj}. 
The fluxes related solely to scalar
perturbations are of order $\mathcal{O}(\epsilon^2q^2)$,
since they involve terms quadratic in the physical scalar
perturbation $\epsilon q\,\phi^{(1,1)}$.
In order to simplify the notation, in the following we
will absorb the factor $\epsilon q$ into the definition
of $\phi^{(1,1)}$. Thus, all expressions presented below
are proportional to $\epsilon^2q^2$.

In a Schwarzschild background, the (averaged) energy fluxes of the perturbed field can be computed using the scalar field's stress-energy tensor~\cite{Teukolsky:1973ha,Teukolsky:1974yv}:
\begin{equation}\label{eq:dotE_Phi}
\begin{aligned}
    \dot E^{\Phi,\infty} &= -\lim_{r\to +\infty} r^2\int d\Omega\, T^{\phi^{(1,1)}}_{\mu r}\xi_{(t)}^{\mu}\,,\\ 
    \dot E^{\Phi,H} &= \lim_{r\to 2M} 4M^2 \int d\Omega\, T^{\phi^{(1,1)}}_{\mu\nu}\xi_{(t)}^{\mu}l^{\nu}\,,
\end{aligned}
\end{equation}
where $\xi_{(t)}^{\mu}:=\partial/\partial t$ is the Killing vector field associated with the time-translation invariance of the BH metric and $l^{\mu} =\partial/\partial t$ is the null vector, normal to the horizon. Analogous expressions for angular momentum fluxes $\dot L^{\Phi,\infty/H}$ are obtained by replacing $\xi_{(t)}^{\mu}$ with $\xi_{(\varphi)}^{\mu}$ in Eq.~\eqref{eq:dotE_Phi}, where $\xi_{(\varphi)}^{\mu}:=\partial/\partial \varphi$ is the Killing vector field associated with the axisymmetry of the BH geometry.

To compute the rates at which the mass $M_b$ and spin $J_b$ of the
cloud change, we use~\cite{Brito:2023pyl}
\begin{equation}
\dot{M}_b=\omega\,\dot{Q},\quad \dot{J}_b=m_i\,\dot{Q}\,.
\end{equation}
Using the conservation of the current defined in Eq.~\eqref{eq: current}, together with the divergence theorem, to compute the (averaged) rate of change of the scalar charge $\dot{Q}$, with
\begin{equation}
\begin{aligned}
 & \dot{Q}^{\infty}=-\lim_{r\to+\infty}r^{2}\int d\Omega\, j_{r}, \\
 & \dot{Q}^{H}=\lim_{r\to2M}4M^{2}\int d\Omega\, j_{\mu} l^{\mu}.
\end{aligned}
\end{equation}

The gravitational perturbation induced by the secondary acts as a forcing term for the scalar field, leading to an exchange of energy and angular momentum between the secondary and the scalar cloud. In a steady-state regime, this exchange is encoded in the fluxes of energy and angular momentum carried by the scalar perturbations to spatial infinity and through the BH horizon~\cite{Clough:2021qlv, Croft:2022gks, Annulli:2020lyc}. By conservation of the system’s total energy and angular momentum, as shown in Ref.~\cite{Brito:2023pyl}, the particle's energy $\mathcal{E}$ and angular momentum $\mathcal{L}$ are expected to change according to
\begin{equation}\label{eq:balance_law}
\begin{aligned}
    &\dot{\mathcal{E}}=-m_p^{-1}\left(\dot{E}^{g,\infty}+\dot{E}^{g,H}+\dot{E}^{s,\infty}+\dot{E}^{s,H}\right),\\
    &\dot{\mathcal{L}}=-m_p^{-1}\left(\dot{L}^{g,\infty}+\dot{L}^{g,H}+\dot{L}^{s,\infty}+\dot{L}^{s,H}\right),
\end{aligned}
\end{equation}
where $\dot{E}^{g,\infty/H}$ and $\dot{L}^{g,\infty/H}$ are the GW energy and angular momentum fluxes at infinity and at the BH horizon, respectively, and we defined
\begin{align}
    &\dot{E}^{s,\infty/H}\:=\:\dot{E}^{\Phi,\infty/H}+\omega\:\dot{Q}^{\infty/H}\:,\label{eq:energy_loss}\\
    &\dot{L}^{s,\infty/H}\:=\:\dot{L}^{\Phi,\infty/H}+m_i\:\dot{Q}^{\infty/H}\:.\label{eq:angular_momentum_loss}
\end{align}
The quantities $\dot{E}^s=\dot{E}^{s,\infty}+\dot{E}^{s,H}$ and $\dot{L}^s=\dot{L}^{s,\infty}+\dot{L}^{s,H}$ represent the total rate of change of the energy and angular momentum of the secondary due to the presence of the scalar field configuration. We note that a first-principles derivation of these balance laws is still lacking. Such a derivation may be possible using the results of Ref.~\cite{Dyson:2026ddd}. 

Using the flux expressions above, together with Eqs.~\eqref{eq: (1,1) order scalar field ansatz} and~\eqref{eq: scalar_pert_amps}, one finds~\cite{Brito:2023pyl, Duque:2023seg}
\begin{equation}
\begin{aligned}
    &\dot{E}_{\ell_f,m_f,n}^{s,\infty}=2\sigma_{mn}\,{\rm sgn}(\omega_+) \Re\left[\sqrt{\omega_+^2-\mu^2}\right]\left|{Z}^{\ell_fm_fn}(\sigma_{mn})\right|^2,\\
    &\dot{E}_{\ell_f,m_f,n}^{s,H}=2\sigma_{mn}\omega_+ \left|{Z}^{\ell_fm_fn}(\sigma_{mn})\right|^2\,,\label{eq:energy loss modes}
\end{aligned}
\end{equation}
To avoid clutter, we omit the explicit evaluation limits for ${Z}^{\ell_fm_fn}(\sigma_{mn})$, i.e., $r\to\infty$ for fluxes at infinity and $r\to 2M$ for horizon fluxes. Similarly, for the angular momentum loss rates, one finds~\cite{Brito:2023pyl, Duque:2023seg}
\begin{equation}
\begin{aligned}
    &\dot{L}_{\ell_f,m_f,n}^{s,\infty} = 2m\,{\rm sgn}(\omega_+) \Re\left[\sqrt{\omega_+^2-\mu^2}\right]\left|{Z}^{\ell_fm_fn}(\sigma_{mn})\right|^2\,,\\
    &\dot{L}_{\ell_f,m_f,n}^{s,H}= 2m\,\omega_+ \left|{Z}^{\ell_fm_fn}(\sigma_{mn})\right|^2\,.\label{eq:angular momentum loss modes}
\end{aligned}
\end{equation}
These expressions can be easily related to the (mode-by-mode) decomposition of the Noether-charge fluxes through $\dot E^{s,\infty/H}_{\ell_f m_f n}=-\sigma_{mn}\dot Q^{\infty/H}_{\ell_f m_f n}$ and $\dot L^{s,\infty/H}_{\ell_f m_f n}=-m\dot Q^{\infty/H}_{\ell_f m_f n}$. The  quantity $-\dot Q_{\ell_f m_f n}=-\dot Q^{\infty}_{\ell_f m_f n}-\dot Q^H_{\ell_f m_f n}$ can then be interpreted as the rate at which scalar particles are transferred from the background state into the $\{\ell_f, m_f\}$ state through the $n$-th radial harmonic, with each scalar particle depositing energy $-\sigma_{mn}$ and angular momentum $-m$ into the orbit.

In the Newtonian limit, the quantity $\dot{E}^{s,\infty}$ is equivalent to the {\it ionization power} first computed in Ref.~\cite{Annulli:2020lyc,Baumann:2021fkf,Tomaselli:2023ysb}, whereas $\dot{L}^{s,\infty}$ is equivalent to the {\it ionization torque}, defined as the rate of angular momentum transferred into the continuum in Ref.~\cite{Annulli:2020lyc, Tomaselli:2023ysb}. To make this connection explicit, in App.~\ref{App:Revisiting Newtonian fluxes} we revisit the computation of the orbital loss rates in the Newtonian limit using a formalism that closely follows the relativistic framework. By contrast, as discussed in Refs.~\cite{Brito:2023pyl,Tomaselli:2025jfo}, the horizon fluxes $\dot{E}^{s,H}$ and $\dot{L}^{s,H}$ encode information about on- and off-resonant transitions between quasi-bound states of the cloud~\cite{Baumann:2018vus,Baumann:2019ztm}. 
As noted in Ref.~\cite{Tomaselli:2025jfo}, the relativistic horizon loss rates computed in Refs.~\cite{Brito:2023pyl,Dyson:2025dlj} cannot be reproduced using a Newtonian approximation where one assumes that $\dot{E}^{s,H}$ is only due to transitions between quasi-bound states. 

We note that throughout this work we neglect $\mathcal{O}(\epsilon^2)$ conservative corrections from the cloud's self-gravity, as well as $\mathcal{O}(\epsilon^2q^2)$ corrections to the gravitational-wave fluxes~\cite{Brito:2023pyl,Keijzer:2026vul}, neither of which has yet been computed for dipolar clouds. For spherical clouds, these corrections have been shown to be approximately degenerate with a redshift factor at small $M\mu$ and with a mass shift at large $M\mu$~\cite{Duque:2023seg,Keijzer:2026vul}. Their relative importance at intermediate $M\mu$ for dipolar clouds is unclear, but they should not affect the main qualitative features discussed below and in our companion Letter~\cite{Xu:2026cky}, in particular the appearance of eccentricity-induced resonances. We further note that recent work~\cite{Dyson:2026ddd} has derived a relativistic local torque-balance equation from first principles, identifying---alongside the usual matter-induced torques---an additional contribution that they termed \textit{geometric torques}, which is absent in the Newtonian limit. This term arises from couplings between the background matter environment and quadratic metric perturbations. Although~\cite{Dyson:2026ddd} indicates that geometric torques are subdominant, their quantitative impact in our setup remains to be assessed.

\subsection{Results}\label{sec:results_equatorial}
In order to compute $\dot{E}^{s,H}$ and $\dot{L}^{s,H}$, we implemented the procedure described above in \texttt{Julia} (details of the numerical procedure are given in App.~\ref{App:Numerics}). While our implementation is generic, the results presented in this section focus specifically on the case where the background scalar field is in a prograde dipolar state, characterized by the quantum numbers $\{n_i, \ell_i, m_i\}=\{0,1,1\}$. Since $\dot{E}^{s,H}$ and $\dot{L}^{s,H}$ scale linearly with the cloud mass fraction $M_b/M$ and quadratically with the mass ratio $q$, we present all results normalized by $q^2 M_b/M$. For concreteness, and since our aim is mainly to understand the impact of eccentricity, we restrict our attention to configurations with $M\mu=0.2$ unless otherwise stated. In this section, we also focus on {\it prograde} orbits, meaning the orbital angular momentum points in the same direction as the cloud's spin. Generic orbits will be discussed in Sec.~\ref{Sec:Point Particle in Generic Orbits}.

\begin{figure*} 
    \centering
    \includegraphics[width=0.48\textwidth]{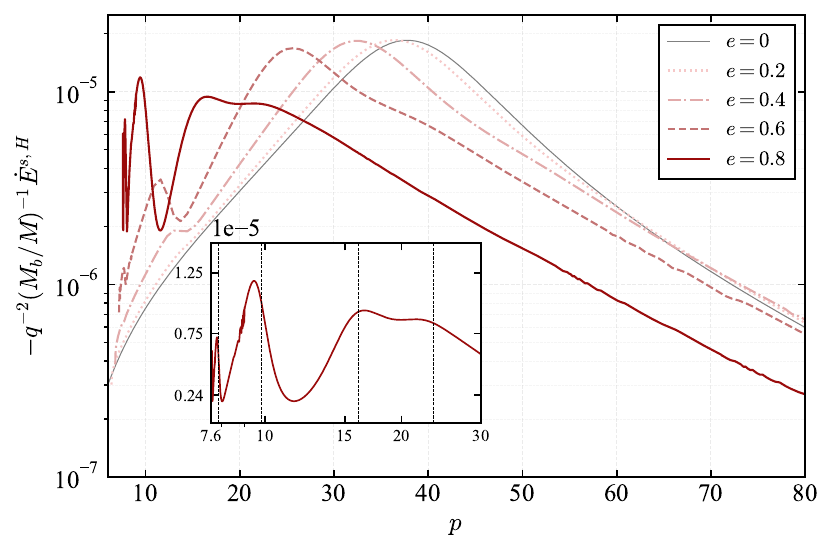}
     \includegraphics[width=0.48\textwidth]{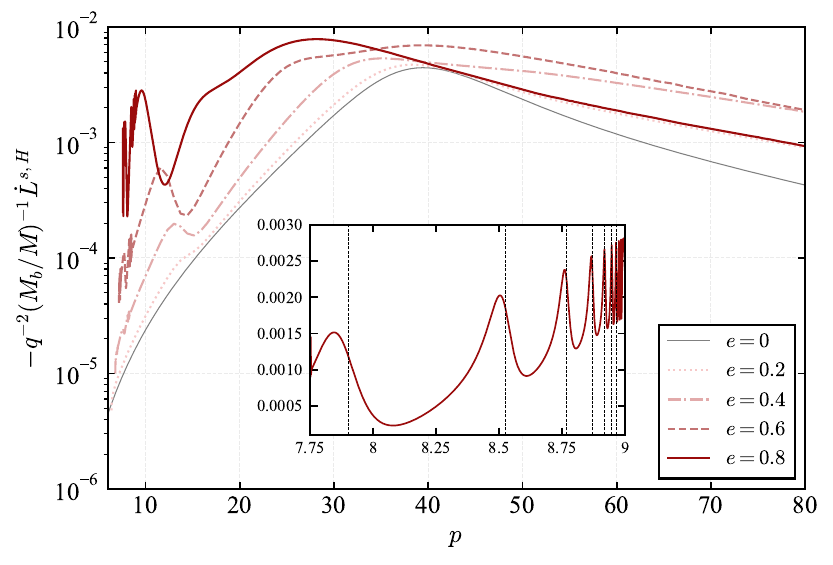}
    \caption{Horizon loss rates $\dot E^{s,H}$ ({\it left panel}) and $\dot L^{s,H}$ ({\it right panel}) as a function of the semi-latus rectum $p$, for different values of eccentricity $e$. We consider a prograde dipolar $n_i=0$, $\ell_i=m_i=1$ scalar cloud, with $M\mu=0.2$.}
    \label{fig:HorFlux}
\end{figure*}

\subsubsection{Horizon loss rates}
We first present the loss rates $\dot E^{s,H},\dot L^{s,H}$ due to fluxes at the horizon.
For the horizon loss rates, the $\{\ell_{f},m_{f}\}=\{0,0\}$ mode is the dominant contribution~\cite{Brito:2023pyl} (see App.~\ref{App:Comparison of different modes}). Since this mode couples only to $\{l,m\}=\{1,-1\}$ metric perturbations, for which there is a fully analytic solution (App.~\ref{app: Monopole and dipole perturbations}), in what follows we include only this mode in the computation of the horizon loss rates. This allows us to obtain results even at very high eccentricities.

The main results are summarized in Fig.~\ref{fig:HorFlux}, where we show $\dot L^{s,H}$ and $\dot E^{s,H}$ as a function of the semi-latus rectum $p$, for different values of eccentricity $e$. A prominent feature that can be observed is a series of resonance peaks, also discussed in our companion letter~\cite{Xu:2026cky}. For the values of eccentricity shown here, the loss rates increase with eccentricity in the strong-field regime, and larger eccentricities typically lead to a greater number of resonance peaks.
For a cloud with generic quantum numbers $\{n_i,\ell_i,m_i\}$, these resonances occur at an infinite set of orbital frequencies, which we label by $n_f$, given by~\cite{Brito:2023pyl,Xu:2026cky}:
\begin{equation}\label{eq:resonances}
    \sigma^{n_f}_{mn}=\Re(\omega_{n_f \ell_f m_f}) - \Re(\omega_{n_i\ell_i m_i})\,,
\end{equation}
where $\omega_{n \ell m}$ denotes an eigenfrequency of the KG equation with quantum numbers $\{n,\ell,m\}$ (see Sec.~\ref{sec: Scalar cloud}). 

In the inset of the left panel of Fig.~\ref{fig:HorFlux}, we show that, for $e=0.8$, some of the peaks in $\dot E^{s,H}$ are consistent with the location predicted by Eq.~\eqref{eq:resonances} for the resonant excitation of the fundamental mode $\{n_f,\ell_f,m_f\} = \{0,0,0\}$. In particular, the dashed vertical lines, ordered from left to right, indicate the resonance positions associated with the Fourier harmonic indices $n = 2, 1, 0$, and $-1$, respectively.
Zooming in closer to the LSO, we also find a clear resonance substructure associated with the excitation of different overtones $n_f$. This is most apparent at high eccentricities. 

In the inset of the right panel of Fig.~\ref{fig:HorFlux}, we show a magnified view of the angular momentum loss rate for $e=0.8$, where a substructure associated with resonances in the $n=2$ harmonic is visible. The dashed vertical lines, ordered from left to right, indicate the resonance positions corresponding to the excitation of the overtones $n_{f} = 0, 1, 2, \dots$. Interestingly, the peak position of the loss rate does not coincide exactly with the location of the expected resonance frequency. This behavior, which has been observed in previous work~\cite{Brito:2023pyl}, is typical of resonances involving modes with a large damping rate, as explained in Ref.~\cite{Santos:2026lzq}. As we discuss below, the same resonant pattern is also present for higher-$n$ modes, but at smaller values of $p$.

These results show that orbital eccentricity can excite a dense spectrum of resonances in the vicinity of the LSO. Importantly, one can easily check that these resonances arise because $\Omega_{\rm \varphi}\neq \Omega_{r}$ near the LSO. As a consequence, these resonances could not have been predicted from a purely Newtonian analysis, where $\Omega_{\rm \varphi}= \Omega_{r}$.

The accumulation of resonances close to the LSO can be understood directly
from the limiting behavior of the orbital frequencies. As the separatrix is approached, the radial period diverges and therefore $\Omega_r\to0$, while $\Omega_\varphi$ approaches a finite value associated with the whirl motion. For the prominent $m=-1$ resonance family, Eq.~\eqref{eq:resonances} can be written as $-\Omega_\varphi+n\Omega_r=\Delta\omega$, where $\Delta\omega$ is the relevant
cloud level splitting. Thus, $n\simeq(\Omega_\varphi+\Delta\omega)/\Omega_r$, and progressively larger values
of $n$ are required as $\Omega_r$ decreases. This explains the tower of higher-$n$ resonances accumulating toward the LSO and makes explicit why this structure has no Newtonian counterpart.

\begin{figure*}
    \centering
    \includegraphics[width=0.8\textwidth]{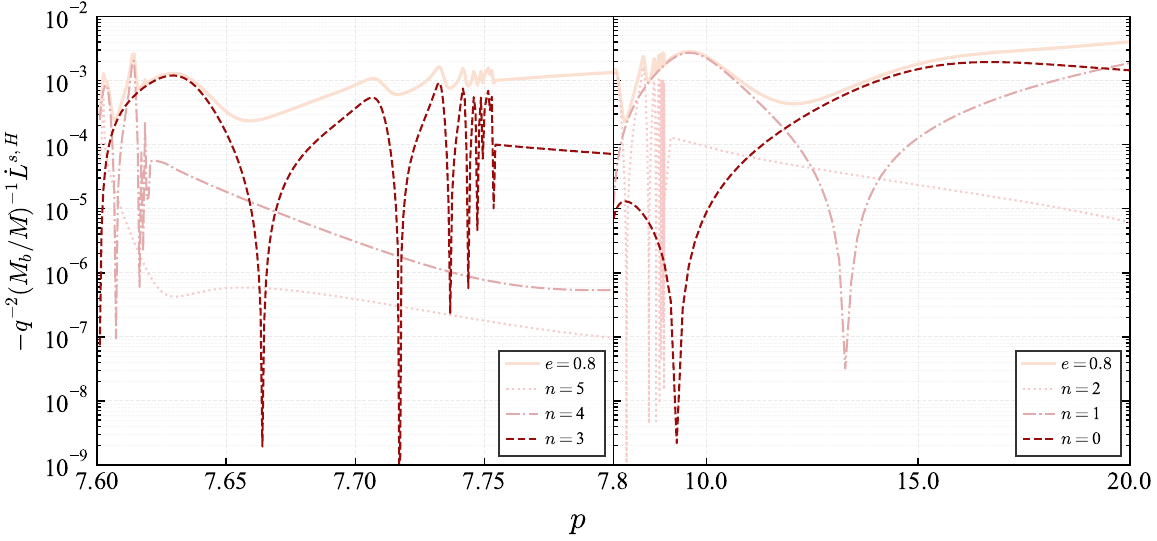}    
    \caption{Angular momentum loss rate $\dot L^{s,H}$ as a function of the semi-latus rectum $p$ for eccentric orbits with $e=0.8$. We zoom over two regions in the range $p\in [7.6,20]$. The total loss rate is compared with individual mode contributions: $n=3, 4, 5$ in the left subplot and $n=0, 1, 2$ in the right subplot.} 
    \label{fig:CompareHorFluxe08_angular momentum}
\end{figure*}

To better understand the resonant substructure, in  Fig.~\ref{fig:CompareHorFluxe08_angular momentum} we show the contributions from the individual $n$ modes for the $e=0.8$ case, together with the total loss rate (solid line) obtained by summing over modes from $n=-15$ to $n=20$. One can clearly see that the peaks slightly below $p\sim 10$ are associated with resonances in the $n=2$ mode. Their locations are consistent with the resonant excitation of different quasi-bound states with quantum numbers $\{n_f,\ell_f,m_f\}=\{n_f,0,0\}$. In addition, the $n=1$ contribution exhibits a broad peak around $p\sim 10$, corresponding to the resonance with the $\{n_f,\ell_f,m_f\}=\{0,0,0\}$ state. As $p$ decreases further, as illustrated in the left panel of Fig.~\ref{fig:CompareHorFluxe08_angular momentum}, resonances begin to appear in progressively higher-$n$ modes. In particular, the larger the value of $n$, the closer the corresponding resonant peaks lie to the LSO. 

\begin{figure*}
\centering
\includegraphics[width=0.8\textwidth]{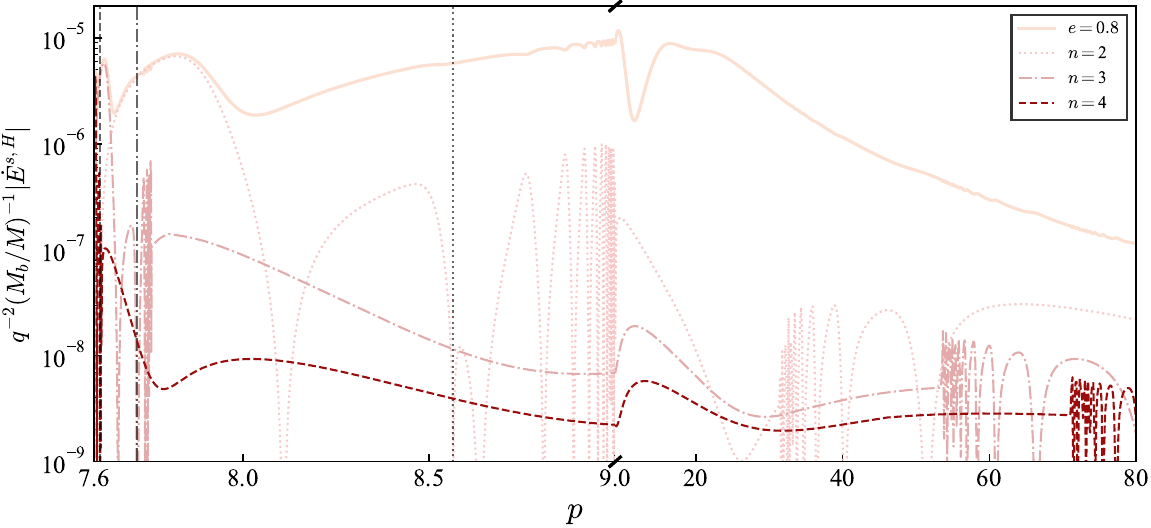}
\caption{Absolute value of the energy loss rate $\dot E^{s,H}$ as a function of the semi-latus rectum $p$ for eccentric orbits with $e=0.8$. We zoom over two regions in the interval $p \in [7.6, 80]$. The solid line (labeled $e=0.8$) represents the total loss rate summed over Fourier modes from $n=-15$ to $n=20$. To illustrate the origin of the oscillations seen in the total rate, we also show individual mode contributions from $n=2, 3, 4$. Vertical lines indicate the points where individual $n$-mode contributions change from negative (to the left) to positive values (to the right), with the line style identifying the corresponding $n$ mode.}\label{fig:CompareHorFluxe08_energy}
\end{figure*}

Another noteworthy feature is that the angular momentum loss rate $\dot L^{s,H}$ is always negative, at least within the region of parameter space considered here. This can be understood from the fact that the horizon loss rates are dominated by modes with $\{\ell_f,m_f\}=\{0,0\}$. Since the initial cloud is in the $\{\ell_i,m_i\}=\{1,1\}$ state, the process necessarily reduces the angular momentum carried by the cloud.
Moreover, because $m_f=0$, one can directly verify that $\dot L_{\Phi,m_f=0}^{H}=0$ from Eq.~\eqref{eq:angular_momentum_loss}; see also \cite[Eq.~(85)]{Brito:2023pyl}.
Conservation of angular momentum therefore implies that angular momentum is transferred directly from the cloud to the orbiting particle; note that GW emission, when taken into account, will still cause the particle to inspiral, as it generically dominates over the exchange with the scalar field in the strong-field regime~\cite{Xu:2026cky}. 
This process is physically analogous to tidal acceleration in binary systems, as can be seen more explicitly from Eq.~\eqref{eq:angular momentum loss modes}. For $m_f=0$ and $m_i=1$, one finds $\dot L_{00n}^{s,H}\propto -(\omega-\Omega_{\varphi}+n\Omega_{r})$; hence, whenever $\omega>\Omega_{\varphi}-n\Omega_r$, we have $\dot L^{s,H}_{00n}<0$. This condition is always satisfied for the $n$ modes that dominate the loss rate. The mechanism is analogous to tidal acceleration in the Earth--Moon system: since the Earth's rotational angular frequency exceeds the Moon's orbital frequency, $\Omega_{\rm Earth}>\Omega_{\rm Moon}$, tidal interactions transfer angular momentum from the Earth's rotation to the Moon's orbit. Here, an analogous transfer occurs from the scalar cloud to the orbiting secondary.


The behavior of the energy loss rate $\dot E^{s,H}$ is slightly more involved. While the total $\dot E^{s,H}$ remains negative throughout the parameter space considered here, the contributions from individual $n$ modes can change sign in certain regions. This is illustrated in Fig.~\ref{fig:CompareHorFluxe08_energy}, where the total energy loss rate (solid line) for $e=0.8$ is compared with the separate $n$-mode contributions. Focusing on the region $p<9$, the vertical lines indicate the points at which the individual $n\geq 2$ contributions switch from positive (to the right of the lines) to negative (to the left). The linestyle of each vertical marker identifies the corresponding $n$ mode; for instance, the dotted vertical line denotes the sign change of the $n=2$ contribution. The sign of each contribution is governed by $\sigma_{-1n}=-\Omega_{\varphi}+n\Omega_r$ [see Eq.~\eqref{eq:energy loss modes}]. In particular, $\sigma_{-1n}$ changes sign precisely at the locations of the vertical lines for $n=2,3,4$ (from right to left, respectively), and is positive to their right. By contrast, $\sigma_{-1n}<0$ for $n\leq 1$, which explains why the total $\dot E^{s,H}$ remains negative, since those modes dominate at larger $p$.

Resonances of a given $n$ mode with $\{n_f\geq 2,\ell_f=0,m_f=0\}$ states occur precisely in regions where the corresponding individual $n$ contribution satisfies $\dot E^{s,H}_{00n}>0$, while the opposite holds for resonances involving states with $n_f\leq 1$. This behavior is consistent with the expected on-resonance energy transfer: states with $\{n_f\leq 1,\ell_f=0,m_f=0\}$ have lower energy than the initial $\{n_i,\ell_i,m_i\}=\{0,1,1\}$ state [see~\cite[Fig.~3]{Baumann:2019eav}], so these resonances extract energy from the cloud and transfer it to the secondary, whereas the opposite occurs for resonances involving $n_f\geq 2$ states.

Besides the more prominent resonances discussed above, the energy loss rates for $e=0.6$ and $e=0.8$ shown in Fig.~\ref{fig:HorFlux} also exhibit small fluctuations at large $p$. These features are not numerical artifacts; rather, they arise from resonances associated with positive-$n$ Fourier modes. This is illustrated in Fig.~\ref{fig:CompareHorFluxe08_energy}, where the total energy loss rate (solid line) for $e=0.8$ is compared to individual $n$-mode contributions.
The fact that the $n\geq 2$ modes exhibit resonances both at large $p$ and close to the LSO can be understood from the behavior of $\sigma_{mn}$. For $n\geq 2$ and $m=-1$ (recall that $m=m_f-m_i$), $\sigma_{mn}$ is a non-monotonic function of $p$ for fixed eccentricity. As a result, the resonance condition~\eqref{eq:resonances} can be satisfied at two distinct orbital locations for the same set of quantum numbers. In particular, the oscillations observed in the range $p\sim 30$--$80$ arise from resonances of different $n\geq 2$ modes with quasi-bound states characterized by $\{n_f\geq 2,\ell_f=0,m_f=0\}$, with larger values of $n$ producing resonances at correspondingly larger values of $p$.


More generally, the appearance of resonances can be anticipated by examining the sign of $\omega_+^2-\mu^2$ for a given mode. In the regime where $\omega_+^2-\mu^2>0$ (for example, $p\in(9,32)$ for the $n=2$ mode), the mode's wavefunction behaves as a propagating wave at infinity. By contrast, when $\omega_+^2-\mu^2<0$ ($p\lesssim 9$ or $p\gtrsim 32$ for $n=2$), the mode becomes exponentially bound at infinity, allowing quasi-bound states of the cloud to be resonantly excited. As the threshold $\omega_+^2-\mu^2\to 0^-$ is approached, an infinite tower of states can in principle be excited, explaining the large number of resonances observed here (see also Ref.~\cite{Tomaselli:2025jfo} for a related discussion).

\begin{figure}
    \centering
    \includegraphics[width=1.02\linewidth]{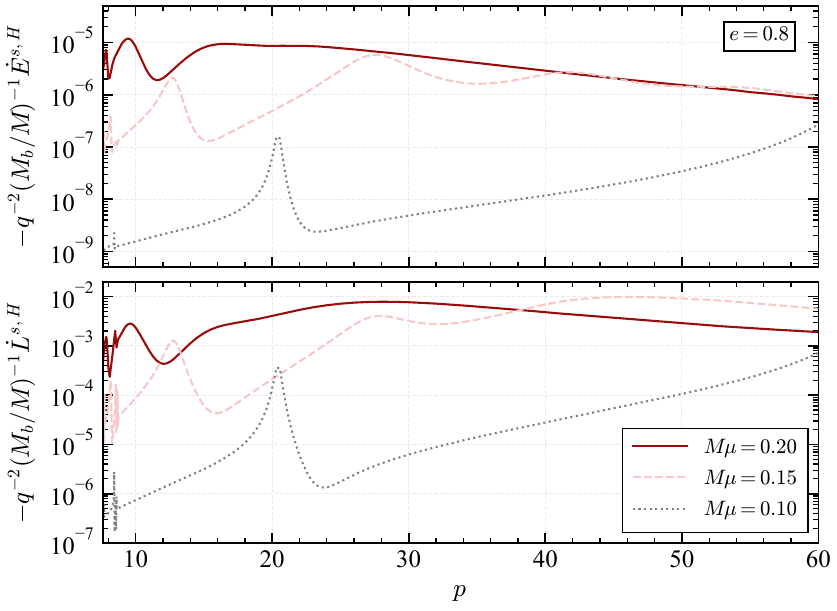}    
    \caption{Horizon energy (\textit{top}) and angular-momentum (\textit{bottom}) loss rates as functions of the semi-latus rectum $p$ for eccentric orbit with $e=0.8$, for different configurations $M\mu=0.2,\,0.15$, and $0.1$. We consider a prograde dipolar scalar cloud with $\{n_i,\ell_i,m_i\}=\{0,1,1\}$.} 
    \label{fig:alpha_dependence}
\end{figure}

To assess how the features discussed above depend on the parameter $M\mu$, in Fig.~\ref{fig:alpha_dependence} we compare the energy and angular-momentum loss rates for $M\mu=0.1,\,0.15$ and $0.2$, keeping the eccentricity fixed at $e=0.8$.
As $M\mu$ increases, the overall magnitude of the horizon loss rates gets larger.  Nevertheless, based on previous results with circular orbits, scalar clouds
with couplings $M\mu\sim 0.1$ should still produce detectable dephasings in LISA EMRIs for astrophysically relevant cloud masses~\cite{Khalvati:2024tzz}. At the same time, the resonant features shift towards larger values of $p$ as $M\mu$ decreases, in agreement with the trend previously found for circular orbits~\cite{Brito:2023pyl}. Importantly, our results show that the eccentricity-induced resonant structure persists also at smaller couplings, although both the locations and amplitudes of the resonances change.

Finally, we note that the impact of eccentricity on the detectability of scalar clouds with future space-based detectors like LISA is studied in our companion Letter~\cite{Xu:2026cky}, where we show that the amplification of the orbital loss rates due to resonances can significantly improve the detectability of scalar clouds.

\subsubsection{Infinity loss rates}
\begin{figure*}
    \centering
    \includegraphics[width=0.48\textwidth]{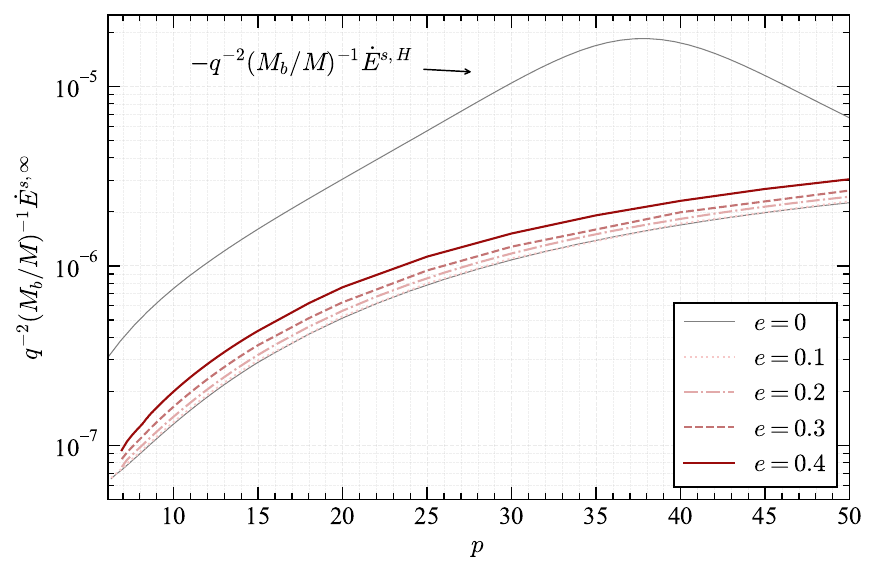}
     \includegraphics[width=0.48\textwidth]{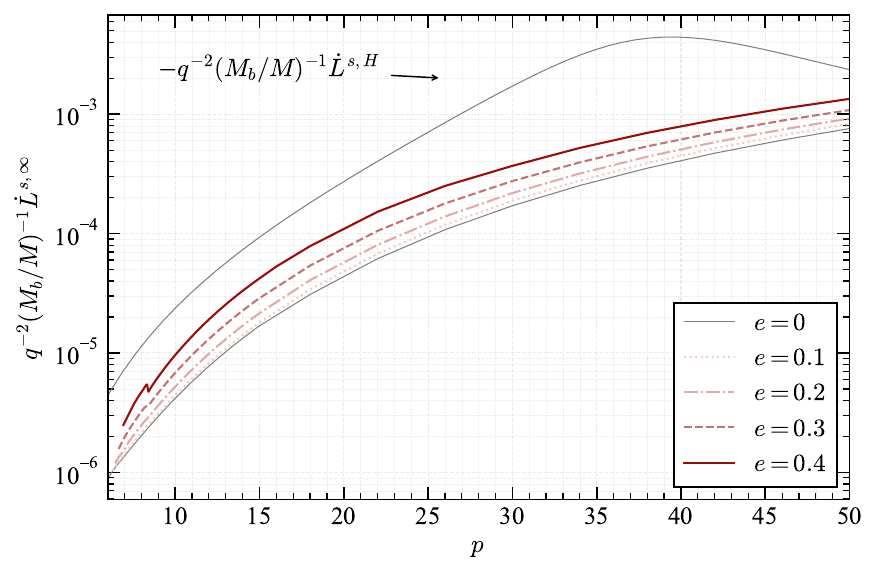}
    \caption{Energy $\dot E^{s,\infty}$ (\textit{left panel}) and angular momentum $\dot L^{s,\infty}$ (\textit{right panel}) loss rates as a function of the semi-latus rectum $p$ for different values of eccentricity $e$, when considering a dipolar $\ell_i=m_i=1$ scalar cloud in the state $n_i=0$, with $M\mu=0.2$. For reference, the corresponding horizon loss rates for $e = 0$ are also shown.}
    \label{fig:InfFlux}
\end{figure*}
We now turn to the loss rates associated with the scalar waves propagating to spatial infinity, $\dot E^{s,\infty}$ and $\dot L^{s,\infty}$. The results are summarized in Fig.~\ref{fig:InfFlux}. We note that, as the eccentricity increases, achieving convergence in the $n$-mode summation requires a substantially larger number of terms, which makes the numerical evaluation increasingly demanding at high eccentricities. For this reason, we restrict the results presented here to $e \le 0.4$, although improving the numerical efficiency and convergence of the code remains an important goal for future work (see also App.~\ref{App:Comparison of different modes} for a convergence study with increasing $\ell_f$). Results for $e= 0.6$ over a smaller range of $p$'s are also shown in the companion letter~\cite{Xu:2026cky}.

Our results show a clear trend: both $\dot E^{s,\infty}$ and $\dot L^{s,\infty}$ increase with eccentricity $e$, while remaining subdominant to the horizon loss rate over the range of parameters shown in Fig.~\ref{fig:InfFlux}. This is consistent with previous findings for circular orbits~\cite{Brito:2023pyl,Dyson:2025dlj,Li:2025ffh}. At sufficiently large $p$, however, the fluxes to infinity eventually dominate~\cite{Brito:2023pyl,Dyson:2025dlj,Li:2025ffh}.

A particularly striking feature expected in the infinity loss rates is the emergence of a staircase-like structure, in which the rates increase sharply (as $p$ decreases) at specific threshold values of $p$~\cite{Baumann:2022pkl,Baumann:2021fkf,Tomaselli:2023ysb,Brito:2023pyl,Dyson:2025dlj}. This behavior arises because, as $p$ decreases and the orbital frequencies increase, successive modes cross their threshold for propagation to infinity. In other words, higher-energy unbound modes become kinematically allowed when $\omega_{+}^2-\mu^2>0$ for a given mode [see Eq.~\eqref{eq:energy loss modes}]. 

In the circular case, the smallest value of $p$ for which this transition occurs (for $M\mu=0.2$) is approximately $p\sim 97$~\cite{Tomaselli:2023ysb,Brito:2023pyl} (see also Sec.~\ref{Sec:Point Particle in Generic Orbits}). As the eccentricity increases, these transitions shift to smaller values of $p$ and become more numerous due to the additional sum over eccentric harmonics $n$. While these discrete transitions are present for all eccentricities, they are expected to become progressively more pronounced as $e$ increases~\cite{Tomaselli:2023ysb}, since at higher eccentricities, the power is distributed across a broader range of $n$-harmonics. For the values of $e$ that we considered here, this is visible in the $e=0.4$ case, where a sharp jump, which is absent for lower eccentricities, appears near $p \approx 8.3$. We expect this structure to become even more complex for larger eccentricities. The location of this jump also indicates where one set of the resonances discussed previously accumulate as $n_f$ increases (in this case, the resonances in the $n=2$ harmonic). In particular, the resonances studied here accumulate on the left of this jump, unlike what happens for (non-eccentricity induced) resonances that occur at larger $p$, which instead always accumulate to the right of the jumps~\cite{Tomaselli:2025jfo}.

Finally, we note that the infinity energy and angular momentum loss rates discussed here are equivalent to the ionization power and torque introduced in Refs.~\cite{Baumann:2021fkf,Baumann:2022pkl,Tomaselli:2023ysb}. To make this connection clearer, in App.~\ref{App:Revisiting Newtonian fluxes} we revisit the computation of the ionization power and torque using a formalism that closely parallels our relativistic treatment, and compare the resulting Newtonian rates with the relativistic ones (see Fig.~\ref{fig:NvsR} in App.~\ref{App:Revisiting Newtonian fluxes}). Overall, we find the expected behavior that, as the semi-latus rectum $p$ increases, the relativistic results seem to approach the Newtonian ones, although for the values of $p$ and $M\mu$ we considered here there are still $\mathcal{O}(1)$ differences at $p=50$, which is consistent with previous results in Refs.~\cite{Dyson:2025dlj,Tomaselli:2025jfo} where similar comparisons in the circular case were done.

\section{Point Particle in Generic Orbits}\label{Sec:Point Particle in Generic Orbits}
Realistic astrophysical scenarios suggest that EMRIs can originate from generic initial configurations. In particular, standard formation channels predict that the secondary compact object may possess both non-negligible eccentricity and a significant inclination relative to the primary's equatorial plane~\cite{Sun:2025lbr}. 

In vacuum, the notion of orbital inclination is meaningful only in a Kerr background, where the BH spin axis uniquely selects the equatorial plane with respect to which the orbital inclination is defined. However, in the present work, we instead focus on a Schwarzschild BH background surrounded by a {\it non-spherically symmetric} scalar cloud. This setup is motivated by the fact that scalar clouds formed through superradiant instabilities are generically non-spherically symmetric configurations carrying non-zero angular momentum (see Sec.~\ref{sec: Scalar cloud}).

A key advantage of the Schwarzschild background is its spherical symmetry, which ensures that generic geodesic motion remains strictly planar. As a result, instead of studying inclined orbits, one can equivalently consider equatorial Schwarzschild geodesics while rotating the scalar cloud with respect to the orbital plane, as done in Ref.~\cite{Tomaselli:2023ysb} within a Newtonian approximation. This perspective allows us to retain much of the formalism and many of the computational tools introduced in the previous sections.

Although a similar construction is not feasible for generic Kerr geodesics, 
which are in general not planar, our setup provides a useful benchmark for future investigations in Kerr backgrounds (particularly for low spins).

\subsection{Rotation of the scalar cloud}\label{sec:Rotation of scalar cloud}

To describe the most general configuration where the orbit of the secondary is both inclined and eccentric, we must account for the relative orientation between the scalar cloud's symmetry axis and the orbital plane. We fix the orbital plane as the reference $x$-$y$ plane, with the (initial) periapsis aligned along the $+x$ direction. The orientation of the scalar cloud is then determined by rotating the initial bound state using the full Wigner $D$-matrix, $D_{m_i, m_i'}^{(\ell_i)}(\alpha, \beta, \gamma)$, where the Euler angles $(\alpha, \beta, \gamma)$ are mapped to the orbital elements as follows:
\begin{itemize}
\item $\beta$ represents the relative inclination (\textit{obliquity angle}) between the orbital plane and the cloud's equatorial plane.

\item In the Newtonian limit, $\gamma$ represents the argument of periapsis, which defines the rotation of the elliptic orbit within the orbital plane relative to the line of nodes. Due to relativistic apsidal precession, here we interpret $\gamma$ as an initial argument of periapsis. 

\item $\alpha$ represents the longitude of the ascending node.
\end{itemize}
Specifically, we rotate the initial spherical harmonic $Y_{\ell_i m_i}$ from the cloud's frame to the orbital frame using:
\begin{equation}
\begin{aligned}
Y_{\ell_{i} m_{i}}(\theta^{A}) &= \sum_{m_i^{\prime}=-\ell_i}^{\ell_i} D_{m_i, -m_i^{\prime}}^{(\ell_i)}(\alpha, \beta, \gamma) Y_{\ell_i, -m_i^{\prime}}\left(\theta^{A} \right) \\
&= \sum_{m_i^{\prime}=-\ell_i}^{\ell_i} e^{-i m_{i} \alpha} d_{m_i, -m_i^{\prime}}^{(\ell_i)}(\beta) e^{i m_{i}^{\prime} \gamma} Y_{\ell_i, -m_i^{\prime}}\left(\theta^{A}\right),
\end{aligned}
\end{equation}
where, following the conventions in Ref.~\cite{Tomaselli:2023ysb}, the Wigner small $d$-matrix is given by:
\begin{equation}
\begin{aligned}
&d_{m_i,m_i^{\prime}}^{(\ell_i)}(\beta)=\mathcal{N}\sum_{s=s_{\mathrm{min}}}^{s_{\mathrm{max}}}(-1)^{m_i-m_i^{\prime}+s}\\
&\quad\times\frac{\left(\cos\frac{\beta}{2}\right)^{2\ell_i+m_i^{\prime}-m_i-2s}\left(\sin\frac{\beta}{2}\right)^{m_i-m_i^{\prime}+2s}}{(\ell_i+m_i^{\prime}-s)!s!(m_i-m_i^{\prime}+s)!(\ell_i-m_i-s)!}\,,
\end{aligned}
\end{equation}
with $s_{\min} = \max (0, m_i^{\prime}-m_i)$, $s_{\max}=\min (\ell_i+m_i^{\prime}, \ell_i-m_i)$ and the normalization factor is given by 
$$\mathcal{N}=\sqrt{(\ell_i+m_i)!(\ell_i-m_i)!(\ell_i+m_i^{\prime})!(\ell_i-m_i^{\prime})!}~.$$



\subsection{Energy and angular momentum loss rates for generic orbits}

After applying the rotation described above, the source term in Eq.~\eqref{eq:source in TD} transforms to
\begin{equation}
\begin{aligned} 
 & S_{lm}^{T}(t,r,\theta, \varphi, \alpha, \beta, \gamma)= \sum_{m_i^{\prime}=-\ell_i}^{\ell_i}D_{m_i, -m_i^{\prime}}^{(\ell_i)}(\alpha, \beta, \gamma)\\
 &\times\Bigg[P_{lm}^{T}(t,r)Y_{lm}Y_{\ell_i,-m_i^{\prime}} \\
 &\qquad +\hat{P}_{lm}^{T}(t, r)\left(Y_{,\theta}^{lm}Y_{,\theta}^{\ell_i,-m_i^{\prime}}+\frac{Y_{,\varphi}^{lm}Y_{,\varphi}^{\ell_i,-m_i^{\prime}}}{\sin^2\theta}\right) \\
 &\qquad +A_{lm}^{T}(t,r)\frac{Y_{,\theta}^{lm}Y_{,\varphi}^{\ell_{i},-m_i^{\prime}}-Y_{,\varphi}^{lm}Y_{,\theta}^{\ell_{i},-m_i^{\prime}}}{\sin\theta}\Bigg],
\end{aligned}
\end{equation}
which, after projecting onto spherical harmonics, implies the selection rule $-m_i^{\prime}+m-m_f=0$ [cf. Eq.~\eqref{eq:spherical harmonics integrals}]. Unlike the equatorial case, due to the sum over $m_i'$ in the source term, for a given $m_f$ the source term will have support over the discrete set of frequencies $\sigma_{mn}$ with $m$ in the range $m\in [m_f-\ell_i, m_f+\ell_i]$. As such, $Z_{\ell_f m_f}^{T}(t, r)$ will take the form
\begin{equation}\label{eq:ZT_generic}
Z_{\ell_f m_f}^{T}(t, r) = \sum_{n,m}Z^{\ell_fm_fn}(r)e^{-i\sigma_{mn} t}\,,
\end{equation}
where $m\in [m_f-\ell_i, m_f+\ell_i]$. Following the same steps as we did to obtain Eq.~\eqref{eq:ordinary differential equation}, we now get a radial equation for each pair $\{\ell_f, m_f\}$ given by
\begin{equation}
\begin{aligned}
    &\left[\frac{d^2}{dr_*^2}+(\omega+\sigma_{mn})^2-V_f\right]Z^{\ell_fm_fn}=\\
    &\qquad=
    D_{m_i, -m_i^{\prime}}^{(\ell_i)}(\alpha, \beta, \gamma){S}_{m_f,-m_i^{\prime};n}^{\ell_f,\ell_i}\,,
\end{aligned}
\end{equation}
where $m=m_f+m_i^{\prime}$. Importantly, this equation is identical to Eq.~\eqref{eq:ordinary differential equation}, apart from the constant factor $D_{m_i, -m_i^{\prime}}^{(\ell_i)}(\alpha, \beta, \gamma)$ in the source term. Therefore to compute the Fourier coefficients $Z^{\ell_fm_fn}$ it is useful to instead solve Eq.~\eqref{eq:ordinary differential equation} for each quadruplet $\{\ell_f,m_f,m_i',n\}$ using the same methods and numerical code we used for equatorial orbits, and simply multiply the resulting Fourier coefficients by the corresponding Wigner $D$-matrix. 

Following the same procedure as in Sec.~\ref{sec: scalar fluxes}, we then find that, for generic non-equatorial orbits, the expressions for the energy loss rates are given by
\begin{equation}
\begin{aligned}\label{eq:generic energy fluxes}
    \dot{E}_{\ell_f,m_f,n}^{s,\infty}=&\sum_{m_i^{\prime}=-\ell_i}^{\ell_i} 2\sigma_{mn}\,{\rm sgn}(\omega_+) \Re\left[\sqrt{\omega_+^2-\mu^2}\right]\\ 
    &\times \left|D_{m_i, -m_i^{\prime}}^{(\ell_i)}{Z}^{\ell_fm_fn}(\sigma_{mn})\right|^2,\\
    \dot{E}_{\ell_f,m_f,n}^{s,H}=&\sum_{m_i^{\prime}=-\ell_i}^{\ell_i} 2\sigma_{mn}\omega_+ \left|D_{m_i, -m_i^{\prime}}^{(\ell_i)}{Z}^{\ell_fm_fn}(\sigma_{mn})\right|^2\,.
\end{aligned}
\end{equation}
The corresponding angular momentum loss rates $\dot L^{s,\infty/H}$ can be simply obtained by multiplying each term inside the sum over $m_i'$ by a factor $m/\sigma_{mn}$. Notice that this is the angular-momentum loss rate along the direction of the unperturbed orbital angular momentum, $\dot L\equiv-\dot{\boldsymbol L}\cdot\hat{\boldsymbol L}_{\rm orb}$. The transverse torque components, which would drive a secular change of the orientation of the orbital plane and of the inclination angle, are not computed in the present work. They could in principle be computed following the approach in Ref.~\cite{Tomaselli:2023ysb}, however a more useful approach when thinking about extensions to generic orbits around a Kerr BH, would be to derive an evolution equation for the Carter constant in the presence of clouds (see e.g. Ref.~\cite{Hughes:2021exa}), which we leave for future work.


Some remarks should be made about Eq.~\eqref{eq:generic energy fluxes}. Firstly, we note that the averaged loss rates do not depend on the rotation $\alpha$. This is expected, given that the energy density of a (complex) scalar cloud is axisymmetric. Therefore, the rotation $\alpha$ around the cloud's symmetry axis only contributes a global phase factor $e^{-im_i \alpha}$, which does not contribute to the fluxes. Secondly, we notice that there are no cross terms between different $m_i^{\prime}$ in Eq.~\eqref{eq:generic energy fluxes} because those vanish upon averaging over several radial librations. This implies that averaged loss rates do not depend on the initial argument of periapsis $\gamma$, since the $\gamma$-dependent phase factors cancel when taking the absolute value of the Wigner $D$-matrix. This follows from the fact that generic non-resonant orbits, characterized by incommensurable fundamental frequencies ($\Omega_\varphi/\Omega_r \notin \mathbb{Q}$), are not closed, instead undergoing relativistic apsidal precession that causes the orbit to densely and ergodically fill the orbital annulus bounded by the periapsis and apoapsis (i.e. with $r$ limited in the range $r\in [r_{\rm min},r_{\rm max}]$). Consequently, the dependence on the initial argument of periapsis $\gamma$ averages out in the fluxes. This approximation is valid as long as the (averaged) apsidal precession timescale $t_{\rm prec}\sim 1/\langle\dot \gamma\rangle=1/(\Omega_\varphi-\Omega_r)\sim M p^{5/2}(1-e^2)^{-3/2}$~\cite{Manzini:2025gjx} is much shorter than the radiation-reaction timescale $t_{\rm RR}$. Assuming a regime where inspiral is driven by GW emission, this is always true, given that $t_{\rm RR}\sim 5Mp^4/(256q\sqrt{1-e^2})\gg t_{\rm prec}$~\cite{Peters:1964zz,Zwick:2021ayf}, for $q\ll 1$ (neglecting eccentricity-dependent factors of $\mathcal{O}(1)$). 
This is in contrast to what happens in a Newtonian approximation since in this case there is no relativistic apsidal precession and the averaged loss rates for generic orbits do depend on $\gamma$ (see App.~\ref{App:Revisiting Newtonian fluxes}). 

\subsection{Results}
Using the formalism developed above, we now study how the obliquity angle $\beta$ affects the loss rates $\dot{E}^{s,\infty/H}$ and $\dot{L}^{s,\infty/H}$ (details of the numerical procedure are given in App.~\ref{App:Numerics}). For concreteness, we focus again on a dipolar scalar cloud ($\ell_i=m_i=1$) in the 
state ($n_i=0$), with $M\mu=0.2$.

\subsubsection{Non-equatorial, circular orbits}

\begin{figure*}
    \centering
    \includegraphics[width=0.48\textwidth]{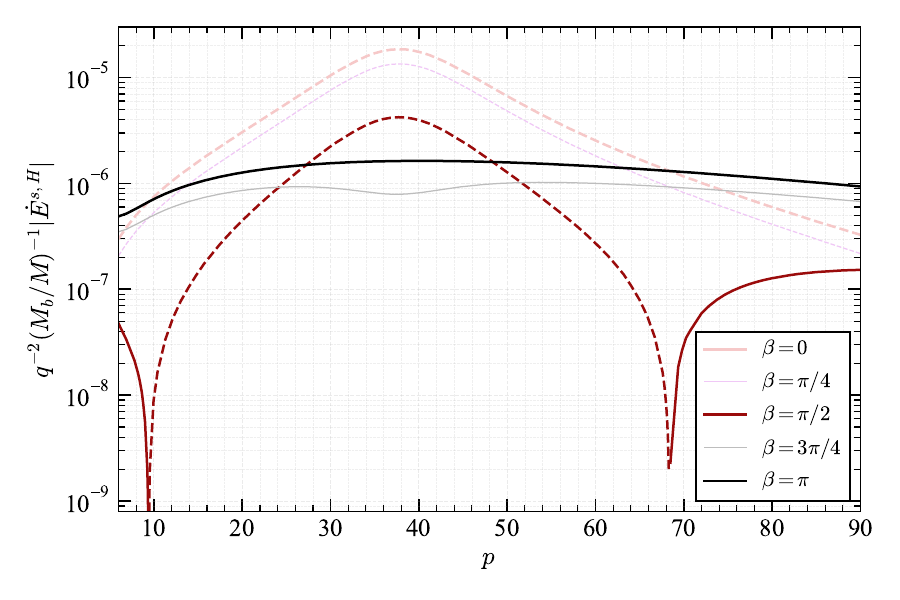}
     \includegraphics[width=0.48\textwidth]{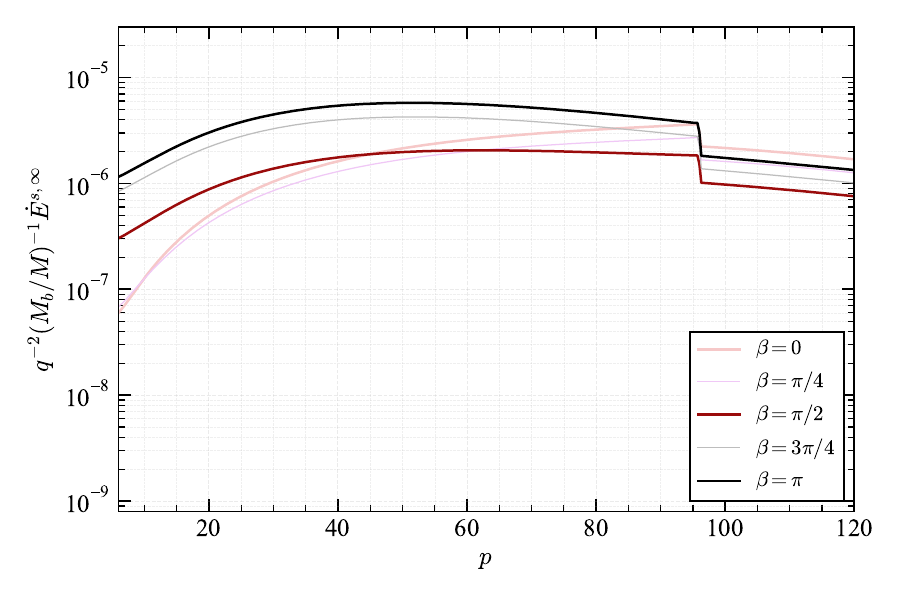}
    \caption{Scalar energy loss rates at the horizon $\dot E^{s,H}$ (\textit{left panel}) and at infinity $\dot E^{s,\infty}$ (\textit{right panel}) as functions of the orbital radius $p=r_{p}/M$, for circular orbits. Different curves correspond to distinct obliquity angles $\beta$, as indicated in the legend. Solid (dashed) lines denote positive (negative) loss rates. Intermediate obliquities ($\beta = \pi/4, 3\pi/4$) are plotted with thinner lines to clearly distinguish them from the purely equatorial ($\beta = 0, \pi$) and polar ($\beta= \pi/2$) configurations.}
    \label{fig:InclinedFlux}
\end{figure*}
\begin{figure}[h!]
    \centering
    \includegraphics[width=1.02\linewidth]{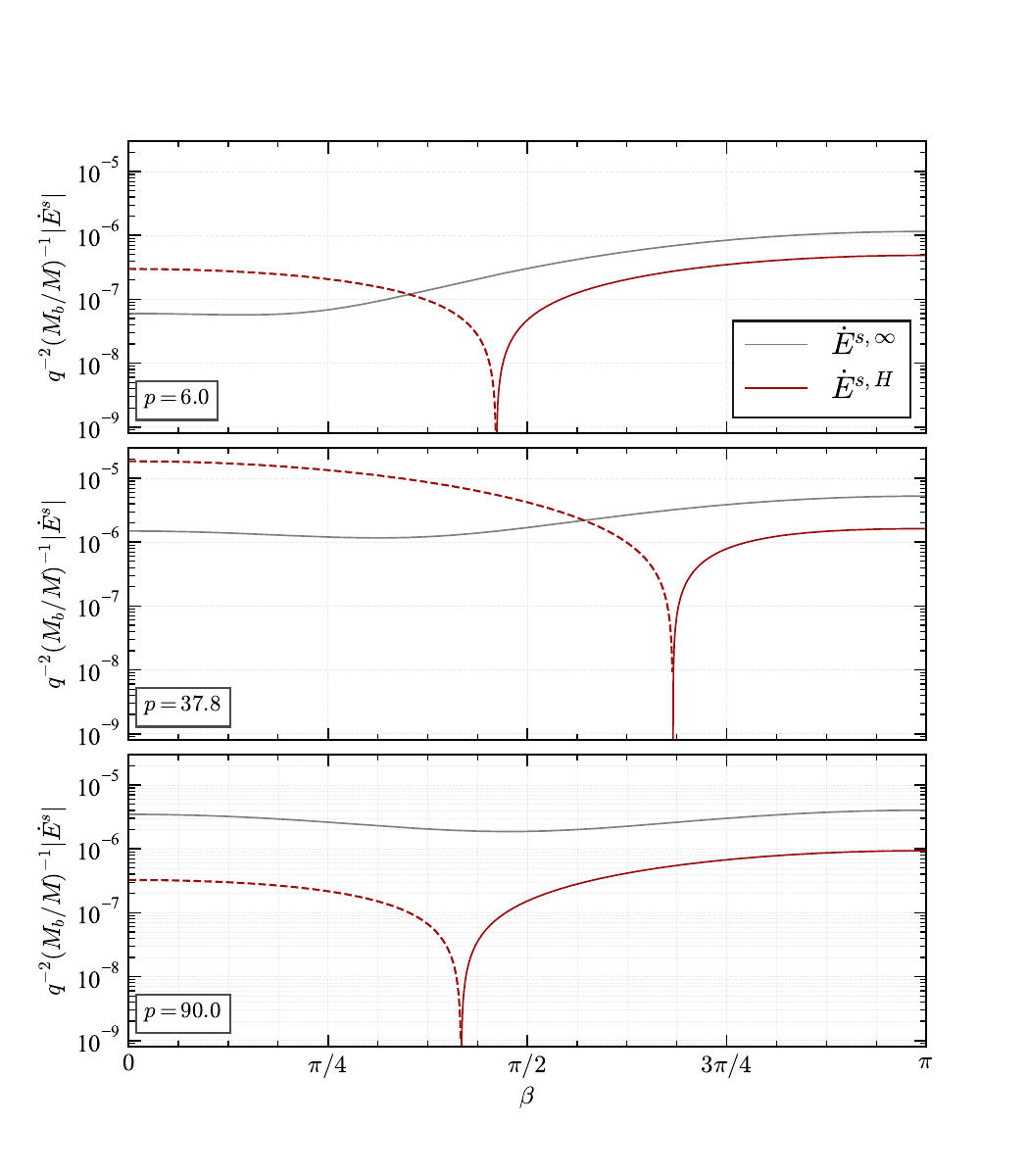}
    \caption{Dependence of the scalar energy loss rates at infinity $\dot{E}^{s, \infty}$ (gray) and at the horizon $\dot{E}^{s, H}$ (red) on the obliquity angle $\beta$ for circular orbits. The panels illustrate three representative regimes: the strong-field limit ($p=6.0$, \textit{top panel}), an intermediate region near the maximum of the energy loss rate ($p=37.8$, \textit{middle panel}), and the weak-field limit ($p=90.0$, \textit{bottom panel}). Solid (dashed) curves indicate positive (negative) loss rates. The scalar cloud is initialized in the 
    dipolar state ($n_i=0, \ell_i=m_i=1$) with $M\mu=0.2$.}
    \label{fig:Flux_vs_Beta}
\end{figure}
Let us start by considering inclined circular orbits. For this case, the radial overtone index is fixed to $n=0$ and $\sigma_{m0}=m\Omega_{\rm \varphi} = m\sqrt{M/r_{p}^{3}}$, with $r_p=pM$. Furthermore, for circular orbits, the angular momentum and energy loss rates satisfy the relation $\dot{E}^{s,\infty/H} =\Omega_{\varphi}\dot{L}^{s,\infty/H}$ and therefore we focus the discussion on $\dot{E}^{s,\infty/H}$. 

Our results are summarized in Figs.~\ref{fig:InclinedFlux} and~\ref{fig:Flux_vs_Beta}, where we show how the loss rates for circular orbits depend on the semi-latus rectum $p$ and the obliquity $\beta$. For comparison, we also show the limiting cases $\beta=0$, corresponding to  prograde equatorial orbits, and $\beta=\pi$, corresponding to retrograde equatorial orbits. These limiting cases have also been studied in Ref.~\cite{Brito:2023pyl}.

As expected, given the cloud's energy density doughnut-like shape~\cite{Santos:2020pmh}, the loss rates have a strong dependence on the orbit's obliquity, also in agreement with Newtonian calculations~\cite{Tomaselli:2023ysb}. However, the horizon loss rates $\dot{E}^{s,H}$ show particularly distinctive features, as shown in the left panel of Fig.~\ref{fig:InclinedFlux}. For prograde orbits with low obliquity (e.g., $\beta=0, \pi/4$), $\dot{E}^{s,H}$ is negative for all the parameter space covered here ($p\in [6,90]$), meaning that the cloud is transferring energy/angular momentum to the orbiting particle, as we already discussed in the previous section. Conversely, for retrograde orbits with high obliquity ($\beta=\pi,  3\pi/4$), $\dot{E}^{s,H}$ remains positive throughout, meaning that the particle is losing energy/angular momentum, which is consistent with the findings of Ref.~\cite{Brito:2023pyl}. On the other hand, for high inclinations ($\beta=\pi/2$) we instead find a behavior that falls in between these two limiting cases. In particular, in the region under study we find two zero-crossing points where $\dot{E}^{s,H}$ changes sign.
This is to be contrasted with the asymptotic loss rate $\dot{E}^{s,\infty}$ (right panel) which is always positive. Also worth noticing is the fact that $\dot{E}^{s,\infty}$ exhibits a sharp increase (when going from higher to smaller orbital separations) at $p\sim 97$, as already mentioned in Sec.~\ref{Sec:Point Particle in Eccentric, Equatorial Motion}. As can be seen, the location of these ``sharp'' changes in the asymptotic loss rate does not depend on $\beta$, which is also in agreement with Newtonian calculations~\cite{Tomaselli:2023ysb}.


In order to better understand the dependence of the loss rates with $\beta$ and the relative importance of the infinity and horizon loss rates, in Fig.~\ref{fig:Flux_vs_Beta} we show $\dot{E}^{s,H}$ and $\dot{E}^{s,\infty}$ as a function of $\beta$, for three representative orbital separations, representing large, intermediate and small orbital separations. In particular:
\begin{itemize}
\item Weak-field regime ($p=90.0$, \textit{bottom panel}): The energy loss is dominated by radiation to infinity ($\dot{E}^{s,\infty}$, gray curve), which shows a relatively weak dependence on $\beta$. The horizon loss rate is sub-dominant.
\item Strong-field regime ($p=6.0$, \textit{top panel}): For prograde orbits ($\beta \lesssim \pi/3$), $\dot{E}^{s,H}$ dominates over $\dot{E}^{s,\infty}$, implying that the cloud transfers energy into the orbit. For $\beta \gtrsim \pi/3$, on the other hand, $\dot{E}^{s,\infty}$ dominates and orbiting particles lose energy.
\item Intermediate regime ($p=37.8$, \textit{middle panel}): This is the point where the absolute value of $\dot{E}^{s,H}$ reaches its maximum for prograde orbits. Therefore, the value of $\beta$ at which $\dot{E}^{s,H}$ changes sign in this panel corresponds to the maximum value of $\beta$ for which $\dot{E}^{s,H}<0$ in the interval of orbital separations we consider here (at much larger orbital separations, $\dot{E}^{s,H}$ should also be negative around floating resonances~\cite{Baumann:2019ztm,Tomaselli:2024bdd,Boskovic:2025ixx}). 
\end{itemize}
In all three cases, we see a $p$-dependent critical inclination where $\dot{E}^{s,H}$ vanishes. Moreover, for small enough $p$, there is also an inclination for which $\dot{E}^{s,\infty}+\dot{E}^{s,H}=0$, meaning there is a specific boundary in the $(p,\beta)$ parameter space for which the scalar-induced torques exactly vanish, marking a transition between an overall energy transfer from the cloud to the orbit to a purely dissipative drag.


\subsubsection{Generic orbits}
For generically inclined and eccentric orbits, we focus primarily on the horizon loss rates, mainly due to the high computational cost of evaluating infinity loss rates, but also because, as we discussed so far, the horizon loss rates show overall the most interesting features, such as the resonances discussed in Sec.~\ref{Sec:Point Particle in Eccentric, Equatorial Motion}. Given that $\dot E^{s,H}$ is strongly dominated by the $\ell_f=m_f=0$ (see App.~\ref{App:Comparison of different modes}) we can, without much difficulty, compute $\dot E^{s,H}$ only considering this mode. We leave a complete study of the parameter space as well as orbital evolutions for generic orbits for future work.

\begin{figure}
    \centering
    \includegraphics[width=0.99\linewidth]{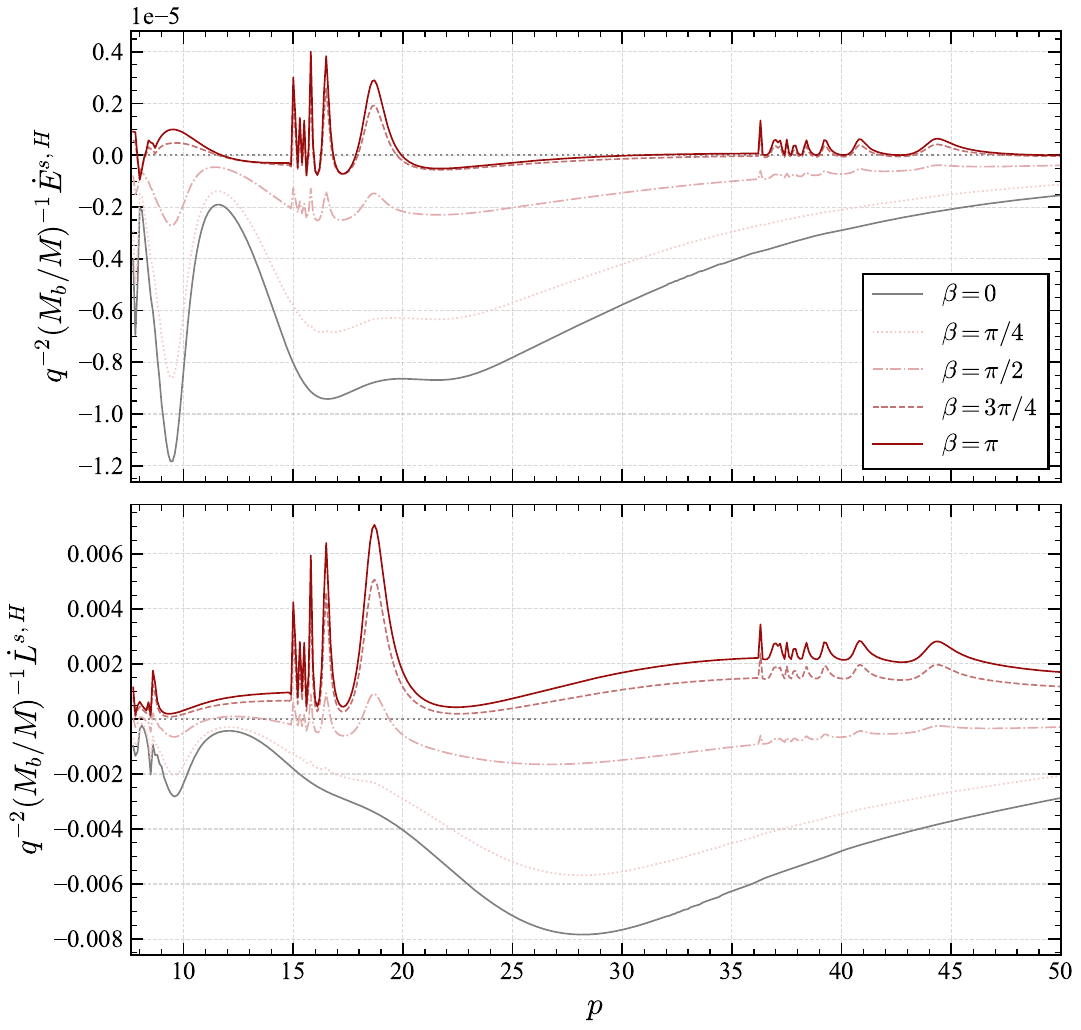}
    \caption{Energy $\dot{E}^{s,H}$ ({\it top panel}) and angular momentum $\dot{L}^{s,H}$ ({\it bottom panel}) loss rates as functions of the semi-latus rectum $p$ for fixed eccentricity $e=0.8$ and varying obliquity $\beta$.}
    \label{fig:HorFluxEccInclined}
\end{figure}

As illustrated in Fig.~\ref{fig:HorFluxEccInclined}, for an eccentric orbit the trend of how the horizon loss rates change with increasing $\beta$ is similar to that of circular orbits discussed above: energy and angular momentum loss rates generally transition from negative to positive as $\beta$ increases. Notably, new resonances associated with retrograde orbits ($\beta \geq \pi/2$) emerge and become increasingly prominent with growing obliquity. In particular, the peaks appearing in the region $p \in (36,45)$ arise from resonant excitations in the $n=0$ Fourier harmonic of $\{n_f,\ell_f,m_f\}=\{n_f\geq 2,0,0\}$ states, with $n_f=2$ corresponding to the rightmost peak. Likewise, the peaks in the region $p \in (15,20)$ are produced by resonant excitations in the $n=-1$ Fourier harmonic of the same quasi-bound states, while the small peaks seen closer to the LSO correspond to resonances in $n< -1$ harmonics.

Aside from the resonances, another distinctive feature emerging at high eccentricities is the fact that even for equatorial retrograde orbits ($\beta=\pi$), the energy loss rate can be negative at specific values of the semi-latus rectum $p$, unlike in the circular case (cf.~Fig.~\ref{fig:InclinedFlux}). This behavior arises from the relative contributions of different Fourier harmonics $n$. In the regions where $\dot E^{s,H}<0$ for $\beta=\pi$, the loss rate is dominated by harmonics with $n<-1$. According to Eq.~\eqref{eq:energy loss modes}, these harmonics contribute negatively to $\dot E^{s,H}$ throughout most of the parameter space because the loss rate is dominated by the $\ell_f=m_f=0$ mode, which for $\beta=\pi$ couples only to gravitational perturbations with $m=1$. Consequently, the sign of $\dot E^{s,H}$ is determined by the sign of $\sigma_{1n}$, which is negative almost everywhere for $n< -1$, except in the vicinity of the LSO. It is worth noticing though that, for equatorial retrograde orbits, infinity loss rates should dominate (at least outside resonances) for all values of $p$ we are considering here (cf.~Fig.~\ref{fig:Flux_vs_Beta}), so we expect the total scalar loss rate $\dot E^s=\dot E^{s,\infty}+\dot E^{s,H}$ to remain positive throughout.

Finally, we notice that for inclined orbits, the dependence on the inclination angle originates from the Wigner rotation coefficients appearing in Eq.~\eqref{eq:generic energy fluxes}, whose angular structure is independent of $M\mu$. We therefore expect the qualitative dependence on the inclination discussed in this section to persist at smaller $M\mu$. However, changing $M\mu$ modifies the radial profile of the cloud and the relative amplitudes of the different azimuthal components. Consequently, the magnitude of the loss rates and the precise locations of resonances can change, as already discussed in Sec.~\ref{sec:results_equatorial}.

\section{Conclusions and Outlook}\label{Sec:Conclusions and Outlook}

In this work, we developed a fully relativistic framework to study EMRIs evolving on eccentric and inclined orbits within a scalar cloud environment around a Schwarzschild BH, extending previous works that were restricted to circular, equatorial orbits~\cite{Brito:2023pyl, Duque:2023seg, Dyson:2025dlj, Li:2025ffh}. This approach also overcomes the limitations of Newtonian approximations, which fail to capture the strong-field dynamics relevant for LISA observations, as evidenced by the inability of Newtonian calculations to accurately capture horizon fluxes~\cite{Tomaselli:2025jfo} as well as up to $\mathcal{O}(10)$ quantitative deviations we find when comparing relativistic results for the {\it ionization power}~\cite{Tomaselli:2023ysb} against Newtonian approximations (see App.~\ref{App:Revisiting Newtonian fluxes}). 

Using this framework, we computed the orbital energy and angular momentum loss rates due to the presence of the cloud and uncovered a rich resonance structure present in the strong-field region, due to the excitation of the cloud's modes.  Most notably, we identified a rich substructure of resonances in the strong-field regime (near the LSO) that are completely absent when considering circular orbits or Newtonian approximations. Physically, these resonances arise due to a large apsidal precession in the strong-field regime resulting in a split of the azimuthal and radial orbital frequencies ($\Omega_\varphi \neq \Omega_r$). A key outcome of our analysis is that increasing orbital eccentricity increases the importance of the resonances. Our results highlight the necessity of a relativistic treatment in the strong-field regime and demonstrate that eccentricity could be essential to firmly identify scalar cloud environments~\cite{Xu:2026cky}.

Furthermore, by exploiting the spherical symmetry of the BH background, we modelled generic inclined orbits through a rotation of the scalar cloud with respect to the orbital plane and derived general expressions for the orbital loss rates as functions of the obliquity angle $\beta$. Our results show that, for circular orbits, the system exhibits a critical inclination threshold separating two distinct dynamical regimes in the strong-field region: retrograde orbits generally experience a purely dissipative drag from the scalar environment, whereas low-obliquity prograde orbits can extract energy and angular momentum from the cloud. Moreover, we demonstrated that the resonant structure of the scalar torque depends sensitively on the orbital inclination, although the broader picture remains unchanged: scalar resonances continue to arise predominantly in the strong-field regime for sufficiently high eccentricities, independently of the inclination angle.
We expect this general picture to persist for generic orbits in a Kerr background. In that case, however, the tri-periodic structure of the orbital motion is likely to generate an even richer spectrum of resonances and more intricate substructures, as implied by Eq.~\eqref{eq:resonances}.

Looking ahead, the construction of faithful EMRI waveforms requires incorporating not only the dissipative scalar loss rates computed here, but also conservative effects, as well as corrections to the gravitational-wave fluxes induced by the scalar cloud~\cite{Brito:2023pyl,Keijzer:2026vul}. Moreover, the relative importance of the \textit{geometric torque} term derived in Ref.~\cite{Dyson:2026ddd}, not included in our calculation, remains unclear at the moment.
Other sub-dominant environmental effects that could be incorporated include for example the shift in the cloud's eigenfrequencies due to the metric perturbations induced by the secondary and the cloud's self-gravity~\cite{Cannizzaro:2023jle,Brito:2023pyl,Takahashi:2026ruw}. The systematic computation of these missing pieces as well as connecting the torque-balance framework of Ref.~\cite{Dyson:2026ddd} with the approach taken here, based on the computation of energy and angular-momentum fluxes induced by scalar perturbations, is the subject of ongoing work and remains an important goal for developing a first-principles description of an EMRI evolution in matter environments.

Ultimately, the rich phenomenology unveiled in this work, specifically the strong-field scalar resonances induced by eccentricity as well as the complex dependence of scalar torques on orbital inclination, strongly motivates the necessity of including generic orbital parameters in future waveform models for EMRIs in bosonic environments.

\begin{acknowledgments}
We are grateful to Seth Hopper for kindly sharing his code developed in Ref.~\cite{Hopper:2010uv} with us.
We are grateful to Conor Dyson, Robrecht Keijzer and Thomas Spieksma for discussions.
We acknowledge financial support provided by FCT – Fundação para a Ciência e a Tecnologia, I.P., through the ERC-Portugal program Project ``GravNewFields''. We also thank the Fundação para a Ciência e a Tecnologia (FCT), Portugal, for the financial support to the Center for Astrophysics and Gravitation (CENTRA/IST/ULisboa) through grant No.~\href{https://doi.org/10.54499/UID/PRR/00099/2025}{UID/PRR/00099/2025} and grant No.~\href{https://doi.org/10.54499/UID/00099/2025}{UID/00099/2025}.
QX gratefully acknowledges support from FCT through Grant No.~\href{https://doi.org/10.54499/2025.01546.BD}{2025.01546.BD}.
R.V. gratefully acknowledges the support of the Dutch Research Council (NWO) through an Open Competition Domain Science-M grant, project number OCENW.M.21.375.
C.Y. acknowledges the financial support provided under the European Union's H2020 ERC Advanced Grant ``Black holes: gravitational engines of discovery'' grant agreement no.~Gravitas--101052587. Views and opinions expressed are however those of the author only and do not necessarily reflect those of the European Union or the European Research Council. Neither the European Union nor the granting authority can be held responsible for them. C.Y. acknowledges support from the Villum Investigator program supported by the VILLUM Foundation (grant no.~VIL37766) and the DNRF Chair program (grant no.~DNRF162) by the Danish National Research Foundation.
\end{acknowledgments}

\appendix

\section{Revisiting Newtonian energy loss rates}\label{App:Revisiting Newtonian fluxes}
The orbital energy and angular momentum loss rates discussed in the main paper have been first computed in the non-relativistic limit in Refs.~\cite{Baumann:2021fkf, Baumann:2022pkl,Tomaselli:2023ysb}. In this limit, the bound-state solutions mathematically map to those of the hydrogen atom in quantum mechanics. In this appendix, we revisit this approximation using an approach that closely follows our relativistic framework.

\subsection{Gravitational atom}
To obtain the non-relativistic limit of the Einstein-Klein-Gordon system, it is convenient to factor out high-frequency oscillations, employing the following ansatz:
\begin{equation}
    \mathbf{\Phi}(t,r,\theta^A) = e^{-i\mu t}\hat{\Phi}(t,r,\theta^A),
\end{equation}
where $\hat{\Phi}(t,r,\theta^A)$ is assumed to vary on timescales much longer than $\mu^{-1}$.
Using this ansatz, it can be shown that in the non-relativistic limit, the Einstein-Klein-Gordon system \eqref{eq: EKG field equation} reduces itself to the Schrödinger-Poisson system (see e.g.~\cite{Annulli:2020lyc}): 
\begin{align}
    \nabla^{2}U &= 4\pi \left[M\delta(\textbf{r})+2 \mu^2|\hat{\Phi}|^{2}\right],\label{eq:Poisson equation}\\
    i\frac{\partial\hat{\Phi}}{\partial t} &= -\frac{1}{2\mu}\nabla^{2}\hat{\Phi}+\mu U\hat{\Phi},
\end{align}
with a weak gravitational potential $U(r)\ll 1$. Here, we add a point-like source with density $M\delta(\textbf{r})$ that models the central BH with $M$ around which the cloud forms~\cite{Arana:2024kaz}. Recall that the mass of a single boson is given by $\hbar\mu$ so that all factors of $\hbar$ in the Schrödinger-Poisson system are absorbed into the definition of $\mu$.

We now adopt the same perturbative scheme introduced in Sec.~\ref{sec:Perturbation scheme}. Specifically, we assume that $|\hat\Phi|\sim \mathcal{O}(\epsilon)$ with $\epsilon\ll 1$. To leading order, the gravitational potential then reduces to the spherically symmetric Newtonian potential generated by a point mass $M$, $U_0= -{M}/{r}$, where the subscript $0$ denotes leading-order background quantities. The scalar field in this background, denoted as $\phi_0$, is governed by the leading-order Schrödinger equation with a Coulomb potential,
\begin{equation}
    i\frac{\partial\phi_0}{\partial t} =-\frac1{2\mu}\nabla^{2}\phi_0+\mu U_{0}\phi_0.
\end{equation}
As is well known, this equation admits bound-state solutions given by
%
\begin{equation}
    \phi_0(t,r,\theta^A)=e^{-iE_{n_i\ell_i}t}R^{h}_{n_i\ell_i}(r)Y_{\ell_im_i}(\theta^{A})\,,
    \label{eq:hydrogen-like-eigenfunction}
\end{equation}
where $R^{h}_{n_i\ell_i}(r)$ is the hydrogenic radial function
\begin{equation}
\begin{aligned}\label{eq:hydrogenic radial function}
    R^{h}_{n_i\ell_i}(r)=&\frac{\sqrt{M_b}}{\sqrt{2}\mu}\sqrt{\left(\frac{2\mu\bm{\alpha}}{\bar{n}}\right)^3\frac{(\bar{n}-\ell_i-1)!}{2\bar{n}(\bar{n}+\ell_i)!}}\left(\frac{2\bm{\alpha}\mu r}{\bar{n}}\right)^{\ell_i}\\
    \times&L_{\bar{n}-\ell_i-1}^{2\ell_i+1}\left(\frac{2\mu\bm{\alpha} r}{\bar{n}}\right)\exp\left(-\frac{\mu\bm{\alpha} r}{\bar{n}}\right),
\end{aligned}
\end{equation}
$\bar{n} := n_i + \ell_i+1 $ is the principal quantum number, $L_{\bar{n}-\ell_i-1}^{2\ell_i+1}(x)$ is the associated Laguerre polynomial and we defined $\bm{\alpha}:=M\mu$. Notice that
we have normalized $\phi_0$, such that the cloud's mass is given by $M_{b}=\int 2\mu^2 |\phi_0|^2dV$. At this order, the energy eigenvalue in Eq.~\eqref{eq:hydrogen-like-eigenfunction} is given by $E_{n_i\ell_i}=-\mu\bm{\alpha}^2/(2\bar{n}^2)$. Going beyond the Newtonian limit, the first corrections to the energy eigenvalues appear at order $\mathcal{O}(\bm{\alpha}^{4})$ and read~\cite{Baumann:2019eav}
\begin{equation}\label{eq:energy level}
    E_{n_i\ell_i}\equiv\omega-\mu\simeq -\mu\left(\frac{\bm{\alpha}^2}{2\bar{n}^2}+\frac{\bm{\alpha}^4}{8\bar{n}^4}+\frac{3\bar{n}-2\ell_i -1}{\ell_i +1/2}\frac{\bm{\alpha}^4}{\bar{n}^{4}} \right)\,.
\end{equation}

\subsection{Binary system}
As we did in the main text, we now consider a secondary orbiting object. In the Newtonian regime, this object follows a Keplerian trajectory, generically characterized by closed elliptical orbits. This is in contrast to the relativistic framework in which eccentric orbits experience apsidal precession. It is therefore useful to outline the binary system considered in this section. 

We work in the reference frame of the central BH, described by spherical coordinates $\{r, \theta, \varphi\}$. The coordinates of the secondary are denoted by $\{r_*, \theta_*, \varphi_*\}$, where $r_*$ represents the instantaneous binary separation and $\theta_*$ is the polar angle (an equatorial orbit corresponds to $\theta_* = \pi/2$), while $\varphi_*$ is the azimuthal angle, which coincides with the true anomaly in the Newtonian limit.

On an eccentric Keplerian orbit, both $r_{*}$ and $\varphi_*$ vary with time. A useful parametrization is given in terms of the eccentric anomaly $E_a$ and eccentricity $e$,
\begin{align}
    &r_*=a_{s}(1-e\cos E_a),\\
    &(1-e)\tan^{2}\frac{\varphi_{*}}{2}=(1+e)\tan^{2}\frac{E_a}{2},
\end{align}
where $a_{s} = (M/\Omega_{\rm orb}^{2})^{1/3}$ is the semi-major axis, which is related to the orbital frequency $\Omega_{\rm orb}$ through Kepler's third law. The eccentric anomaly as a function of time must then be found by solving Kepler’s equation
\begin{equation}
    M_a=E_a-e\sin E_a\,,\label{eq:mean anomaly}
\end{equation}
where $M_a$ is the mean anomaly which, modulo a constant phase, we can set to $M_a=\Omega_{\rm orb} t$.
Alternatively, one can write the true anomaly $\varphi_*$ in terms of the mean anomaly $M_a$ using the series expansion
\begin{equation}
\begin{aligned}\label{eq:true_anomaly_vs_mean_anomaly}
    \varphi_{*}=&M_a+2\sum_{j=1}^\infty\frac{\sin (jM_a)}{j}\\
    \times&\left\{J_j(je)+\sum_{k=1}^\infty \bm{\beta}^k\left[J_{j-k}(je)+J_{j+k}(je)\right]\right\},
\end{aligned}
\end{equation}
where we defined $\bm{\beta}=(1-\sqrt{1-e^2}\,)/e$ and  $J_j(x)$ denotes the Bessel functions of the first kind. 

In terms of the true anomaly, the binary separation can be rewritten as
\begin{equation}
     r_*=\frac{a_{s}\left(1-e^2\right)}{1+e\cos\varphi_{*}}.
\end{equation}
To directly compare with the relativistic framework it is also useful to note that, in the Newtonian limit, the semi-latus rectum $p$ is related to the semi-major axis by the relation
\begin{equation}
pM=a_{s}\left(1-e^2\right)\,.
\end{equation}

\subsection{Perturbations due to the secondary object}\label{app: Perturbations}
The presence of the secondary introduced in the previous subsection acts as a perturbing source on the gravitational atom system. We can thus expand the gravitational potential and the scalar field perturbatively as
\begin{equation}
\begin{aligned}
    &U = U_0(r) + \delta U(t,r,\theta^A), \\
    &\hat{\Phi} = \phi_0(t,r,\theta^A) + \delta\phi(t,r,\theta^A),
\end{aligned}
\end{equation}
where $\delta U$ is the tidal gravitational perturbation induced by the secondary, and $\delta \phi$ represents the corresponding response of the scalar cloud. 

Working in the reference frame of the central BH, the gravitational perturbation $\delta U$ can be decomposed in a spherical harmonic basis as~\cite{Tomaselli:2023ysb}
\begin{equation}
    \delta U(t,r,\theta^A)=\sum_{l,m}\left[u^T_{lm}(t,r)Y_{lm}(\theta^{A})\right]\,,
\end{equation}
where $u_{lm}^{T}(t,r)$ is given by
\begin{widetext}
\begin{align}
\begin{split}\label{eq:uTlm}
    u_{lm}^{T}(t, r)=\begin{cases}
    \displaystyle-\frac{4\pi m_p}{2l+1}Y^*_{lm}\left(\theta_*,\varphi_*\right)\left[\frac{ r^{l}}{r_{*}^{l+1}}\Theta(r_{*}-r)+\frac{ r_{*}^{l}}{r^{l+1}}\Theta(r-r_{*})\right]&\text{if }l\neq 1, \\-\displaystyle\frac{4\pi m_p}{3}Y^*_{1m}\left(\theta_*,\varphi_*\right)\left(\frac {r_{*}}{r^{2}}-\frac{r}{r_{*}^{2}}\right)\Theta(r-r_{*})&\text{if }l=1,\end{cases}
\end{split}
\end{align}
\end{widetext}
with $\Theta$ the Heaviside step function. The last expression depends on time through $r_*\equiv r_*(t)$ and $\varphi_*\equiv \varphi_*(t)$ and for now we assume equatorial orbits $\theta_*=\pi/2$. Due to the periodic nature of the secondary's orbit, characterized by the orbital frequency $\Omega_{\rm orb}$, $u^T_{lm}(t,r)$ can be expanded in a Fourier series over the harmonic frequencies $\sigma_g \equiv g\Omega_{\rm orb}$ (where $g \in \mathbb{Z}$),
\begin{equation}
u^T_{lm}(t,r)=\sum_{g=-\infty}^{\infty} u^{lmg}(r) e^{-i\sigma_g t},
\end{equation}
with the Fourier coefficients $u^{lmg}(r)$ given by
\begin{equation}
\begin{aligned}\label{eq:ulmg}
    {u^{lmg}}(r) &= \frac{1}{2\pi}\int^{2\pi}_0 u_{lm}^{T}(t,r)e^{ig M_a}dM_a\\
    &=\frac{1}{2\pi}\int^{2\pi}_0 u_{lm}^{T}(t,r)e^{ig(E_a-e\sin E_a)}\frac{dM_a}{dE_a}dE_a\,,
\end{aligned}
\end{equation}
where $\frac{dM_a}{dE_a}=(1-e\cos E_a)$ according to Eq.~\eqref{eq:mean anomaly}.
Similarly, we can decompose the scalar perturbations as [cf.~Eq.~\eqref{eq: (1,1) order scalar field ansatz}],
\begin{equation}
    \delta\phi=\frac{1}{r}\sum_{\ell_{f},m_{f}}\left[ \mathcal{Z}^T_{\ell_{f}m_{f}}(t, r)Y_{\ell_{f}m_{f}}(\theta^{A})e^{-iE_{n_i\ell_{i}}t}\right],
\end{equation}
where we can again decompose $\mathcal{Z}^T$ in Fourier
components:
\begin{equation}
    \mathcal{Z}^T_{\ell_{f}m_{f}}(t, r) = \sum_{g} \mathcal{Z}_{\ell_{f}m_{f}g}(r)e^{-i\sigma_g t}\,.
\end{equation}

Expanding the Schrödinger equation up to linear order in the perturbations $\delta U$ and $\delta \phi$, we find that $\delta\phi$ satisfies~\cite{Annulli:2020lyc} 
\begin{align}
    i\frac{\partial\delta\phi}{\partial t}& =-\frac1{2\mu}\nabla^{2}\delta\phi+\mu U_{0}\delta\phi+\mu\delta U\phi_0\,.
\end{align}
Inserting the above Fourier series expansion into this equation gives
\begin{equation}
\begin{aligned}
    &\sum_{\ell_{f},m_{f},g}\left[\mathcal{D}_{+}{\mathcal{Z}}^{\ell_{f}m_{f}g}(g\Omega_{\rm orb},r)\right]Y_{\ell_{f}m_{f}}(\theta^{A})e^{-i\sigma_g t}\\=&~2\mu^{2}rR_{n_i\ell_{i}}^{h}(r)\sum_{l,m,g{'}}{u}^{lmg{'}}(r)Y_{lm}(\theta^{A})Y_{\ell_{i}m_{i}}(\theta^{A})e^{-i\sigma_{g{'}} t}\,,\label{eq:diff eq for the eiomegat}
\end{aligned}
\end{equation}
with the differential operators defined by
\[\begin{aligned}\mathcal{D}_{+}\equiv\frac{d^{2}}{dr^{2}}+2\mu(E_{n_i\ell_{i}}+\sigma_g)+\frac{2\mu^{2}M}{r}-\frac{\ell_{f}(\ell_{f}+1)}{r^{2}}.\end{aligned}\]
Since both sides of the equation contain sums over $g$ (and $g'$) with exponential factors $e^{-i\sigma_g t}$, we can simply equate each term with the same frequency components $(g' = g)$, and after projecting the resulting set of equations onto spherical harmonics we obtain
\begin{equation}
\begin{aligned}\label{eq: diff eq with same g}
    &\mathcal{D}_{+}{\mathcal{Z}^{\ell_{f}m_{f}g}}(g\Omega_{\rm orb},r) \\=&2\mu^{2}rR_{n_i\ell_{i}}^{h}(r)\sum_{l,m}{u}^{lmg}(r)\int Y_{lm}Y_{\ell_{i}m_{i}}Y_{\ell_{f}m_{f}}^{*} d\Omega\\
    =&2\mu^{2}rR_{n_i\ell_{i}}^{h}(r)\sum_{l,m}{u}^{lmg}(r)C\left(l,m;\ell_{i},m_{i};\ell_{f},m_{f}\right),
\end{aligned}
\end{equation}
where the angular coupling coefficients can be expressed in terms of the Wigner 3-$j$ symbols as
\begin{equation}
\begin{aligned}
&C \left(l,m;\ell_{i},m_{i};\ell_{f},m_{f}\right)\\\equiv&(-1)^{m_{f}}\sqrt{\frac{(2l+1)(2\ell_i+1)(2\ell_{f}+1)}{4\pi}} \\
\times&
\begin{pmatrix}l&\ell_i&\ell_{f}\\0&0&0\end{pmatrix}\begin{pmatrix}l&\ell_i&\ell_{f}\\m&m_i&-m_{f}\end{pmatrix}\,.
\end{aligned}
\end{equation}
These coefficients obey the usual selection rules, meaning that the integral above vanishes unless the following conditions are satisfied:
\begin{subequations}
\begin{align}
& m=m_{f}-m_i, \label{eq: equatorial selection rule}\\
& \left|l-\ell_i\right| \leq \ell_{f} \leq l+\ell_i, \\
& l+\ell_i+\ell_{f}\ \text{is even.}
\end{align}
\end{subequations}
Resorting to the Green's function method, one gets
\begin{equation}\label{eq:sol_perts}
\begin{aligned}
    \mathcal{Z}(\sigma_g,r) =&\frac{\mathcal{Z}_{+}(\sigma_g,r)}{\mathcal{W}(\sigma_g)}\int_0^r\mathcal{Z}_{-}(\sigma_g,r')\mathcal{S}(r')dr' \\+&\frac{\mathcal{Z}_{-}(\sigma_g,r)}{\mathcal{W}(\sigma_g)}\int_r^\infty\mathcal{Z}_{+}(\sigma_g,r^{\prime})\mathcal{S}(r^{\prime})dr^{\prime},
\end{aligned}
\end{equation}
where
\begin{equation}\label{eq:source_equatorial}
\begin{aligned}
    \mathcal{S}&(r;\ell_{f},m_{f},\ell_{i},m_{i},g) \\
    \equiv& 2\mu^{2}rR_{n_i\ell_{i}}^{h}(r)\sum_{l,m}{u}^{lmg}(r)C\left(l,m;\ell_{i},m_{i};\ell_{f},m_{f}\right)\,,
\end{aligned}
\end{equation}
and $\mathcal{Z}_{\pm}$ are two linearly independent solutions to the homogeneous equation~\cite{Arana:2024kaz},
\begin{equation}
\begin{aligned}
    \mathcal{Z}_{-}=h_{1}M_{\kappa,\ell_{f}+\frac12}\left(\sqrt{-8\mu(E_{n_i\ell_i}+\sigma_g)}r\right),\\\mathcal{Z}_{+}=h_{2}W_{\kappa,\ell_{f}+\frac12}\left(\sqrt{-8\mu(E_{n_i\ell_i}+\sigma_g)}r\right)\,.
\end{aligned}
\end{equation}
In the last expression, $\kappa = 2\mu^2M/\sqrt{-8\mu(E_{n_i\ell_i}+\sigma_g)}$, $h_{1}$ and $h_{2}$ are integration constants and $M$ and $W$ correspond to Whittaker functions. Their Wronskian is given by 
\begin{equation}
\begin{aligned}
    &\mathcal{W}\left[(\mathcal{Z}_{-})^{\ell_{f},g},(\mathcal{Z}_{+})^{\ell_{f},g}\right] =-h_{1}h_{2}\\&\quad\frac{\Gamma(2+2\ell_{f})\sqrt{8\mu|E_{n_i\ell_i}|}\sqrt{1+\sigma_g/E_{n_i\ell_i}}}{\Gamma\left(\ell_{f}+1-(n_i+\ell_i+1)/\sqrt{1+\sigma_g/E_{n_i\ell_i}}\right)}.
\end{aligned}
\end{equation}
For the solution at large distances,
\begin{equation}
\begin{aligned}\label{eq:Z_inf}
     \mathcal{Z}^{\infty} (\ell_{f},m_{f},g)=\lim_{r\to\infty}\frac{\mathcal{Z}_{+}(r)}{\mathcal{W}}\int_0^r\mathcal{Z}_{-}(r')\mathcal{S}(r')dr'.
\end{aligned}
\end{equation}

\subsection{Orbital loss rates for equatorial orbits}
The scalar field fluctuations, induced by the secondary, cause a perturbation to its stress-energy tensor, which, at leading order and asymptotically, is given by
\begin{equation}
\begin{aligned}
    &\delta T_{\mu\nu}^{S}(r\to\infty) \\
    &\quad\sim~ 2\partial_{(\mu}\delta\mathbf{\Phi}^{*}\partial_{\nu)}\delta\mathbf{\Phi} - \eta_{\mu\nu}\left[\partial_{\alpha}\delta\mathbf{\Phi}^{*}\partial^{\alpha}\delta\mathbf{\Phi}+\mu^{2}|\delta\mathbf{\Phi}|^{2}\right],
\end{aligned}
\end{equation}
with $\delta\mathbf{\Phi} \equiv e^{-i\mu t}\delta \phi$.
The (outgoing) flux of (averaged) energy through a 2-sphere at infinity is~\cite{Annulli:2020lyc}
\begin{equation}
\begin{aligned}
    \dot{E}^\mathrm{\Phi,\infty} &=-\lim_{r\to\infty}r^2\int d\theta d\varphi\sin\theta \delta T_{rt}^S\\
    &=2\sum_{\ell_{f},m_{f},g}\left|\omega+\sigma_g\right|\Re\left[\sqrt{(\omega+\sigma_g)^2-\mu^2}\right]\left|\mathcal{Z}^{\infty}\right|^{2}.
\end{aligned}
\end{equation}
We can also compute the rate of change of the scalar charge~\cite{Annulli:2020lyc}
\begin{equation}
    \dot{Q}^{\rm\infty}=-\lim\limits_{r\to\infty}r^2\int d\theta d\varphi\sin\theta \delta j_r ,
\end{equation}
with
\begin{equation}
    \delta j_r(r\to\infty)\sim\,2\,\mathrm{Im}\left(\delta\mathbf{\Phi}^*\partial_r\delta\mathbf{\Phi}\right)\,,
\end{equation}
which gives
\begin{equation}
    \dot{Q}^{\infty}=-2\sum_{\ell_{f},m_{f},g}\Re\left[\sqrt{(\omega+\sigma_g)^2-\mu^2}\right]\left|\mathcal{Z}^{\infty}\right|^{2}.
\end{equation}
Therefore, the energy lost per unit of time by the orbiting object is,
\begin{equation}
\begin{aligned}
     \dot{E}^{s,\infty} &= \dot{E}^{\Phi,\infty} + \omega\dot{Q}^{\infty}\\
     &= 2\sum_{\ell_{f},m_{f},g}\sigma_g\,\Re\left[\sqrt{(\omega+\sigma_g)^2-\mu^2}\right]\left|\mathcal{Z}^{\infty}\right|^{2}
\end{aligned}
\end{equation}
In the limit of a high-frequency, but still non-relativistic excitations, we have $\mu \gg \Omega_{\rm orb} \gg |E_{n_i\ell_{i}}|$, and the energy loss rate can be approximated by
\begin{equation}
     \dot{E}^{s,\infty} ={2}\sum_{\ell_{f},m_{f},g}g\Omega_{\rm orb}\Re\left[\sqrt{2\mu(g\Omega_{\rm orb}+E_{n_i\ell_i})}\right]\left|\mathcal{Z}^{\infty}\right|^{2}.
\end{equation}
This is equivalent to the {\it ionization power} defined in Ref.~\cite{Tomaselli:2023ysb}.

In the limit of circular orbits, gravitational perturbations induced by the secondary oscillate at a single harmonic frequency $\sigma_g \to m\Omega_{\rm orb}$. Using the selection rule in Eq.~\eqref{eq: equatorial selection rule}, this reduces to $(m_f - m_i)\Omega_{\rm orb}$. We can thus define, $\sigma_{m_{f}}=(m_{f} -m_i)\Omega_{\rm orb}$ and write the energy lost by the orbit for circular orbits as
\begin{equation}
\begin{aligned}
     \dot{E}^{s,\infty}_{\rm circular} =2\sum_{\ell_{f},m_{f}}\sigma_{m_{f}}\Re\left[\sqrt{2\mu(\sigma_{m_{f}}+E_{n_i\ell_i})}\right]\left|\mathcal{Z}^{\infty}(\sigma_{m_{f}})\right|^{2},
\end{aligned}
\end{equation}
where we now have dropped the sum over $g$.

\subsection{Generically inclined orbits}
For generically inclined orbital configurations (with respect to the cloud's equatorial plane), we adopt the same approach as in Sec.~\ref{sec:Rotation of scalar cloud} of the main text. Namely, we rotate the initial spherical harmonic $Y_{\ell_i m_i}$ from the cloud's frame to the orbital frame.

Following the same steps as in App.~\ref{app: Perturbations}, we obtain an equation similar to Eq.~\eqref{eq: diff eq with same g}, but now with a source term that contains a linear combination of cloud states with different azimuthal numbers:
\begin{equation}
\begin{aligned}\label{eq: (generically inclined) diff eq with same g}
    &\mathcal{D}_{+}{\mathcal{Z}^{\ell_{f}m_{f}g}}(g\Omega_{\rm orb},r)=2\mu^{2}rR_{n_i\ell_{i}}^{h}\\
    \times&\sum_{l,m}{u}^{lmg}\sum_{m^{\prime}_i=-\ell_i}^{\ell_i}D_{m_i,-m^{\prime}_i}^{(\ell_i)}\int Y_{lm}Y_{\ell_i,-m^{\prime}_i}Y_{\ell_{f}m_{f}}^{*} d\Omega .
\end{aligned}
\end{equation}
Thus, for inclined orbits, the selection rule~\eqref{eq: equatorial selection rule} is modified to
\begin{equation}\label{eq: inclined orbit selection rule}
    m=m_{f}+m^{\prime}_i,\quad \text{for}  -\ell_i\leq m^{\prime}_i \leq \ell_i\,,
\end{equation}
which implies that the sum over $m$ can be dropped. The source term of Eq.~\eqref{eq: (generically inclined) diff eq with same g} can then be written as
\begin{equation}
\begin{aligned}\label{eq:generically inclined source term}
    &\sum_{m^{\prime}_i=-\ell_i}^{\ell_i}D_{m_i,-m^{\prime}_i}^{(\ell_i)}\mathcal{S}(r;\ell_{f},m_{f},\ell_{i},-m^{\prime}_i,g)
    \\
    =&~2\mu^{2}rR_{n_i\ell_{i}}^{h}\sum_{m^{\prime}_i=-\ell_i}^{\ell_i}D_{m_i,-m^{\prime}_i}^{(\ell_i)}\sum_{l}{u}^{l,m_f+m^{\prime}_i,g}\\
    &\times C \left(l,m_f+m^{\prime}_i;\ell_{i},-m^{\prime}_i;\ell_{f},m_{f}\right)\,,
\end{aligned}
\end{equation}
where $\mathcal{S}$ takes the same form as in Eq.~\eqref{eq:source_equatorial} but now with $m_i$ replaced by $-m^{\prime}_i$. Therefore, the asymptotic solution of Eq.~\eqref{eq: (generically inclined) diff eq with same g} at large distances can be expressed as a linear combination given by
\begin{equation}
\begin{aligned}\label{eq:generically inclined Z}
     &\sum_{m^{\prime}_i=-\ell_i}^{\ell_i}D_{m_i,-m^{\prime}_i}^{(\ell_i)}\mathcal{Z}^{\infty} (r;\ell_{f},m_{f},\ell_{i},m^{\prime}_i,g)\\=&\sum_{m^{\prime}_i=-\ell_i}^{\ell_i}D_{m_i,-m^{\prime}_i}^{(\ell_i)}\lim_{r\to\infty}\frac{\mathcal{Z}_{+}(r)}{\mathcal{W}}\int_0^r\mathcal{Z}_{-}(r')\mathcal{S}(r')dr'.
\end{aligned}
\end{equation}

Finally, the orbital energy loss rate for a generic inclined (and possibly eccentric) orbit can be written as 
\begin{equation}
\begin{aligned}\label{eq:generic_energy_loss}
    \dot{E}^{s,\infty}_{\rm generic}& 
    = {2}\sum_{\ell_{f},m_{f},g}g\Omega_{\rm orb}\Re\left[\sqrt{2\mu(g\Omega_{\rm orb}+E_{n_i\ell_i})}\right]\\
    &\times\left|\sum_{m^{\prime}_i=-\ell_i}^{\ell_i}D_{m_i,-m^{\prime}_i}^{(\ell_i)} \mathcal{Z}^{\infty}(r;\ell_{f},m_{f},\ell_{i},m^{\prime}_i,g)\right|^2.
\end{aligned}
\end{equation}
Notice that, unlike in the relativistic case, the orbits in the Newtonian limit are always closed, namely there's no apsidal precession for eccentric orbits. Because all $m_i^{\prime}$ modes oscillate at the degenerate frequency $g\Omega_{\rm orb}$, they coherently superpose. Consequently, cross terms multiplying different $m_i^{\prime}$ in Eq.~\eqref{eq:generic_energy_loss} do not average to zero, and the energy loss rate for non-equatorial eccentric orbits retains an explicit dependence on the argument of periapsis $\gamma$. This shows one caveat of assuming Keplerian orbits without taking relativistic corrections into account. As we argued in Sec.~\ref{Sec:Point Particle in Generic Orbits}, in the small mass-ratio limit, the typical timescale for apsidal precession is much shorter than the radiation reaction timescale, and therefore apsidal precession should not be neglected in a full adiabatic evolution of the orbit under radiation reaction. One way to take this into account in the Newtonian picture is to average Eq.~\eqref{eq:generic_energy_loss} over $\gamma$, in which case all cross terms average to zero and the sum over $m_i^{\prime}$ can be taken out of the absolute value squared. Doing this, Eq.~\eqref{eq:generic_energy_loss} recovers a similar format to its relativistic counterpart [cf. Eq.~\eqref{eq:generic energy fluxes}].

\subsubsection{Inclined circular orbits}
Similarly to the equatorial case, for inclined circular orbits (i.e. $g\to m$), because of the selection rule in Eq.~\eqref{eq: inclined orbit selection rule}, for a given $m$ and $m_f$, there exists only one value of $m^{\prime}_i$ that yields a non-vanishing contribution, meaning that in the circular limit the sum over $m^{\prime}_i$ drops in Eq.~\eqref{eq: (generically inclined) diff eq with same g} (see below for more details). 
We then find that the total energy loss rate associated with inclined circular orbits can be written as
\begin{equation}
\begin{aligned}
     &\dot{E}^{s,\infty}_{\rm incl, circ} 
     ={2}\sum_{m^{\prime}_i=-\ell_i}^{\ell_i}\left[d_{m_i,-m^{\prime}_i}^{(\ell_i)}(\beta)\right]^{2}\\ &\times \sum_{\ell_{f},m_{f}}\sigma_{m_{f},m^{\prime}_i}\Re\left[\sqrt{2\mu(\sigma_{m_{f},m^{\prime}_i}+E_{n_i\ell_i})}\right]\left|Z^{\infty}(\sigma_{m_{f},m^{\prime}_i})\right|^{2},
\end{aligned}
\end{equation}
where $\sigma_{m_{f},m^{\prime}_i}=(m_{f} +m^{\prime}_i)\Omega_{\rm orb}$ and we remind that $d_{m_i,-m^{\prime}_i}^{(\ell_i)}(\beta)$ is the Wigner small d-matrix.

To understand why the cross terms appearing in Eq.~\eqref{eq:generic_energy_loss} exactly vanish in the limit of a circular orbit, we provide here a detailed analytical explanation. Combining Eqs.~\eqref{eq:ulmg} and \eqref{eq:uTlm}, we have:
\begin{equation}
    u^{lmg}(r) \propto \int^{2\pi}_0 e^{i[g(E_a-e\sin{E_a})-m\varphi_s(E_a)]} \frac{dM_a}{dE_a}dE_a,
    \label{eq:ulmg_integral}
\end{equation}
where $m = m_f + m'$. In the limit of vanishing eccentricity ($e \to 0$), the true anomaly $\varphi_s$ approaches the eccentric anomaly $E_a$ and also the mean anomaly $M_a$. Keeping terms up to $\mathcal{O}(e)$, the phase expansion is:
\begin{equation}
    \varphi_s(E_a) \approx E_a + e \sin(E_a) + \mathcal{O}(e^2).
\end{equation}
Consequently, the oscillatory phase in the integrand simplifies:
\begin{equation}
    g(E_a - e\sin E_a) - m\varphi_s(E_a) \approx (g-m)E_a.
\end{equation}
Thus, in the strict circular limit $e \to 0$, where $M_a=E_a$, the integral becomes an orthogonality relation for Fourier modes:
\begin{equation}
    \lim_{e\to 0} u^{lmg}(r) \propto \int^{2\pi}_0 e^{i(g-m)E_a } dE_a = 2\pi \delta_{g,m}.
    \label{eq:selection_rule}
\end{equation}
Eq. \eqref{eq:selection_rule} establishes a strict selection rule: for a given $g$, the source term is non-vanishing only if the gravitational azimuthal quantum number satisfies $m = g$.

The full source term involves a summation over the cloud's initial azimuthal quantum number $m'_i$ (where $m = m_f + m'_i$). Applying the selection rule derived in Eq.~\eqref{eq:selection_rule}, the factor $u^{l, m_f+m'_i, g}$ is proportional to $\delta_{g, m_f+m^{\prime}_i}$:
\begin{equation}
\begin{aligned}
    &\sum_{m^{\prime}_i=-\ell_i}^{\ell_i}D_{m_i,-m^{\prime}_i}^{(\ell_i)}\mathcal{S}(r;\ell_{f},m_{f},\ell_{i},m^{\prime}_i,g)\\
    =&~2\mu^{2}rR_{n_i\ell_{i}}^{h}\sum_{m^{\prime}_i=-\ell_i}^{\ell_i}D_{m_i,-m^{\prime}_i}^{(\ell_i)}\sum_{l}{u}^{l}(\delta_{g,m_f+m^{\prime}_i})\\
    &\times C \left(l,m_f+m^{\prime}_i;\ell_{i},-m'_i;\ell_{f},m_{f}\right).
\end{aligned}
\end{equation}
This implies that for a fixed $g$ and $m_f$, only the value $m^{\prime}_i = g - m_f$ contributes to the sum over $m^{\prime}_i$.
Let this specific value be denoted by $m^{\prime}_{i,{\rm res}}$. The asymptotic amplitude $\mathcal{Z}^{\infty}(m^{\prime}_i)$ inherits this property:
\begin{equation}
    \mathcal{Z}^{\infty}(m^{\prime}_i) \propto \delta_{m^{\prime}_i, m^{\prime}_{i,{\rm res}}}.
\end{equation}
Expanding the square modulus term in Eq.~\eqref{eq:generically inclined Z}, we have
\begin{equation}
\begin{aligned}
    &\left| \sum_{m^{\prime}_i} D_{m_i,-m^{\prime}_i}^{(\ell_i)} \mathcal{Z}^{\infty}(m^{\prime}_i) \right|^2 \\
    &= \sum_{m^{\prime}_i} \sum_{m_i''} D_{m_i,-m^{\prime}_i}^{(\ell_i)} \left[D_{m_i,-m_i''}^{(\ell_i)}\right]^* \mathcal{Z}^{\infty}(m^{\prime}_i) \left[\mathcal{Z}^{\infty}(m_i'')\right]^*\,,
\end{aligned}
\end{equation}
Due to the selection rule above, the only non-zero terms in this double sum are then the ones with:
\begin{equation}
    m_i' = m'_{i,{\rm res}} \quad \text{and} \quad m_i'' = m'_{i,{\rm res}}.
\end{equation}
This implies that $m'$ must strictly equal $m''$. Consequently, all off-diagonal cross terms (where $m_i' \neq m_i''$) vanish identically. This collapse of the double sum onto the diagonal terms ($m_i'=m_i''$) ensures the exact cancellation of the phase factors in the Wigner $D$-matrix. 
The flux expression thus reduces to a single incoherent summation:
\begin{equation}
    \sum_{m_i'} \left| d_{m_i,-m_i'}^{(\ell_i)} \right|^2 \left| \mathcal{Z}^{\infty}(m_i') \right|^2,
\end{equation}
which matches the formula of inclined circular orbits. This proves that in the circular limit, different $m'$-modes do not interfere because they correspond to distinct, non-degenerate orbital frequency harmonics (i.e., different values of $m\Omega_{\rm orb}$).

Finally, we note that all the calculations above can also be used to compute the Newtonian limit of the angular momentum loss rate, which is equivalent to {\it ionization torque} computed in Ref.~\cite{Tomaselli:2023ysb}.

\subsection{Numerical procedure}
To compute the Newtonian energy loss rates, we have implemented the procedure discussed above in a code written in the \texttt{Julia} programming language.
This section details the numerical strategies employed to ensure accuracy and convergence.

\paragraph*{\underline{Evaluation of special functions.}} 
The computation of the solution~\eqref{eq:sol_perts} relies on the precise evaluation of the Whittaker functions $M_{\kappa, \mu}(z)$ and $W_{\kappa, \mu}(z)$. Since standard \texttt{Julia}  libraries lack a direct implementation of these functions for complex parameters, we evaluate them via their relations to the Confluent Hypergeometric functions, ${}_1F_1(a;b;z)$ and $U(a,b,z)$, which are implemented in the \texttt{HypergeometricFunctions.jl} library:
\begin{align}
    M_{\kappa, \mu}(z) &= e^{-z/2} z^{\mu+1/2} {}_1F_1(\mu-\kappa+1/2;\, 1+2\mu;\, z) \,, \nonumber \\
    W_{\kappa, \mu}(z) &= e^{-z/2} z^{\mu+1/2} U(\mu-\kappa+1/2;\, 1+2\mu;\, z) \,\nonumber .
\end{align}
The spherical harmonics $Y_{lm}$ and associated selection rules are handled using the \texttt{SphericalHarmonics.jl} package.

\paragraph*{\underline{Fourier series coefficients.}}
To perform the integration over the eccentric anomaly $E_a \in [0, 2\pi]$, needed to compute the Fourier series coefficients $u^{lmg}(r)$ [see Eq.~\eqref{eq:ulmg}], we employ the composite trapezoidal rule over a uniform grid of $N=100$ points. This method provides a robust and efficient approach to resolving the phase oscillations, driven by the combination of the mean anomaly and the orbital phase $\varphi_s$, which become increasingly difficult to resolve for higher harmonic numbers $g$ and large eccentricities $e$.


\paragraph*{\underline{Radial integration and asymptotic amplitudes.}}
The central numerical task is the computation of the asymptotic amplitude $\mathcal{Z}^{\infty}$, defined in Eq.~\eqref{eq:Z_inf}. The source term $\mathcal{S}(r)$ decays exponentially at large distances given its linear dependence on the hydrogenic bound state function $R_{n_i \ell_i}^h(r)$. Consequently, the theoretically infinite integration domain is numerically truncated to a finite range $[0, R_{\text{inf}}]$.
To optimize computational efficiency while maintaining accuracy, we define the truncation radius $R_{\text{inf}}$ dynamically based on wavelength of the outgoing scalar radiation. Specifically, we set $R_{\text{inf}} \propto k^{-1}$, where $k$ is the wave number of the radiated scalar perturbations, ensuring the integration covers a sufficiently large number of wavelengths. 
The radial integral is evaluated using a high-order (order 25) adaptive Gauss-Kronrod scheme.

\paragraph*{\underline{Summation and convergence.}}
The total energy loss rate $\dot{E}^{s,\infty}$ is computed by summing over the final quantum numbers $\{\ell_f, m_f\}$ and the Fourier harmonic index $g$.
\begin{itemize}
    \item Final quantum numbers: We sum over $\ell_f$ up to a cutoff $\ell_{f,\text{max}}$; we typically set $\ell_{f,\text{max}} = 10$ to maintain accuracy across the parameter space. For each $\ell_f$, the gravitational multipole $l$ is strictly constrained by the selection rule $|\ell_f - \ell_i| \le l \le \ell_f + \ell_i$ in the source term.
    \item Fourier harmonics: For eccentric orbits, the summation over $g$ is formally infinite. However, the power radiated in higher harmonics is suppressed by the Bessel functions $J_g(ge)$ appearing in the series expansion of the true anomaly, Eq.~\eqref{eq:true_anomaly_vs_mean_anomaly}. We truncate the series at a cutoff $g_{\text{max}}$, chosen such that the contribution of the neglected terms is numerically small. As expected, we find that larger eccentricities require summing over a larger number of $g$ modes. For the $e=0.6$ case presented below, we explicitly checked convergence by calculating the summation up to $g_{\text{max}}=11$, finding a maximum relative truncation error of $\approx 0.4\%$ with respect to setting $g_{\text{max}}=10$. Therefore for the results presented here, we set $g_{\text{max}}=10$.
\end{itemize}

\subsection{Comparison with previous work and with relativistic results}





To validate the numerical and analytical framework presented above, we compute the total energy loss for binaries across a range of separations and orbital geometries, benchmarking our results against those reported in Ref.~\cite{Tomaselli:2023ysb}. We find that our results are in excellent agreement with their findings across the majority of the parameter space. However, we find noticeable deviations emerging as the separation and eccentricity increase, especially in the vicinity of sharp discontinuities where we found differences that can reach up to $20\%$. At this stage, it remains unclear whether these differences stem from intrinsic differences between the two frameworks or are artifacts of numerical resolution. Nonetheless, given the agreement between both frameworks far from sharp discontinuities, we proceed to compare the Newtonian results against the relativistic ones presented in the main text.

\begin{figure}
    \centering
    \includegraphics[width=0.99\linewidth]{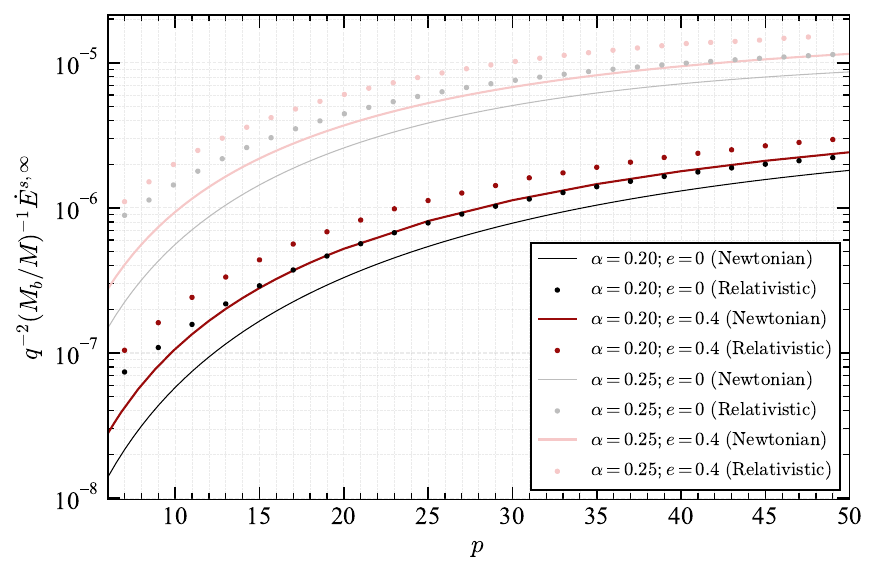}
    \caption{Comparison of the energy loss rate at infinity $\dot{E}^{s,\infty}$ (i.e. the {\it ionization power}) obtained using the relativistic framework (scattered points) and the Newtonian framework (solid lines). The results are shown for coupling strengths $\bm{\alpha} = 0.20$ (dark colours) and $\bm{\alpha} = 0.25$ (light colours), with orbital eccentricities $e=0$ (black series) and $e=0.4$ (red series) and prograde equatorial orbits. We consider a dipolar $\ell_i=m_i=1$ scalar cloud in the state $n_i=0$.}
    \label{fig:NvsR}
\end{figure}

Figure~\ref{fig:NvsR} displays the energy loss rate $\dot{E}^{s,\infty}$ as a function of $p$ for prograde equatorial orbits. As in the main text, we consider a dipolar $\ell_i=m_i=1$ scalar cloud in the state $n_i=0$. We perform the evaluation for two scalar coupling strengths, $\bm{\alpha}=0.20$ and $\bm{\alpha}=0.25$, and compare circular orbits ($e=0$) against eccentric configurations ($e=0.4$). The solid curves represent the predictions of our Newtonian approximation, while the scattered points denote fully relativistic results computed using the framework of the main text. For circular orbits, the differences between the two approaches are in agreement with previous work~\cite{Dyson:2025dlj,Tomaselli:2025jfo}. The comparison suggests that the relativistic results converge towards the Newtonian ones as the separation $p$ increases, with a similar trend seen for both the circular and eccentric orbits, confirming that the Newtonian framework provides an increasingly accurate description in the weak-field regime (large $p$).



\subsection{Remarks regarding generic orbits}

\begin{figure}
    \centering
    \includegraphics[width=0.99\linewidth]{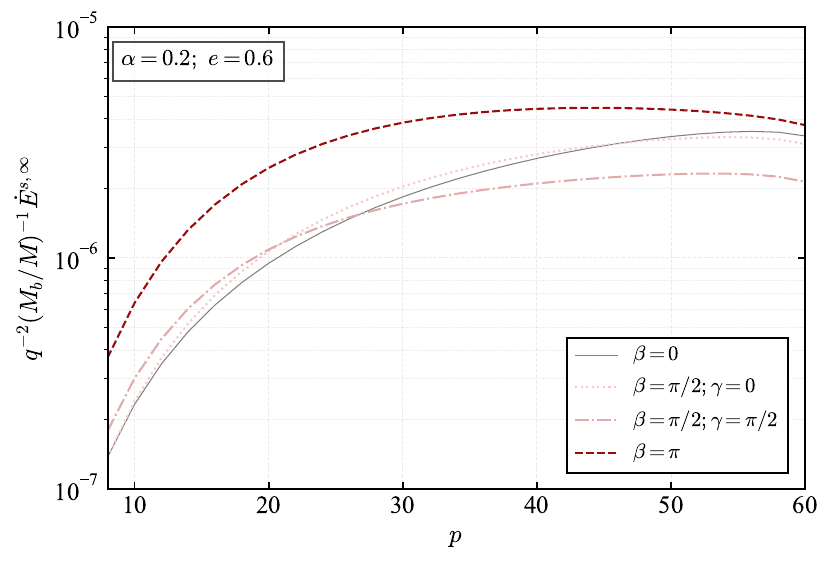}
    \caption{Energy loss rate at infinity $\dot E^{s,\infty}$ as a function of the semi-latus rectum $p$ for orbits with a fixed eccentricity $e=0.6$, obtained using the Newtonian framework. The grey solid and dark-red dashed lines represent prograde ($\beta=0$) and retrograde ($\beta=\pi$) equatorial orbits, respectively. For polar configurations ($\beta = \pi/2$), we compare two distinct orientations: $\gamma=0$ (dotted line), where the orbit is rotated about its major axis, and $\gamma=\pi/2$ (dash-dotted line), where it is rotated about its minor axis. We consider a dipolar $\ell_i=m_i=1$ scalar cloud in the state $n_i=0$ with coupling strength $\bm{\alpha} = 0.20$.}
    \label{fig:NewtonGeneric}
\end{figure}

As already explained above, in the Newtonian approximation, the loss rates for generically inclined and eccentric orbits depend on both the obliquity $\beta$ and on the argument of periapsis $\gamma$ [see Eq.~\eqref{eq:generic_energy_loss}]. Since such generic results have never been presented in the literature, for completeness, here we show an example of the dependence of the loss rate on $\gamma$ for an eccentric polar orbit ($\beta = \pi/2$), comparing it with the case of an equatorial retrograde ($\beta=\pi$) and prograde orbit ($\beta=0$). The results are presented in Fig.~\ref{fig:NewtonGeneric} for an eccentricity $e=0.6$ and the same cloud configuration considered before. For polar orbits, the energy loss exhibits a dependence on the argument of periapsis $\gamma$. 
As we argued above, a sensible comparison between Newtonian and relativistic results requires averaging over $\gamma$ to take into account relativistic apsidal precession. Due to the numerical cost of evaluating relativistic infinity loss rates for non-equatorial eccentric orbits, we leave such detailed comparisons for generic orbits to future work.

\section{Numerical procedure for the relativistic case}\label{App:Numerics}

In this appendix we provide details about the numerical implementation used to compute the relativistic scalar energy and angular momentum loss rates discussed in the main text. 

The general steps employed to solve the inhomogeneous differential Eq.~\eqref{eq:ordinary differential equation} are the following:

\begin{itemize}
    \item First, we determine the eigenfrequency $\omega_{n_i\ell_i}$ and the radial profile $R^{b}_{n_i\ell_i}(r)$ of the scalar cloud using the continued-fraction method of Ref.~\cite{Dolan:2007mj}.

    \item We then reconstruct the metric perturbation amplitudes, specifically the set of radial functions $\{H_0, H_1, H_2, K, h_0, h_1\}$, by implementing the metric reconstruction equations provided in Refs.~\cite{Hopper:2010uv, Hopper:2015icj}. For $l \geq 2$, this requires evaluating the quantities $\Psi^\pm_{lm}(t,r) = \sum_{n'} R^\pm_{lmn'} e^{-i\sigma_{mn'} t}$. Since this step involves an infinite summation over $n'$, we employ an adaptive truncation that automatically includes all relevant $n'$ modes while maintaining a user‑controlled accuracy. For a given $n$, with $n$ the scalar field Fourier harmonic index [see Eq.~\eqref{eq: scalar_pert_amps}], we start from the mode with $n'=n$, and sum symmetrically to both higher and lower values of $n'$. 
    Jointly with the scalar cloud quantities computed in the previous step, we can then compute all the source terms in Eq.~\eqref{eq: radial functions of source}.

    
    \item Finally, the solution to Eq.~\eqref{eq:ordinary differential equation} with appropriate boundary conditions is obtained by integrating the homogeneous scalar field equation to first obtain $Z_{\rm in/up}$, which are then used in Eq.~\eqref{eq:solution of inhomogeneous equation}. In practice, due to the form of the metric perturbation amplitudes [see Eq.~\eqref{eq: generic_metric_amp}], we separate the source terms~\eqref{eq: radial functions of source} into pieces involving $\delta(r-r_p(t))$, $\delta'(r-r_p(t))$, $\Theta(r - r_p(t))$ and $\Theta(r_p(t) - r)$. After doing that, we perform the radial integration in Eq.~\eqref{eq:solution of inhomogeneous equation} analytically for the terms involving the Dirac delta function and its derivative, while the radial integrals involving the Heaviside functions are computed numerically. Moreover, since we only need to know the asymptotic behavior of $Z(r)$ either at infinity or at the horizon to compute the loss rates, only one of the integrals in Eq.~\eqref{eq:solution of inhomogeneous equation} needs to be computed in practice, i.e. only the first term proportional to $Z_{\rm up}$ contributes when $r\to\infty$ while at the horizon ($r\to 2M$), only the second term proportional to $Z_{\rm in}$ contributes.
\end{itemize}

\paragraph*{\underline{Differential equation solver.}}

To obtain the homogeneous solutions of the governing ODE in Eq.~\eqref{eq:ordinary differential equation} we have implemented a high-precision solver in \texttt{Julia}. 

The integration is performed up to a large outer boundary $r_{\infty}$ that needs to be chosen with care. For modes with $\omega_+^2 - \mu^2 > 0$, $r_{\infty}$ is placed many wavelengths away from the BH to capture the asymptotic behavior of the solutions correctly. 
However, in the regime where $\omega_+^2 - \mu^2 < 0$, the scalar perturbations decay exponentially. Integrating the scalar equation far beyond the characteristic decay length scale $1/\sqrt{\mu^2 - \omega_+^2}$ is numerically impractical. 
Therefore we determine $r_{\infty}$ dynamically, which distinguishes two regimes based on the sign of $\omega_+^2 - \mu^2$. For decaying modes ($\omega_+^2 - \mu^2 < 0$), we define the decay lengthscale
\begin{equation}
\kappa = \sqrt{\mu^2-\omega_+^2},
\end{equation}
and choose
\begin{equation}
r_\infty^{\rm(decay)}
=
\min\!\left[
\frac{100}{\kappa},10^3
\right].
\end{equation}
This ensures that the outer boundary covers a sufficient number of decay e-folds, while at the same time preventing the numerical domain from becoming unnecessarily large when $\kappa$ is very small.
For propagating modes, ($\omega_+^2 - \mu^2 > 0$), the implementation retains a frequency-based prescription tied to the orbital frequency,
\begin{equation}
r_\infty^{\rm(prop)}
=
\min\!\left[
\frac{2000}{|\sigma_{mn}|},
10^6
\right].
\end{equation}

The $Z_{\rm in}$ solution is obtained by integrating outward from $r_h=2M(1+\epsilon_h)$ with $\epsilon_h=10^{-4}$ using a truncated near-horizon expansion as initial conditions (see below), while the outgoing solution is integrated inward from $r=r_\infty$ using a truncated large-$r$ asymptotic expansion as initial conditions. If the initial choice of $r_\infty$ is too large, the asymptotic initial data can become non-finite, producing \texttt{NaN} values in the evaluation of the differential equation. We then reduce $r_\infty$ iteratively by a factor of two until the initial data becomes finite. In this way, we can safely start from a conservatively large outer boundary, which is physically more appropriate, without compromising numerical stability.

We solve the homogeneous radial equation using a dedicated adaptive explicit Runge--Kutta method of order $5(4)$ based on the Dormand--Prince pair, taking the fifth-order formula for step advancement and the embedded fourth-order formula for local error estimation. The error control is imposed simultaneously on the field and its radial derivative, with both the relative and absolute error tolerances set to $10^{-6}$. The step size is adjusted dynamically within the range $\left[(r_\infty-r_h)/10^{13}, (r_\infty-r_h)/10^5\right]$

The accepted mesh points are stored together with the values of the solution and its first two radial derivatives, and the solution at intermediate radii is reconstructed using interpolation. This is crucial for the subsequent source integrations, where the homogeneous solutions and their derivatives are evaluated repeatedly. The Wronskian [Eq.~\eqref{eq:Wronkskian_Z}] is then computed at an interior matching point chosen as the geometric mean of the inner and outer boundaries.

To obtain the gravitational perturbations, for which we need to compute the homogeneous solutions to the Regge-Wheeler and Zerilli equations, the same solver and overall algorithm is used.

\paragraph*{\underline{Integration domain.}}
To obtain the homogeneous solutions $\hat{R}^{\pm}_{\rm e/o}$ and $Z_{\rm in/up}$, we numerically integrate the corresponding differential equations subject to the boundary conditions specified in Eqs.~\eqref{eq:Rhat-}, \eqref{eq:Rhat+}, \eqref{eq:Zin}, and \eqref{eq:Zup}. Since the homogeneous solutions are subsequently evaluated through interpolation of the differential equation output, the source integrals are restricted to the same radial domain on which the differential equation solutions are constructed. Therefore, the integration range is also taken to be $r \in [r_h, r_{\infty}]$.

To ensure numerical stability and efficiently handle orbital eccentricity, all time integrals appearing in the source terms are transformed into integrals over the relativistic anomaly $\chi \in [0, 2\pi]$ via the mapping given in Eq.~\eqref{eq:t(chi)}. In the numerical implementation, small neighbourhoods around $\chi = 0,\pi,2\pi$ are neglected, since some integrand functions can become indeterminate exactly at the boundaries (although note that the limit $\chi\to 0,\pi,2\pi$ of the integrand functions is always well defined). The integrals over $\chi$ are evaluated in the interval $[\epsilon_\chi,\pi-\epsilon_\chi]\cup[\pi+\epsilon_\chi,2\pi-\epsilon_\chi]$ with $\epsilon_\chi=10^{-6}$, which we found to be necessary for the results near the resonant peaks to fully converge.

\paragraph*{\underline{Boundary conditions.}}
Following Ref.~\cite{Brito:2023pyl}, we mitigate numerical truncation errors by employing high-order asymptotic expansions for the boundary conditions at both the horizon and infinity:
\begin{eqnarray}
    \hat{R}^-_{\rm e/o}(r\to 2M)&=& e^{-i \sigma r_*} \sum_{i=0}^{10} a_i (r-2M)^i\,,\\
    \hat{R}^+_{\rm e/o}(r\to \infty)&=& e^{+i \sigma r_*} \sum_{i=0}^{9} \frac{b_i}{r^i}\,,\\
     Z_{\rm{in}}(r\to 2M)&=& e^{-i \omega_+ r_*} \sum_{i=0}^{10} \tilde{a}_i (r-2M)^i\,,\\
    Z_{\rm{up}}(r\to \infty)&=& e^{+i k_+ r_*}r^{-\nu_+} \sum_{i=0}^{9} \frac{\tilde{b}_i}{r^i}\,,
\end{eqnarray}
where the series coefficients $(a_i,b_i,\tilde{a}_i,\tilde{b}_i)$ are determined by substituting these expansions into the homogeneous equations, solving order-by-order in $(r-2M)$ or $1/r$, and fixing the zeroth-order normalization coefficients to unity ($a_0=b_0=\tilde{a}_0=\tilde{b}_0=1$).

Special care needs to be taken, however, close to static ($\sigma_{mn}=0$) modes, which can occur at discrete points in the ($p,e$) parameter space even for $m\neq 0$ and $n\neq 0$ (see e.g. Ref.~\cite{Akcay:2013wfa}).
To ensure numerical stability and avoid the challenges associated with integrating modes with very low frequencies~\cite{Akcay:2013wfa}, our implementation excludes all modes satisfying
\begin{equation}
M|\sigma_{mn}| < 10^{-8}.
\end{equation}
For these cases we do not perform a direct evaluation and instead set their scalar flux contributions to zero.

To evaluate the impact of neglecting nearly static modes, we also performed preliminary tests by imposing a decaying power-law behavior at infinity (for the gravitational perturbations) when $M|\sigma_{mn}| \approx 0$. This follows from the fact that, for exactly static modes, the Regge-Wheeler and Zerilli homogeneous equations admit power-law solutions at infinity, rather than oscillatory ones, via the expansion:
\begin{equation}
    \hat R^{+}_{\rm e/o,~static}(r\to\infty)=r^{-l}\sum_{i=0}^{9}\frac{d_i}{r^i}\,,
\end{equation}
where the coefficients $d_i$ are determined recursively and we use the normalization $d_0=1$. Here, the unphysical growing solution, proportional to $r^{l+1}$, is discarded. Since we did not find significant differences in the loss rates when using this approach compared to neglecting the contribution of nearly static modes, in our final implementation we simply discard their contribution.

\paragraph*{\underline{Validation and normalization.}}
We validated our numerical implementation of the GW fluxes against established benchmarks. In the circular limit, our results are consistent with the high-precision values provided by the \texttt{Black Hole Perturbation Toolkit}~\cite{BHPToolkit}. For eccentric orbits, we reproduced the results reported in Ref.~\cite{Hopper:2010uv}. Specifically, even for a highly eccentric case ($e=0.9$), our results agree with these benchmarks to at least five significant digits, providing sufficient accuracy for the purposes of this work.

\subsection{Additional steps for non-equatorial orbits}\label{app:numerics non-equatorial orbit}
The numerical integration of the radial perturbation equations for non-equatorial orbits follows the general pipeline used for equatorial orbits.
As shown in Eq.~\eqref{eq:generic energy fluxes}, due to the fact that we are considering a Schwarzschild BH background, the inclination enters solely through the mixing of different azimuthal projections $m_i^{\prime}$ of the scalar cloud, weighted by the Wigner coefficients $\left|D_{m_i, -m_i^{\prime}}^{(\ell_i)}\right|^2$. This fact allows us to decouple the numerical integration from the geometric orientation. The procedure is organized as follows:
\begin{itemize}
\item For a fixed orbital configuration, we iterate over the relevant multipole moments $(\ell_f, m_f)$, performing an inner loop over the cloud's azimuthal projection index $m_i^{\prime} \in [-\ell_i, \ell_i]$. For each unique tuple $(\ell_f, m_f, m_i^{\prime})$, we solve the corresponding radial differential equation to obtain the partial contribution to the loss rates. Since the gravitational azimuthal number $m$ is fixed by $m = m_f + m_i^{\prime}$, the orbital frequency $\sigma_{mn}$ and the scalar decay/propagation condition ($\omega_+^2 - \mu^2 \lessgtr 0$) are evaluated dynamically for each component. The dynamic boundary condition at infinity, $r_\infty$, is adjusted accordingly for each $m_i^{\prime}$ to ensure numerical stability.

\item The raw partial energy and angular momentum loss rates for each $m_i^{\prime}$ are stored independently, without assuming a specific value for $\beta$. This step represents the computationally expensive portion of the workflow.

\item To obtain the total loss rates for specific $\beta$, we then perform a post-processing step where we linearly combine the pre-computed partial fluxes, weighting each term by the geometric factor $\left|D_{m_i, -m_i^{\prime}}^{(\ell_i)}\right|^2$. This allows us to explore the entire parameter space of $\beta \in [0, \pi]$ using a single set of numerical solutions.
\end{itemize}

This modular approach ensures that the computationally intensive integration is performed only once per orbital parameter set, while the geometric dependence is handled analytically.

\section{Mode-by-mode contribution to the loss rates}\label{App:Comparison of different modes}
In this appendix we show how different $\{\ell_f,m_f\}$ contribute to the relativistic energy loss rates, discussed in Sec.~\ref{Sec:Point Particle in Eccentric, Equatorial Motion}. All results shown here are for eccentric, equatorial prograde orbits for a dipolar $\{n_i,\ell_i,m_i\}=\{0,1,1\}$ cloud with $M\mu=0.2$.

\begin{figure}
    \centering
    \includegraphics[width=0.99\linewidth]{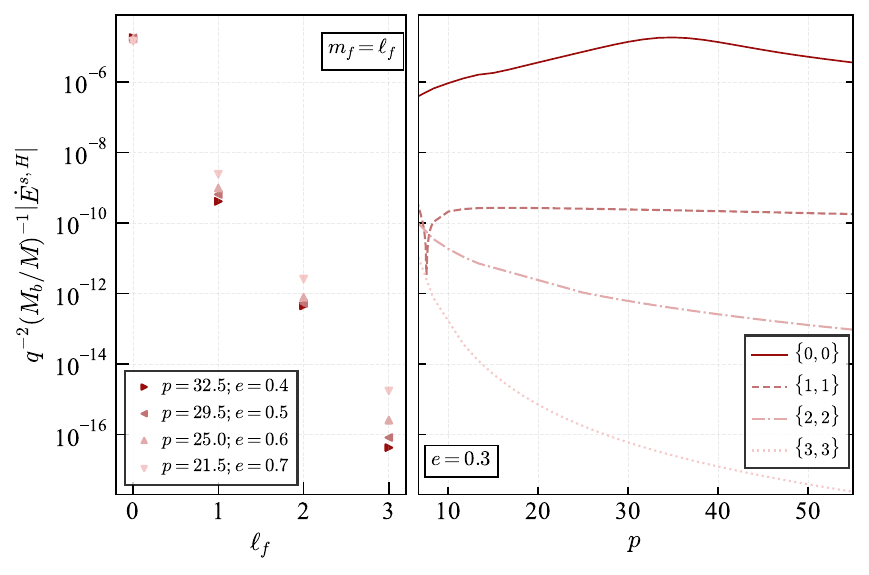}
    \caption{Contributions of individual $\ell_f=m_f$ modes to the horizon energy loss rate $\dot E^{s,H}$. \textit{Left panel}: The first three $\ell_f=m_f$ modes for several representative orbits with different $p$ and $e$. \textit{Right panel}: Dependence of selected $\{\ell_f,m_f\}$ mode contributions on $p$ for a fixed eccentricity $e=0.3$. Note that the $\{\ell_f,m_f\} =\{0,0\}$ contribution is strictly negative, whereas all higher-order modes shown are positive. We consider a dipolar $n_i=0$, $\ell_i=m_i=1$ scalar cloud, with $M\mu=0.2$.}
    \label{fig:HorFluxCompare}
\end{figure}

Fig.~\ref{fig:HorFluxCompare} shows the mode-by-mode contributions to the horizon energy loss rate $\dot E^{s,H}$, for eccentric, equatorial prograde orbits. Clearly, the $\{\ell_f,m_f\} = \{0,0\}$ mode overwhelmingly dominates the horizon loss rate, exceeding higher-order $\ell_f = m_f$ modes (which dominate over $\ell_f\neq m_f$ modes) by $4$ to $5$ orders of magnitude, in agreement with previous work~\cite{Brito:2023pyl}. Therefore, in the main text we evaluated the horizon loss rates only considering the $\{\ell_f,m_f\} = \{0,0\}$ mode.

\begin{figure}
    \centering
    \includegraphics[width=0.99\linewidth]{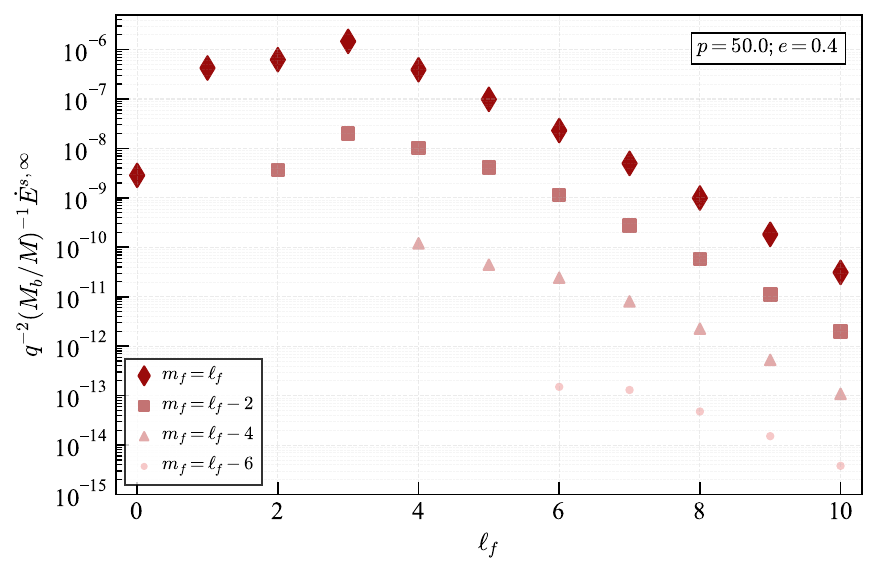}
    \caption{Contributions from different $\{\ell_f,m_f\}$ modes to the infinity energy loss rate $\dot{E}^{s,\infty}$, for an eccentric prograde orbit with $e=0.4$ and $p=50$. Due to the equatorial symmetry of the orbit and the parity of the initial dipolar cloud, modes with odd parity vanish identically; thus, we only need to consider modes satisfying $\ell_f+m_f = 2\mathfrak{p}$, with $\mathfrak{p}\in \mathbb{N}$.}
    \label{fig:different_lm_mode}
\end{figure}

For infinity loss rate $\dot E^{s,\infty}$ the mode-by-mode contribution of different $\{\ell_f,m_f\}$ is illustrated in Fig.~\ref{fig:different_lm_mode} for an eccentric orbit with $e=0.4$.  It is worth noticing that prograde eccentric orbits excite a broader spectrum of modes compared to prograde circular orbits. Notably, the $\ell_f = 0, 1$ modes, which are kinematically forbidden from radiating to infinity in circular configurations, given that those modes satisfy $\omega^2_+ < \mu^2$ for circular orbits, now yield non-zero contributions. This is because the radial orbital harmonics $n\Omega_r$ provide the additional energy required to overcome the potential barrier imposed by the finite boson mass. Nevertheless, we recover the expected trend with the $\ell_f=m_f$-mode contribution decaying exponentially with increasing $\ell_f$. Consequently, for the results shown in the main text we truncate the infinite summation for the infinity loss rates at $\ell_f = 6$.

\section{Source terms for gravitational perturbations}\label{app: Source terms}
In this appendix, we summarize the expressions for ${G}_{l m}$ and ${F}_{l m}$, as given in Ref.~\cite{Hopper:2010uv}, needed to compute the even-parity Zerilli-Moncrief and odd-parity Cunningham-Price-Moncrief master functions. We give their fully evaluated form at $r = r_p(\chi)$, as they appear inside the integral of Eq.~\eqref{eq:Clmn}. As detailed in Ref.~\cite{Martel:2005ir}, these source terms are obtained from combinations of spherical harmonic projections of the particle's stress-energy tensor $T_{p}^{\mu\nu}$.


\subsection{Even parity}
In the even-parity case, we have:
\begin{equation}
\begin{aligned}
 & \frac{{G}_{l m}(\chi)}{m_p} =\mathcal{G}_l^{rr}q_{l m}^{rr}+\mathcal{G}_l^{tt}q_{l m}^{tt}+\mathcal{G}_l^rq_{l m}^r+\mathcal{G}_l^\flat q_{l m}^\flat+\mathcal{G}_l^\sharp q_{l m}^\sharp, \\
 & \frac{{F}_{l m}(\chi)}{m_p} =\mathcal{F}_l^{rr}q_{l m}^{rr}+\mathcal{F}_l^{tt}q_{l m}^{tt},
 \end{aligned}
\end{equation}
where
\begin{equation}
\begin{aligned}
\mathcal{G}_l^{rr}(\chi) &:=\frac{1}{\left(\lambda+1\right)r_{p}\Lambda_{p}^{2}}\left[\left(\lambda+1\right)\left(\lambda r_{p}+6M\right)r_{p}+3M^{2}\right], \\
\mathcal{G}_{l}^{tt}(\chi) &:=-\frac{f_p^2}{(\lambda+1)r_p\Lambda_p^2}\left[\lambda\left(\lambda+1\right)r_p^2+6\lambda Mr_p+15M^2\right], \\
\mathcal{G}_{l}^{r}(\chi) &:=\frac{2f_p}{\Lambda_p},\quad\mathcal{G}_l^\flat(\chi):=\frac{r_pf_p^2}{(\lambda+1)\Lambda_p},\quad\mathcal{G}_l^\sharp(\chi):=-\frac{f_p}{r_p}, \\
\mathcal{F}_l^{rr}(\chi) &:=-\frac{r_p^2f_p}{\left(\lambda+1\right)\Lambda_p},\quad\mathcal{F}_l^{tt}(\chi):=\frac{r_p^2f_p^3}{\left(\lambda+1\right)\Lambda_p},\\
\Lambda(r)&:=\lambda+\frac{3M}{r},
\end{aligned}
\end{equation}
whereas
\begin{equation}
\begin{aligned}\label{eq:qlm(t)}
q^{tt}(\chi)&=8\pi \frac{\mathcal{E}}{r_p^2f_p}Y^*, ~q^{rr}(\chi)=8\pi \frac{f_p}{\mathcal{E}r_p^2}\left(\mathcal{E}^2-U_p^2\right)Y^*,\\
q^{tr}(\chi)&=8\pi \frac{u^r}{r_p^2}Y^*, \\
q^t(\chi)&=\frac{16\pi }{l(l+1)}\frac{\mathcal{L}}{r_p^2}Y^*,~ q^r(\chi)=\frac{16\pi }{l(l+1)}\frac{\mathcal{L}}{\mathcal{E}r_p^2}u^rY_\varphi^*, \\
q^\flat(\chi)&=8\pi \frac{\mathcal{L}^2}{\mathcal{E}}\frac{f_p}{r_p^4}Y^*,~ q^\sharp(\chi)=32\pi \frac{(l-2)!}{(l+2)!}\frac{\mathcal{L}^2}{\mathcal{E}}\frac{f_p}{r_p^2}Y_{\varphi\varphi}^*.
\end{aligned}   
\end{equation}
Here, $Y$, $Y_{\varphi}$, and $Y_{\varphi\varphi}$ are shorthand for the even-parity scalar, vector, and tensor spherical harmonics, respectively, evaluated along the worldline at $\theta = 
\pi/2$ and $\varphi = \varphi_p(\chi)$. Namely,
\begin{equation}
\begin{aligned}
&Y_{\varphi} = \frac{\partial}{\partial\varphi}Y^{lm},\\
&Y_{\varphi\varphi}=\left[\frac{\partial^2}{\partial\varphi^2}+\sin\theta\cos\theta\frac{\partial}{\partial\theta}+\frac{1}{2}l(l+1)\sin^2\theta\right]Y^{lm}.
\end{aligned}   
\end{equation}

\subsection{Odd parity}
In the odd-parity case, we have:
\begin{equation}
\begin{aligned}
&\frac{{G}_{l m}(\chi)}{m_p} =\mathcal{G}_l^{r_1}p_{l m}^r+\mathcal{G}_l^{r_2}\frac{dp_{l m}^r}{d\chi}\frac{d\chi}{dt_p}+\mathcal{G}_l^tp_{l m}^t,\\
&\frac{{F}_{l m}(\chi)}{m_p}=\mathcal{F}_l^rp_{l m}^r+\mathcal{F}_l^tp_{l m}^t, \\
\end{aligned}
\end{equation}
where
\begin{equation}
\begin{aligned}
&\mathcal{G}_l^{r_1}(\chi) :=\frac{\dot{r}_p}{\lambda},\quad 
\mathcal{G}_l^{r_2}(\chi):=\frac{r_p}{\lambda},\quad 
\mathcal{G}_l^t(\chi):=-\frac{f_p}{\lambda},\\
&\mathcal{F}_l^r(\chi):=-\frac{r_p\dot{r}_p}{\lambda},\quad  \mathcal{F}_l^t(\chi):=\frac{r_pf_p^2}{\lambda},
\end{aligned}
\end{equation}
whereas
\begin{equation}
\begin{aligned}
&p^t_{lm}(\chi)=\frac{16\pi}{l(l+1)}\frac{\mathcal{L}}{r_p^2}X_\varphi^*,\\
&p^r_{lm}(\chi)=\frac{16\pi}{l(l+1)}\frac{\mathcal{L}}{\mathcal{E}}\frac{f_p}{r_p^2}u^rX_\varphi^*.\\
\end{aligned}
\end{equation}
Here, $X$, $X_{\varphi}$, and $X_{\varphi\varphi}$ are shorthand for the odd-parity scalar, vector, and tensor spherical harmonics, respectively, evaluated along the worldline at $\theta = 
\pi/2$ and $\varphi = \varphi_p(\chi)$. Namely,
\begin{equation}
\begin{aligned}
&X_\varphi=\sin\theta\frac{\partial}{\partial\theta}Y^{lm},\\ &X_{\varphi\varphi}=\left(\sin\theta\frac{\partial^2}{\partial\theta\partial\varphi}-\cos\theta\frac{\partial}{\partial\varphi}\right)Y^{lm}.
\end{aligned}
\end{equation}

\section{Expressions of metric perturbation amplitudes}\label{app: metric perturbation amplitudes}
In this appendix, we summarize the expressions for the metric perturbation amplitudes used in the main text, as given in Refs.~\cite{Hopper:2010uv, Hopper:2015icj}. These amplitudes are used to build the source term for scalar perturbations~\eqref{eq:source in TD}. To ease the notation, we suppress the $l$ and $m$ indices in the amplitudes, although all quantities below refer to individual mode amplitudes. We collect all necessary ingredients required for reconstructing the metric perturbation amplitudes, including the associated jump conditions across the particle's worldline.

Since the point-particle acts as a distributional source in the RWZ master equations, the master functions inherit discontinuities 
at the particle's radial position. 
Using the method of extended
homogeneous solutions, the jumps in $\Psi$ and $\partial_{r}\Psi$ across the worldline are defined as
\begin{equation}
\begin{aligned}
    [\![\Psi]\!]_p(t)&:=\Psi^+\left(t,r_p(t)\right)-\Psi^-\left(t,r_p(t)\right)\,,\\
     [\![\partial_r\Psi]\!]_p(t)&:=\partial_r\Psi^+\left(t,r_p(t)\right)-\partial_r\Psi^-\left(t,r_p(t)\right)\,,
\end{aligned}
\end{equation}
and take the form~\cite{Hopper:2010uv}
\begin{equation}
\begin{aligned}
    [\![\Psi]\!]_p(t)=&\frac{\mathcal{E}^2}{f_p^2U_p^2}{F}_{lm}(t),\\
    [\![\partial_r\Psi]\!]_p(t)=&\frac{\mathcal{E}^2}{f_p^2U_p^2}\left[{G_{lm}}(t)-2\frac{d{r}_p}{dt}\frac{d[\![\Psi]\!]_p}{dt}\right.\\
    &\left.+\frac{1}{U_p^2r_p^2}\left(3M-\frac{\mathcal{L}^2}{r_p}+\frac{5M\mathcal{L}^2}{r_p^2}\right){F_{lm}}(t)\right]\,,
\end{aligned}
\end{equation}
where $G_{lm}(t)$ and $F_{lm}(t)$ are evaluated at $r=r_p(t)$ and can be computed using the expressions in App.~\ref{app: Source terms}.

\subsection{Even parity}
The even parity $l\geq 2$ metric perturbation amplitudes are related to Zerilli-Moncrief master function and the even-parity source terms by
\begin{equation}
\begin{aligned}
 & K(t,r) =~f\partial_r\Psi^{\rm e}+A\Psi^{\rm e}-\frac{r^2f^2}{(\lambda+1)\Lambda}Q^{tt}, \\
 & H_{2}(t,r) =~\frac{\Lambda}{f^2}\left[\frac{\lambda+1}{r}\Psi^{\rm e}-K\right]+\frac{r}{f}\partial_rK, \\
 & H_{1}(t,r) =~r\partial_t\partial_r\Psi^{\rm e}+rB\partial_t\Psi^{\rm e}-\frac{r^2}{\lambda+1}\left[Q^{tr}+\frac{rf}{\Lambda}\partial_tQ^{tt}\right], \\
 & H_{0}(t,r) =~f^2H_{2}+fQ^\sharp,
\end{aligned}
\end{equation}
where
\begin{equation}
\begin{aligned}
    &A(r):=\frac{1}{r\Lambda}\left[\lambda(\lambda+1)+\frac{3M}{r}\left(\lambda+\frac{2M}{r}\right)\right],\\
    &B(r):=\frac{1}{rf\Lambda}\left[\lambda\left(1-\frac{3M}{r}\right)-\frac{3M^2}{r^2}\right]\,,
\end{aligned}
\end{equation}
and 
\begin{equation}
    Q^{tt/tr/\sharp}(t,r):= q^{tt/tr/\sharp}(t)\delta[r-r_p(t)]\,,
\end{equation}
with the corresponding magnitude $q^{tt/tr/\sharp}$ given in Eq.~\eqref{eq:qlm(t)}. Using the functional form for $\Psi^{\rm e}$ given in Eq.~\eqref{eq:TD extended homogeneous solutions}, the metric perturbation amplitudes take the form given by Eq.~\eqref{eq: generic_metric_amp} with
\begin{align}
    &K^\pm=f\partial_r\Psi^\pm+A\Psi^\pm,\quad K^S=0,\\
    &H_{0}^\pm=f^2H_{2}^\pm,\quad H_{0}^S=f_p^2H_{2}^S+f_pq^\sharp,\\
    &H_{1}^\pm=r\partial_t\partial_r\Psi^\pm+rB\partial_t\Psi^\pm ,\quad H_{1}^S=\mathcal{E}^2\frac{\dot{r}_p}{f_pU_p^2}q^\sharp,\\
    &H_{2}^\pm=\frac{\Lambda}{f^2}\left[\frac{\lambda+1}{r}\Psi^\pm-K^\pm\right]+\frac{r}{f}\partial_rK^\pm,\\
    &H_{2}^S=r_p[\![\partial_r\Psi]\!]_p+\frac{r_pA_p}{f_p}[\![\Psi]\!]_p\,.
\end{align}

\subsection{Odd parity}
The odd-parity $l\geq 2$ metric perturbation amplitudes can be reconstructed via
\begin{align}
    &h_0(t,r)=\frac{f}{2}\partial_r\left(r\Psi^{\rm o}\right)-\frac{r^2f}{2\lambda}p^t_{lm}(t)\delta[r-r_p(t)],\\
    &h_1(t,r)=\frac{r}{2f}\partial_t\Psi^{\rm o}+\frac{r^2}{2\lambda f}p^r_{lm}(t).
\end{align}
Using the functional form of $\Psi^{\rm o}$ given in Eq.~\eqref{eq:TD extended homogeneous solutions}, and plugging the relevant expressions into the above equations, the odd-parity metric perturbation amplitudes again take the form of Eq.~\eqref{eq: generic_metric_amp} with 
\begin{align}
    h_0^\pm&=\frac{f}{2}\partial_r\left(r\Psi^\pm\right)\,,\quad h_0^\mathrm{S}=0\,,\\
    h_1^\pm&=\frac{r}{2f}\partial_t\Psi^\pm\,,\quad h_1^\mathrm{S}=0\,.
\end{align}

\subsection{Monopolar and dipolar gravitational perturbations}\label{app: Monopole and dipole perturbations}
For the monopole case, $l = m = 0$, we adopt the gauge choice $H^{00}_{2} = K ^{00} = 0$, following Refs.~\cite{Zerilli:1970wzz, Hopper:2015icj}. By solving the linearized Einstein equations, one then obtains that the remaining non-zero metric perturbation amplitudes are given by~\cite{Hopper:2015icj}
\begin{align}
H_{0}^{00}&=4\sqrt{\pi} m_p\left[\frac{\mathcal{E}}{r}-\frac{f}{\mathcal{E}f_pr_p}\left(2\mathcal{E}^2-U_p^2\right)\right]\Theta[r-r_p(t)]\,,\\
H_{1}^{00}&=\frac{4\sqrt{\pi} m_p\mathcal{E}}{f^2r}\Theta[r-r_p(t)]\,.
\end{align}

For the even-parity dipole case $\{l, m\} = \{1, 1\}$, we use the Zerilli gauge choice $K^{11} = 0$. The non-zero amplitudes are then given by~\cite{Hopper:2015icj}
\begin{widetext}
\begin{align}
\begin{split}
 H_{0}^{11}&=-2 m_p\sqrt{\frac{2\pi}{3}}\frac{rf_p}{f} \left[\frac{\mathcal{E}r_p}{r^3}+\frac{6\mathcal{L}^2M+6Mr_p^2-3\mathcal{L}^2r_p+\left(2\mathcal{E}^2-3\right)r_p^3}{\mathcal{E}r_p^5}-\frac{6i\mathcal{L}\dot{r}_p}{r_p^3}\right]e^{-i\varphi_p(t)}\Theta[r-r_p(t)],\\
 H_{2}^{11}&=-\frac{2\sqrt{6\pi} m_p}{rf^2r_p}\left(i\mathcal{L}f_p^2-\mathcal{E}r_p\dot{r}_p\right)e^{-i\varphi_p(t)}\Theta[r-r_p(t)], \\
  H_{1}^{11}&=-\frac{2\sqrt{6\pi} m_p\mathcal{E}r_pf_p}{r^2f^3}e^{-i\varphi_p(t)}\Theta[r-r_p(t)].
\end{split}
\end{align}
\end{widetext}
The amplitudes for the $\{l, m\} = \{1, -1\}$ mode can be obtained via the relation $H_{a}^{1,-1} = -(H_{a}^{11})^*$ (with $a=0, 1,$ or $2$). This property follows from the general symmetry of the metric perturbation amplitudes, $\mathcal{M}^{l,-m} = (-1)^m (\mathcal{M}^{l,m})^*$, which ensures that the reconstructed metric perturbation remains real-valued in the time domain.

In the odd-parity dipole case $\{l, m\} = \{1, 0\}$, the residual gauge freedom is used to set $h^{10}_1 = 0$ \cite{Zerilli:1970wzz}. The only remaining non-zero amplitude is
\begin{equation}
h_0^{10}=4 m_p\mathcal{L}\sqrt{\frac{\pi}{3}}\left(\frac{1}{r}\Theta[r-r_p(t)]+\frac{r^2}{r_p^3}\Theta[r_p(t)-r]\right).
\end{equation}

\newpage
\bibliography{References}

\end{document}